\documentclass[ %
				apj,
				iop,
				revtex4,
				numberedappendix,
				twocolappendix
				]{emulateapj}
\pdfoutput=1
\usepackage{head}
\usepackage{adjustbox}
\usepackage[bordercolor=white,backgroundcolor=yellow,linecolor=black,colorinlistoftodos,textwidth=2.5cm]{todonotes}
\def\Eq#1{Eq.~(\ref{#1})}				
\def\Tab#1{Tab.~\ref{#1}}				
\def\Fig#1{Fig.~\ref{#1}}				
\def\Sec#1{Sec.~\ref{#1}}				
\def\App#1{Appendix~\ref{#1}}			
\def\eq#1{Eq.~(\ref{#1})}				
\def\tab#1{Tab.~\ref{#1}}				
\def\fig#1{Fig.~\ref{#1}}				
\def\app#1{Appendix~\ref{#1}}			
\newcommand{\n}[1]{{\num{#1}}}			
\newcommand{\p}{\partial}				
\newcommand{\de}{{\rm d}}				
\newcommand{\const}{\mathrm{const.}}
\newcommand{\eps}{\epsilon}			
\def\mean#1{\left< #1 \right>}			
\def\t#1{\mathrm{#1}}					
\def\mustBeEq{\overset{!}{=}}		 	

\newcommand{\eX}{\hat{x}}
\newcommand{\eY}{\hat{y}}

\usepackage{environ}
\NewEnviron{equationColor}{%
\begin{equation}{
\setlength{\fboxsep}{6pt}
\setlength{\fboxrule}{1.5pt}
\fcolorbox{MainColor!40}{white!0}{$\BODY$}}
\end{equation}}
\NewEnviron{equationColor*}{%
\begin{equation*}{
\setlength{\fboxsep}{6pt}
\setlength{\fboxrule}{1.5pt}
\fcolorbox{MainColor!40}{white!0}{$\BODY$}}
\end{equation*}}

\DeclareSIUnit\scaleheight{H}
\DeclareSIUnit\astronomicalunit{au}
\DeclareSIUnit\au{au}
\DeclareSIUnit\orbit{\Omega^{-1}}

\newcommand{\pencilCode}{{\textsc{PencilCode}}}

\newcommand{\cs}{c_\mathrm{s}}

\newcommand{\stokes}{\mathrm{St}}					

\newcommand{\scaleheight}{H}						

\newcommand{\Npar}{N_\mathrm{par}}					

\newcommand{\Tmax}{T_\mathrm{max}}
\newcommand{\Torb}{T_{\rm orb}}
\newcommand{\torb}{\Torb}

\newcommand{\tauS}{\tau_{\rm s}}			
\newcommand{\tauF}{\tauS}					
\newcommand{\tauColl}{\tau_{\rm coll}}		
\newcommand{\tauCorr}{\tau_\mathrm{corr}} 	

\newcommand{\lCorr}{l_\mathrm{corr}}	
\newcommand{\lc}{\lCorr}		
\newcommand{\lcorr}{\lCorr}	

\newcommand{\rhoDust}{\rho_\mathrm{d}} 	
\newcommand{\rhoDustInit}{\rho_{\mathrm{d,}0}}
\newcommand{\rhoGas}{\rho_\mathrm{g}} 	
\newcommand{\rhoGasInit}{\rho_{\mathrm{g,}0}}

\newcommand{\rhoInt}{\rho_\bullet}

\newcommand{\epsInit}{\eps_0}
\newcommand{\einit}{\epsInit}
\newcommand{\epsMax}{\epsilon_\mathrm{max}}
\newcommand{\emax}{\epsMax}

\newcommand{\urms}{u_\mathrm{rms}}
\newcommand{\vrms}{v_\mathrm{rms}}		

\newcommand{\domain}{\circ}				
\newcommand{\local}{\Box}	

\newcommand{\sg}{\sigma_\domain}

\newcommand{\slo}{\sigma_\local}

\newcommand{\zg}{\zeta_\domain}

\newcommand{\zlo}{\zeta_\local}

\newcommand{\Sd}{\mathrm{Sc}}					
\newcommand{\Sc}{\Sd}							

\newcommand{\dx}{\delta_{\mathrm{x}}}
\newcommand{\dz}{\delta_{\mathrm{z}}}

\hypersetup{pdfauthor=Andreas Schreiber}
\shorttitle{Azimuthal and vertical streaming instability at high dust-to-gas ratios on the scales of planetesmial formation}
\shortauthors{A. Schreiber et al.}
\begin{document}
\title{Azimuthal and Vertical Streaming Instability at High Dust-to-gas Ratios\\ and on the Scales of Planetesimal Formation}
\author{Andreas Schreiber$^{*}$}
\author{Hubert Klahr} 
\affiliation{Max Planck Institute for Astronomy, K{\"o}nigstuhl 17, 69117 Heidelberg, Germany}
\email{aschreiber@mpia.de}
\altaffiltext{*}{Fellow of the International Max Planck Research School for Astronomy and Cosmic Physics at the University of Heidelberg (IMPRS-HD)}

\begin{abstract}%
The collapse of dust particle clouds directly to $\si{\kilo\meter}$-sized planetesimals is a promising way to explain the formation of planetesimals, asteroids and comets. In the past, this collapse has been studied in stratified shearing box simulations with super-solar dust-to-gas ratio $\epsilon$, allowing for streaming instability (SI) and gravitational collapse. 
This paper studies the non-stratified SI under dust-to-gas ratios from $\epsilon=0.1$ up to $\epsilon=1000$ without self-gravity.
The study covers domain sizes of $L=\SI{0.1}{\scaleheight}$, $\SI{0.01}{\scaleheight}$ and $\SI{0.001}{\scaleheight}$, in terms of gas disk scale height $\si{\scaleheight}$, using the \pencilCode. They are performed in radial-azimuthal (2-d) and radial-vertical (2.5-d) extent. The used particles of $\stokes=0.01$ and $0.1$ mark the upper end of the expected dust growth. 
SI-activity is found up to very high dust-to-gas ratios, providing fluctuations in the local dust-to-gas ratios and turbulent particle diffusion $\delta$. 
We find an SI-like instability that operates in $r$-$\varphi$ even when vertical modes are suppressed. This new azimuthal streaming instability (aSI) shows similar properties and appearance as the SI.
Both, SI and aSI, show diffusivity at $\eps=100$ only to be two orders of magnitude lower than at $\eps=1$, suggesting a $\delta\sim \eps^{-1.}$ relation that is shallow around  $\epsilon \approx 1$. The (a)SI ability to concentrate particles is found to be uncorrelated with its strength in particle turbulence. Finally, we performed a resolution study to test our findings of the aSI. This paper stresses out the importance of properly resolving the (a)SI at high dust-to-gas ratios and planetesimal collapse simulations, leading else wise to potentially incomplete results.
\end{abstract}
\maketitle
%
%
%
%
\section{Introduction}
\label{sec:intro}
Planets form in protoplanetary disks (PPDs) around newborn stars. But, the processes that transform dust to $\si{\kilo\meter}$-sized planetary precursor material, so called \textit{planetesimals}, are still under debate. Planetesimals are defined as the first objects which are gravitationally bound, this typically happens at sizes above several $\si{\kilo\metre}$, see \cite{Benz1999}. A promising formation scenario is the gravitational collapse of dense particle clouds or filaments, originating from the idea by \cite{Safronov1972} and \cite{Goldreich1973} of a gravitational unstable disk mid-plane. Another scenario is one of direct growth from sticky dust-dust-collisions, see \cite{Weidenschilling2000} and \cite{Kataoka2013}, which is not the focus of this paper.

In our picture the coagulation process is stopped at the drift and fragmentation barrier (\cite{Birnstiel2012}), producing $\si{\milli\metre}$- to $\si{\centi\metre}$-sized dust. The gravitational cloud collapse then directly transforms these particles into approximately $\SI{100}{\kilo\meter}$-sized planetesimals (\cite{Morbidelli2009}) via gravitational collapse of massive particle clouds. Robust ways to form such particle clouds are a main research topic in the field of planet formation theory, since the expected dust-to-gas column density ratios of $Z\approx0.01$ for the solar nebula are itself insufficient to trigger collapse (\cite{Bai2010}). Hence, one is in need of mechanisms that accumulate dust efficiently into a local disk patch, increasing the dust-to-gas ratio up to values allowing collapse to happen. But, this collapse can come to a stall by other processes, hindering the final collapse, such as by turbulent diffusion and aerodynamic erosion (\cite{Cuzzi2010}). Already diffusion can define a size criterion on the collapse of a particle cloud as estimated by \cite{Klahr2015}.

The streaming instability (SI), found by \cite{Youdin2004}, hereafter YG04, is a dust-gas instability that emerges once dust-to-gas volume ratio reaches unity. It originates from the velocity difference between dust and gas in their equilibrium state (\cite{Nakagawa1986} and \cite{Weidenschilling1987}) as a result of frictional coupling. The SI should not only be seen as a process enhancing the dust concentration, but also as one that introduces additional turbulence right at the scales of planetesimal formation. 

In the work by \cite{Johansen2015} and by \cite{Simon2016}, cloud collapse to a planetesimal has been achieved by reducing the amount of gas artificially. By that, the sedimentation of the dust to the disk mid-plane is the driving dust concentration mechanism. But, for this scenario to happen, higher dust-to-gas ratios are needed than expected and one is in need for other dust concentration mechanism.
Possible mechanisms are gas flow features, capable of collecting particles locally, acting as particle traps, as can happen azimuthally symmetric in zonal flows (\cite{Dittrich2012}), ice lines (\cite{Kretke2007}), dead zone edges (\cite{Dzyurkevich2010}), or locally in vortices (\cite{Raettig2015}. The trapping mechanism itself is often a result of a bump in the radial gas pressure gradient 
\begin{equation*}
\eta= \frac{1}{2} \left(\frac{H}{R}\right)^2 \frac{\de \ln \rho}{\de \ln R}\, ,
\end{equation*}
with $\si{\scaleheight}=c_\mathrm{s}/\Omega$ the gas disk scale height.
In the case of a zonal flow, where the particle inward drift comes to a halt, this is because the gas is orbiting locally close to and faster than the Keplerian velocity. In the case of a vortex, trapping is the result of attractive net forces (\cite{BargeSommeria1995}). It has been shown that these traps are locations with high dust-to-gas ratio together with a minimum in the gas pressure gradient. For the case of zonal flows, they have only a vanishing gas pressure gradient in a small radial extent, thus the surrounding of this point can be SI-active due to the 2nd derivative of $\nabla P$, see \cite{Laibe2017}.

The collapse of a particle cloud itself is barely investigated. Hence, this work studies the pure SI in an environment similar to what is expected to occur in dust rich regions. As show in e.g., \cite{Johansen2007a}, \cite{Dittrich2012} and \cite{Raettig2015} the SI is active in and near particle traps. Thus, in this work we set a constant, non-vanishing gas pressure gradient $\eta$, see \Sec{sec:sim_setup}, and study the SI at high dust-to-gas volume ratios $\epsilon = \rhoDust/\rhoGas$. For collapse to happen, dust densities at Hill density (\cite{Hamilton1991}) are needed and that is expected to be at a dust-to-gas ratios of 10 to 100. From simulations on dust growth (\cite{Birnstiel2010}) and analyses of particle trapping (see above), one can expect particles with a Stokes number of $\stokes \approx 0.01$ to $0.1$ to be the most prominent dust species to get trappe. Thus, we investigate both here in this paper. The Stokes number is a measure of the stopping time $\tauF$ in terms of orbital frequency $\Omega$:
\begin{equation}
\label{eq:stokes}
\stokes = \tauF \Omega
\end{equation}
The stopping time, sometimes also called friction time, for particles in the Epstein drag regime is given by
\begin{equation}
\label{eq:epstein}
\tauF = \frac{\rhoInt a}{\rhoGas \cs},
\end{equation}
with particle size $a$, internal particle density $\rhoInt$, gas density $\rhoGas$ and speed of sound $\cs$. The Stokes number is a measure of particle size and $\stokes=0.1$ roughly translates to particles sizes of around $\SI{0.1}{\meter}$ at $\SI{5}{\au}$ in a \textit{Minimum Mass Solar Nebula} (\cite{Hayashi1981}).
\subsection{Effects of streaming instability and sedimentation}
The term 'planetesimal formation by streaming instability' has been used quite confusingly lately, so we try to quantify things a little. The origin of understanding the SI lies in the derivation of a dispersion relation instability criterion by YG04 operating in $r$-$z$ direction. The work of YG04 found the SI to not operate in purely radial modes, i.e. where $k_x\ne0$ and  $k_y=0$, $k_z=0$, in which situation they only find radial dispersion. Further analysis of this instability by \cite{Youdin2007} and \cite{Johansen2007} (JY07 in the following) in 2-d simulations with radial-vertical extend, and 3-d simulations, identified a turbulent non-linear behavior that limits the ability of the SI to concentrate particles. Besides \cite{Raettig2015}, the SI has not been found in the $r$-$\varphi$ plane, neither in 2-d simulations nor analytically, specially since the azimuthal shear introduces time dependent radial wave numbers (\cite{Klahr2004}) making things complicated. Moreover, coming from the dispersion relation, the SI has always been thought to depend on the existence of vertical modes, making full 3-d simulations necessary. In this paper we will show that this is not the case.

What has been done recently are 3-d simulations and $r$-$z$ simulations that include dust sedimentation to the disk mid-plane by vertical gravity. \cite{Carrera2015} did a parameter study and looked for non-transient particle clumping as an indicator for SI-activity. But, one has to be careful with calling the increase of dust concentration an effect of the SI only, since without frictional back reaction of the dust onto the gas, the old picture of \cite{Safronov1972} and \cite{Goldreich1973} would be correct. It is the turbulent diffusivity of the SI, or Kelvin Helmholtz instability, see \cite{Bai2010} that is actually prohibiting sedimentation and fragmentation. We show in this paper, that the turbulent strength of the SI decreases with increasing dust-to-gas ratio, allowing the dust disk to become thinner and fragment. But, time stable localised dust clumping itself is not a guarantee for planetesimal formation nor an indicator for active SI.

Recent work by \cite{Squire2017a} found a new way in describing the SI as part of a resonant drag instability (RDI). In which, dust can get unstable in any suspended media with a relative motion, if this media allows for undamped oscillatory modes. The RDI can be used to separate the SI into two instabilities, one acting at low dust-to-gas ratios and one at high. RDI also links the SI to the settling instability, where relative motions between dust and gas are induced by settling of the dust to the disk mid-plane and convective motions, and shows that this settling instability has larger growth rates by an order of magnitude, suggesting the SI, or a very similar RDI instability, to drive the dynamics in collapsing particle clouds.
\subsection{Paper outline}
This work presents a parameter study of the radial-vertical SI and its azimuthal counterpart ($k_x\ne0, k_y\ne 0$, $k_z=0$). For the time being, i.e. as long as no detailed analysis exist, and because the radial-vertical SI and its azimuthal counterpart look so alike, we suggest calling it azimuthal streaming instability. The scope of this work lies on high dust-to-gas ratios and small scales, as expected before and throughout gravitational collapse of a particle cloud to a planetesimal. 

The paper starts in \Sec{sec:methods} by outlining the used model and simulation setup. \Sec{sec:quantities} then introduces the investigated quantities. In \Sec{ch:2druns} the results from all simulations are presented in subsections that each cover a specific set of particle species and domain alignment, i.e., $r$-$\varphi$ or $r$-$z$ for either $\stokes=0.1$ or $0.01$. \Sec{app:res_study} further shows a resolution study for the aSI for $\stokes=0.1$ particles. A discussion of the results and their implication on planetesimal formation ends the paper in \Sec{sec:discussion}. Fruther investigations together with a list of the simulation results can be found in the appendix.
\section{Methodology}
\label{sec:methods}
\subsection{Numerical model and method} %
For our investigation we use the open source  \pencilCode\footnote{\url{http://pencil-code.nordita.org/}}, see \cite{Brandenburg2001}, \cite{Brandenburg2002}, and \cite{Brandenburg2005} for details. The \pencilCode{} is a numerical solver, here used on a finite-difference code using sixth-order symmetric spatial derivatives and a third-order Runge-Kutta time integration. The simulations are done in the shearing-sheet approximation, a Cartesian coordinate system co-rotating with Keplerian frequency $\Omega$ at arbitrary distance $R$ from the star. Thus, all quantities have to be interpreted as being local, i.e., the shear is linearised as in \Eq{eq:linear_shear}, with $x$ the radial simulation frame coordinate. All quantities are dimension free therefore time and length can be chosen arbitrary, e.g., by defining the distance $R$ to the star. Time is expressed in local orbits $\left[t\right]=\Omega^{-1}$. The coordinate system $\left(x, y, z\right)$ can be identified as $\left(r-R_0, R_0 (\varphi - \varphi_0 - \Omega_0 t\right), z)$. We do simulation setups with $N_x$, $N_y$ and $N_z$ grid cells, but for each set of parameters, we either set the number of grid cells in vertical or azimuthal direction to one. We thus suppress modes in that specific direction, see \Sec{sec:sim_setup}. The boundary conditions are periodic in $y$- and $z$-direction, and shear-periodic in $x$-direction.

All particles used in the simulations are Lagrangian super-particles, each representing a swarm of identical dust particles that interacts with the gas as a group. Their properties, e.g., density, is smoothed out to the neighboring grid cells via the Triangular Shaped Cloud (TSC) scheme. See \cite{Youdin2007} for details on the implementation in the \pencilCode.
\subsection{Equation for solving the streaming instability}
The presented simulations solve the Navier-Stokes equation for the gas and the particle motion in a shearing box approximation on a Cartesian grid (\cite{Goldreich1965}, \cite{Hawley1992} and \cite{Brandenburg1995}). The gas velocity $\vec{u}$ relative to the Keplerian shear is evolved via its equation of motion
\begin{align}
	\label{eq:dyn_gas}
	\frac{\p \vec{u}}{\p t} + (\vec{u}\cdot\nabla)\vec{u}+u_{0,y}\frac{\p \vec{u}}{\p y} = \; \left(2\Omega u_y \eX-\frac{1}{2}\Omega u_x \eY \right) \notag \\
	-c_s^2\nabla\ln  \rhoGas - 2\Omega\eta\eX - \frac{\epsilon}{\tauF}\left[\vec{u}-\vec{v}\right] + f_\nu(\vec{u},\rhoGas),
\end{align}
with second and third terms on the left-hand side being the advection terms by the perturbed velocity and by the shear flow, respectively. On the right are the terms for Coriolis force, the pressure gradient with ${\nabla} P =c_s^2{\nabla} \rho$, the centrifugal support due to the global radial pressure gradient inside the PPD, the particle-gas drag interface and the viscosity term. Where we define the dust-to-gas ratio as $\epsilon=\rhoDust / \rhoGas$. The unperturbed Keplerian orbital velocity in the local Keplerian frame is 
\begin{equation}
\label{eq:linear_shear}
u_{0,y}=-\frac{3}{2}\Omega x.
\end{equation}
The gas density is evolved with the continuity equation:
\begin{equation*}
	\frac{\p \rhoGas}{\p t} + \nabla \cdot \left( \rhoGas \vec{u} \right) + u_{0,y} \frac{\p \rhoGas}{\p y}=  f_\mathrm{D}(\rhoGas)
\end{equation*}
The functions $f_\mathrm{D}(\rhoGas)$ and $f_\nu(\vec{u},\rhoGas)$ are the artificial hyper-diffusivity and -viscosity that ensure the stability of the \pencilCode. The latter is also responsible for shock viscosity, see \pencilCode{} manual and \App{app:hyperhyper} for further reading. The particles are evolved via
\begin{equation*}
	\frac{\p \vec{x}}{\p t} = - \frac{3}{2} \Omega x_p \eY + \vec{v}\, ,
\end{equation*}
with particle position $x_p$ and particle velocity $\vec{v}$ similarly to the gas velocity via
\begin{equation}
    \label{eq:dyn_dust}
	\frac{\p \vec{v}}{\p t} =  \; \left( 2\Omega v_y \eX-\frac{1}{2}\Omega v_x \eY \right) - \frac{1}{\tauF} [\vec{v}-\vec{u}(\vec{x})]\, ,
\end{equation}
but without pressure gradient acting on them.
\subsection{Simulation setups} %
\label{sec:sim_setup}
We are interested in the properties of the pure SI, hence we start with a best case scenario. Meaning, self-gravity is switched off for all simulations, as well as the vertical stellar gravity component.
Even in the large simulations with $L=\SI{0.1}{\scaleheight}$ we neglect stellar gravity, since we are interested in the properties of the pure SI, and a gravoturbulent situation would instead lead to dust density gradients and hence a non-homogeneous SI. We further neglect collisions. This is not completely justified, since the collision time scales with dust-to-gas ratio via $\tauColl \sim \stokes / \epsilon$. But including collisions would demand to specify the particle size $a_0/H$ as an additional parameter. Also, cross sections are only poorly defined for 2-d simulations. Still, as long as $\vrms/\cs < (6 \epsilon)^{-1}$ collisions should not be of importance, see \App{app:collTime}. We leave this open for further investigations, as we are interested in the general possibility of resolving SI activity at high dust-to-gas ratios and in $r$-$\varphi$ direction.

Our parameter study covers a dust-to-gas density ratio range from $\epsilon=0.1$ up to $\epsilon=1000$ in equidistant manner in log-space, plus a higher coverage for dust-to-gas ratios below unity. The simulation domain is 2-d in $r$-$\varphi$ and 2.5-d in $r$-$z$. In $r$-$\varphi$, the azimuthal velocity component is vital for the azimuthal streaming instability, because here the buoyancy of the local dust concentration couples via the gas pressure to stellar gravity and centrifugal acceleration, see \cite{Nakagawa1986}. In other words, essential for the streaming instability is the influence of the local dust concentration onto its respective equilibrium azimuthal velocity. 
And, as the instability does not have any 1-d unstable modes, one needs at least one additional spatial component. This can be the vertical component, like for the Kelvin-Helmholtz setup in \citep{Johansen2005}, or the radial component as in this work. The linear analysis \citep{Youdin2004} on the other hand has only been conducted for easier to handle axis-symmetric modes. For non-axissymmetric modes, radial wavenumbers are time-dependent due to winding up spirals in the Keplerian shear, making an analysis complicated. In the case of a radial-vertical setup, one ends up with 3 velocity components, but only 2 spatial dimensions, in our nomenclature these setups are 2.5-d. The simulation domain size is varied between $L=\SI{0.1}{\scaleheight}$, $\SI{0.01}{\scaleheight}$ and $\SI{0.001}{\scaleheight}$.

Each run uses 128 grid cells per covered direction and we initiate our runs with 10 particles per grid cell. This high number is needed, since we do not include sedimentation to a mid-plane and consequently do not concentrate particles into a horizontal SI-active layer. Instead we want the SI to be active in the whole simulation domain, similar to the setups in \cite{Johansen2007}. For each simulation we use a single particle species of $\stokes=0.01$ and $0.1$. Particles are initially randomly distributed matching an average density of $\rhoDustInit$ and are initiated in grid cell wise drag force equilibrium together with the gas. We use the following code units for our runs: Sound speed $c_\mathrm{s}\equiv 1$, calculations are in isothermal approximation via $\gamma\equiv 1$, i.e., $c_s^2 \sim T = \const$, densities are normalized to the mean gas density which is set to $\rho_\mathrm{gas,0}=1$, the orbital frequency is $\Omega\equiv 1$. We further set the gas pressure gradient to $\eta=0.05$, which is a typical value for the inner regions of the \cite{Hayashi1981} protoplanetary disk model, which scales gas pressure with $P\sim R^{-3/2}$.
\section{Investigated Quantities}
\label{sec:quantities}
\subsection{Particle diffusion: $\delta$}
\label{sec:diffusion} 
A main part of this paper is the measurement of the SI diffusivity $D$, which can be expressed in gas disk thickness $H$ and sound speed $c_\mathrm{s}$ as a dimensionless quantity
\begin{equation}
\label{eq:diffusioninorbs}
\delta = \frac{D}{c_\mathrm{s} H}\, ,
\end{equation}
i.e., code units, and we adopt this unit system throughout the paper. The diffusion is measured by tracking the position of a sample of at least $10^4$ super-particles and measuring their travel distance over time. The time derivative of the variance $\sigma_\mathrm{Gauss}^2$ of the resulting travel distance gives directly the diffusion by using 
\begin{equation}
\delta=\frac{1}{2}\frac{\p \sigma_\mathrm{Gauss}^2}{\p t},
\label{eq:diffusion}
\end{equation}
see JY07. This leads to a mean squared distance from the initial positions of $\langle r^2\left({t}\right)\rangle_x=D{t}$ after a time ${t}$. Only radial and vertical diffusivity can be measured by this method since shearing motions dominate in azimuthal direction.
\subsection{Particle dispersion: $\sigma$}
\label{sec:dispersion}
To further quantify the turbulent behavior of the particles, the shear-free root-mean-square (rms) of the deviation from particle mean velocity is measured via
\begin{equation}\label{eq:dispersion}
\sigma \equiv v_\mathrm{rms} = \sqrt{\frac{1}{\Npar} \sum\limits_{j}^{\Npar} \left| \vec{v}_{\mathrm{par},j} - \mean{\vec{v}_{\mathrm{par},j} }_x \right|^2},
\end{equation}
with $\Npar$ the number of particles and $\left< ... \right>_x$ the corresponding mean spatial value. $\vec{v}_{\mathrm{par},j}$ is the particle velocity minus its corresponding shear velocity from \eq{eq:linear_shear}. Hence, $\sigma$ is a measure for the turbulent dispersion. This quantity can be calculated \textit{globally}, meaning for the whole simulation domain and indicated in the following by $\domain$, or \textit{locally}, meaning for a single simulation grid cell, indicated by $\local$.
\subsection{Correlation time: $\tauCorr$}
\label{sec:tcorr}
Comparing the measurment methods for $\delta$ with $\sigma$, one sees that our method for estimating $\delta$ has the drawback of only being able to give a single value for the whole simulation domain and not a local diffusivity at a certain spot. Thus, it would be preferable to measure the local turbulent particle dispersion $\slo$ and link it to a local diffusivity $\delta_\local$ via a correlation time
\begin{equation}
\tauCorr \coloneqq \delta/\sigma^2 \quad \Rightarrow \quad \delta_\local = \tauCorr  \slo^2 .
	\label{eq:tauCorr}
\end{equation}
This is only true under the assumption that $\tauCorr$ is constant on all turbulent scales for a whole SI-active simulation. The correlation time can be derived from assuming the turbulent diffusion to be Fickian process, see \cite{Johansen2006b}. Assuming $u_k$ to be the turbulent velocity amplitude on the length scale $l_k$, over which the turbulent eddy transport $D_k$ is occuring, then $D_k = u_k l_k$. One can approximate $l_k = u_k \tau_k$, with $\tau_k$ the eddy lifetime. In this equation $\sqrt{u^2}/\cs$ is the Mach number, hence the diffusion coefficient should scale with $\sigma^2$. Averaging over all scales leads to \Eq{eq:tauCorr} where $T_k \equiv \tauCorr = \mathrm{const}$. 
\subsection{Particle drift: $\zeta$}
To investigate the behavior of the particles with respect to the underlying gas velocity, we use the particle drift $\zeta$ as rms deviation of the particle velocity $\vec{v}_\mathrm{par}$ from the gas velocity $\vec{u}_\mathrm{gas}$ at the particle location $\vec{x_i}$
\begin{equation}\label{eq:drift}
\zeta = \sqrt{\frac{1}{\Npar} \sum\limits_{i}^{\Npar} \left| \vec{v}_{\mathrm{par},i} - \vec{u}\left(\vec{x}_i \right) \right|^2},
\end{equation}
with $\Npar$ the total number of particles within a grid cell, or the total number of particles when evaluated globally. Thus, we distinguish between local drift $\zlo$ and global drift $\zg$. The interpolation of the gas velocity at particle position is done via the TSC method. The drift then can be compared with the equilibrium drift as calculated in \cite{Nakagawa1986} and \cite{Weidenschilling1987}.
\begin{equation}
\label{eq:nakagawa_solution}
\begin{array}{r@{}l}
\mathrm{gas:} \quad  \vec{u}_\mathrm{N} & = \left[ u_x, u_y, u_z \right]^\intercal = \left[2\stokes \epsilon\lambda,\: -\left(1 + \epsilon + \stokes^2\right) \lambda,\: 0\right]^\intercal \eta v_\mathrm{K}\vspace{2mm}\\
\mathrm{dust:} \quad \vec{v}_\mathrm{N} & = \left[ v_x, v_y, v_z \right]^\intercal = \left[-2\stokes \lambda,\: -\left( 1+\epsilon \right) \lambda,\: 0\right]^\intercal \eta v_\mathrm{K}
\end{array}
\end{equation}
with simplification
\begin{equation*}
\lambda = \frac{1}{(1+\epsilon)^2+\stokes^2}.
\end{equation*}
The Nakagawa drift in this paper is the absolute difference in speed of particles relative to the gas and vice versa:
\begin{equation}
\label{eq:nakagawa_drift}
\zeta_\mathrm{Nakagawa} = \left| \vec{v_\mathrm{N}} - \vec{u_\mathrm{N}} \right|
\end{equation}
\subsection{Viscous stress $\alpha$ and Schmidt number}
The $\alpha$-value is a measure for the turbulent strengh of the disk gas. For our simulations it is calculated by setting the Reynolds stress equal to an artificial equivalent viscous stress
\begin{equation}
\langle u'_x u'_y\rangle = \nu \nabla \cdot \vec{u},
\end{equation}
where we use the perturbation theory notation, i.e., perturbations from the mean flow $u$ are primed, $u'$. This viscosity $\nu$ that originates from Reynolds stress, is canonically written in the form of $\alpha$ viscosity, by defining $\alpha = \nu/(c_\mathrm{s} H)$ (\cite{Shakura1973}). Using Eq.~22 from \cite{Klahr2003} and simplifying via $\langle u'_x u'_y\rangle = \langle u_x u_y\rangle - \langle u_x\rangle \langle u_y\rangle$ and $\nabla \vec{u} = - (3/2) \Omega$ from the linearised shear approximation (\Eq{eq:linear_shear}), $\alpha$ can be expressed in a form valid in a shearing box with underlying linear gas transport (here Nakagawa drift) as
\begin{equation}
\label{eq:alpha}
\alpha = \frac{-2\left( \langle u_x u_y\rangle - \langle u_x \rangle \langle u_y \rangle \right)}{3 c_\mathrm{s}^2},
\end{equation}
if one is assuming a constant gas density. In our case this is valid since our measured gas density fluctuations have at maximum an amplitude of $10^{-3}$, with respect to the mean density. This is in agreement with our measurements of the gas rms speed, i.e. 
\[
\frac{\rho'}{\rho} \approx \frac{\vrms^2}{\cs^2}
\]
with perturbation in dust density $\rho'$. The rms speed never exceeds Mach numbers of more than a few percent. The averaging $\langle  . \rangle$ in \eq{eq:alpha} is done in space and time. The second term in this equation is non-negligible, since one needs to subtract the gas drift motion induced by the particle-gas-interaction, as this is not contributing to the Reynolds stress. For example, in local simulations of the magnetorotational instability (MRI), where $\langle u_x \rangle = 0$, the latter term can be dropped.

Additionally, we investigate the Schmidt number, defined as the ratio of radial momentum transport against radial mass diffusion.
\begin{equation}
\label{eq:schmidt}
\Sd = \alpha / \delta_x
\end{equation}
Note that some authors assume $\delta=\alpha$, which is rarely given, see \cite{Johansen2006b}.
%
%
%
\section{Results of the Parameter Study I:\\$\stokes=0.1$ particles}
\label{ch:2druns}
This section presents the results of the four parameter studies, two of which for $\stokes=0.1$ and two for $\stokes=0.01$ particles. For each Stokes number we performed a set of simulations in $r$-$\varphi$ and $r$-$z$ direction which span from dust-to-gas ratios of $\epsilon=0.1$ to $1000$ and cover shearing box sizes of $L=0.1$, $0.01$ and $0.001$, with a numerical resolution of 128 grid cells.

We find the appearance of the SI in $r$-$z$ as expected and also an apparently similar turbulent dust instability in $r$-$\varphi$ that is so far only known from \cite{Raettig2015}. We formally call it \textit{azimuthal streaming instability} (aSI), since no dispersion relation for this instability has been solved yet and its relatedness to SI and RDI remains to be proven. Still, in this paper we will show strong similarities of the aSI with the SI, leading us to pose the question if it is not the SI in both cases, or another form of the resonant drag instability, as described in \cite{Squire2017}.

For $\epsilon > 1$, the non-stratified SI is expected to grow faster than for $\eps \leq 1$, but on larger wavenumbers, see YG04 and \cite{Youdin2007}. Once reaching a particle dominated environment, one would expect the SI to get eventually suppressed. We found for large $\eps$ that even when on large scales (i.e., {$L\geq\SI{0.1}{\scaleheight}$}) the SI might seem dead, but on smaller scales the SI remains active, providing both dust density fluctuations $\emax/\einit$ and diffusivity $\delta$, with $\emax$ the maximum occurring grid-wise dust-to-gas ratio and $\einit=\rhoDustInit/\rhoGasInit$ the initial dust-to-gas ratio. We find this to be true even for dust-to-gas density ratios above $\epsilon\geq100$. We further find the presence of SI to depend not only on grid resolution but also on possible underlying numerical hyper-viscosity/-diffusivity scheme, see \App{app:hyperhyper}. Caution has to be given not to artificially suppress the SI.

A typical timeseries of $\mathrm{max}(\rhoDust/\rhoDustInit)$ in our parameter studies has clear saturation levels. An example is shown in \fig{fig:01_xy_rhopmax_timeseries}, here plotted as maximum occurring dust-to-gas ratio $\epsMax$, which is equivalent to $\mathrm{max}(\rhoDust/\rhoDustInit)$ since $\rhoGas\approx 1$ for all times and for all simulations. All measurements discussed in the following, e.g., diffusivities or $\alpha$-values, are performed in this saturated state, i.e., after a time stable maximum in the dust density is reached. In the following all error bars mark the standard deviation of the corresponding quantity. The individual simulation results can be found in the appendix for the $r$-$\varphi$ runs in \Tab{tab:r-phi-runs} and for the $r$-$z$ runs in \Tab{tab:r-z-runs}. 
\FIGURE
  {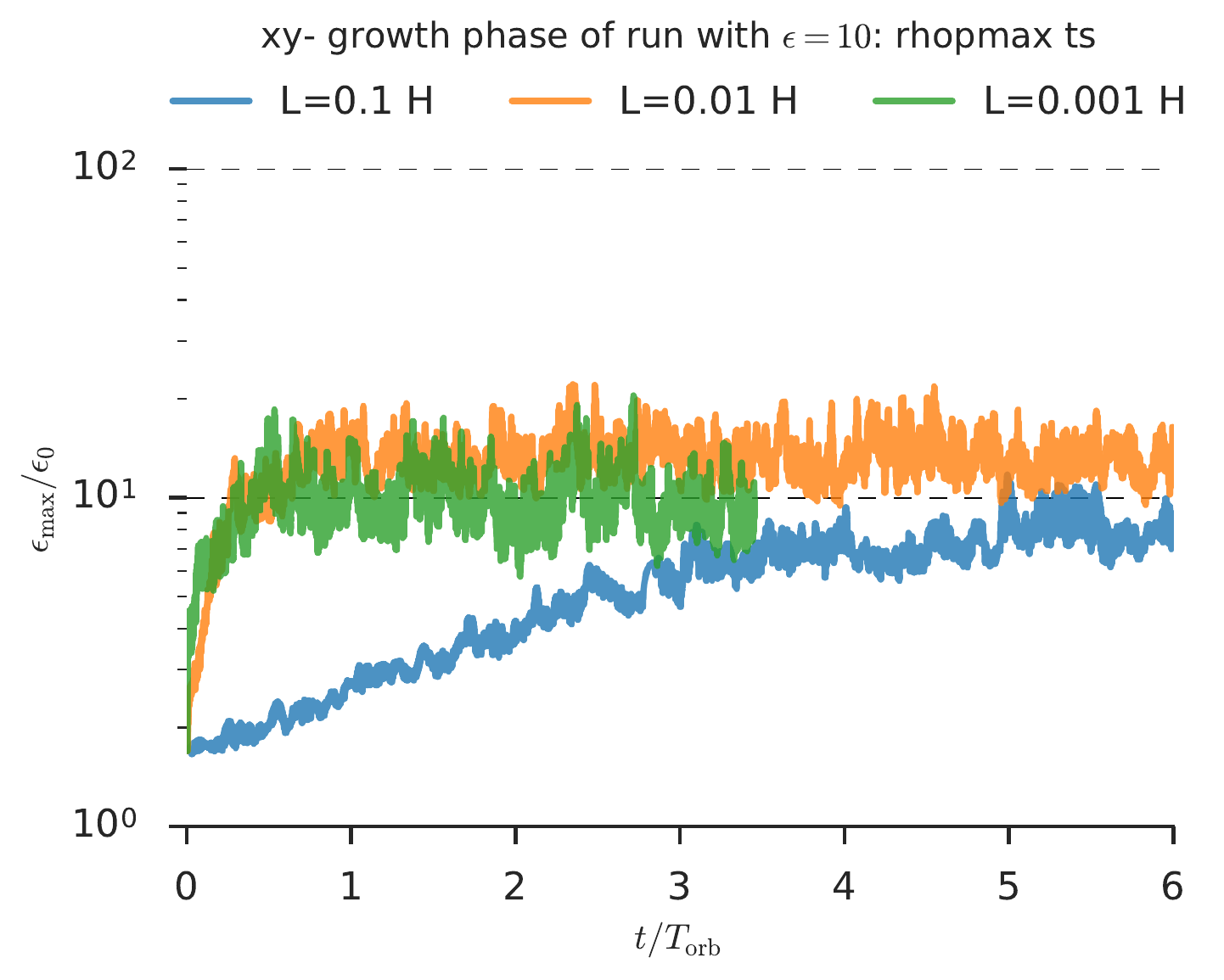}
  {fig:01_xy_rhopmax_timeseries}
  {An example timeseries of the maximum particle concentration for the $r$-$\varphi$ simulation parameter set with $\stokes=0.1$ and $\epsInit=10$. All three simulations show saturation to a dust density fluctuation value around $10$, see \Fig{fig:01_xy_dust_enhancement}. The shown time is restricted for the first few orbits in order to resemble the SI growth phase, which is much faster for the simulations on the smaller length scales (green and orange).}
  {1}{{.3cm, .3cm, .3cm, .7cm}} 
\subsection{$\stokes=0.1$ - $r$-$\varphi$ plane}
\label{sec:01-xy}
The achieved simulation durations range up to 100 orbits, see appendix. The simulations on $L=\SI{e-3}{\scaleheight}$ scales reached only a few orbits, since time-stepping is tiny on these scales. Assuming similar growth rates $s$ for our case study as for linear SI-modes in $x$-$z$ (see \cite{Youdin2007}) of around $s\approx0.1\Omega \dots 1\Omega$, for $\epsilon=1\dots10$ and $\stokes=0.1$, this would indicate that our simulations ran at least on the order of several SI growth rates.
\subsubsection{Dust density fluctuations and growth rates}
\FIGURE%
  {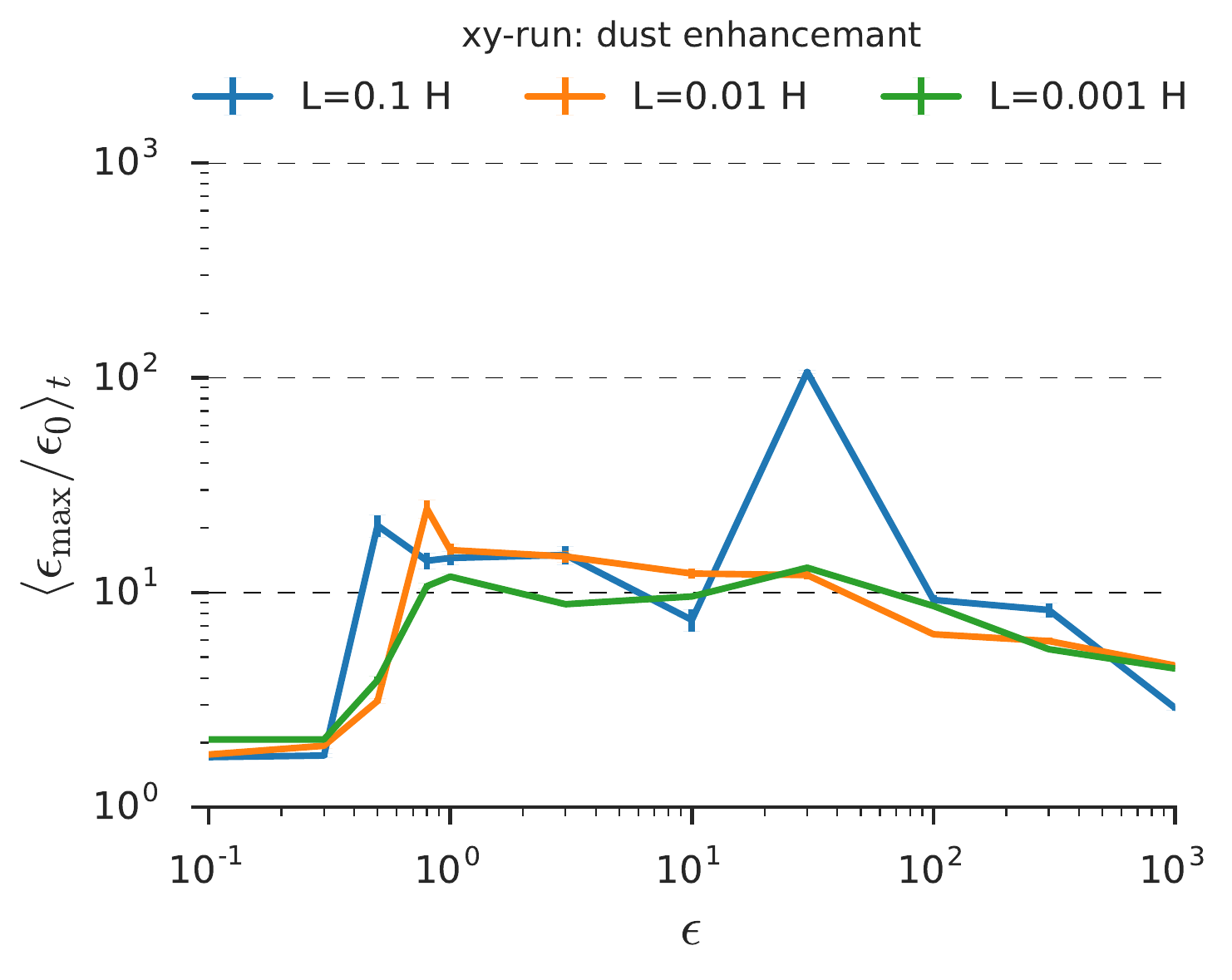}%
  {fig:01_xy_dust_enhancement}%
  {$\stokes=0.1$ - $r$-$\varphi$ plane: Dust density fluctuation values for all simulations. The simulations with active aSI manage to enhance their dust concentration locally to a value around $10$. In the case of ($L=\SI{0.1}{\scaleheight}$, $\epsInit=30$) zonal flows emerge that further concentrate from $\emax/\einit \approx 10.5$ in the saturated state up to a value of $100$.}%
  {1}{{.3cm, .3cm, .3cm, .7cm}}%
The active aSI is enhancing the dust-to-gas ratio locally and likewise creates particle voids. 
We take the $\emax/\einit$ time\-series to calculate the normalized mean maximum dust-to-gas ratio and plot this in \fig{fig:01_xy_dust_enhancement}. For most simulations we find the dust density to be enhanced by a factor of $\approx 10$. Only for the special case of ($\stokes=0.1$, $r$-$\varphi$, $L=\SI{0.1}{\scaleheight}$, $\epsInit=30$) do we find zonal flows to emerge and peak dust densities go up to $\eps \approx 100$, see \Sec{sec:01_xy_zonalFlow}. We further discuss the influence of numerical resolution on this property in \Sec{app:res_study}. In \fig{fig:01_xy_dust_enhancement} one sees that the aSI in $r$-$\varphi$ direction has an active range from $\epsInit=0.8$ up to $\epsInit \geq 300$, meaning the aSI is able to concentrate dust locally significantly higher than the mean value. For $\epsilon=1000$, the aSI seems to be dead by having only dust density fluctuations by a factor of $\approx3$ on the largest scale, but on the two smaller scales remains active by fluctuations in the dust density of $\approx 5$. Furthermore, the aSI has a surprisingly sharp cut off at low dust-to-gas ratios and emerges first on the largest scale (blue line at $\epsilon=0.8$) and then on the smaller scales at $\epsilon=1$. We further find that for $L=\SI{e-2}{\scaleheight}$ and $\SI{e-3}{\scaleheight}$ the saturation level is mostly identical for all $\epsInit$.

\Fig{fig:01_xy_rhopmax_timeseries} shows saturation time to be the fastest on small scales. This confirms what can be found when calculating the analytic growth rates, via \cite{Youdin2004} or \cite{Squire2017a}, for the fastest growing mode. The fastest growing mode gets smaller and growth happens faster with increasing $\eps$. We estimated the growth rate $s$ in units of $\Omega$ by fitting a logistic function
\[ \eps \left( t \right) = A + \frac{B}{1+\mathrm{e}^{-s(t-t_0)}} \]
to our $\emax$ timeseries. The logistic function has an exponential growth for times shorter than the saturation time, i.e., $f(t \ll t_0) \approx A + B \mathrm{e}^{s(t-t_0)}$. We than derive from the function fit the growth rate via 
\[ s = \frac{4}{B} \frac{\de \epsilon}{\de t} \left(t=t_0\right) \,. \]
The measured growth rates for the aSI-active simulations we find to depend on the simulation domain size. In contrast, the growth rates only vary slightly with the initial dust-to-gas ratios. For the aSI-active simulations, they are on average: $s\left( L=\SI{0.1}{\scaleheight} \right) \approx \num{3e-1}\Omega$, $s\left( L=\SI{0.01}{\scaleheight} \right) \approx \num{1.0}\Omega$ and $s\left( L=\SI{0.001}{\scaleheight} \right) \approx \num{2.0}\Omega$. The individual growth rates can be found in \tab{tab:r-phi-runs}.
\subsubsection{End-state snapshots}
\FIGURE%
  {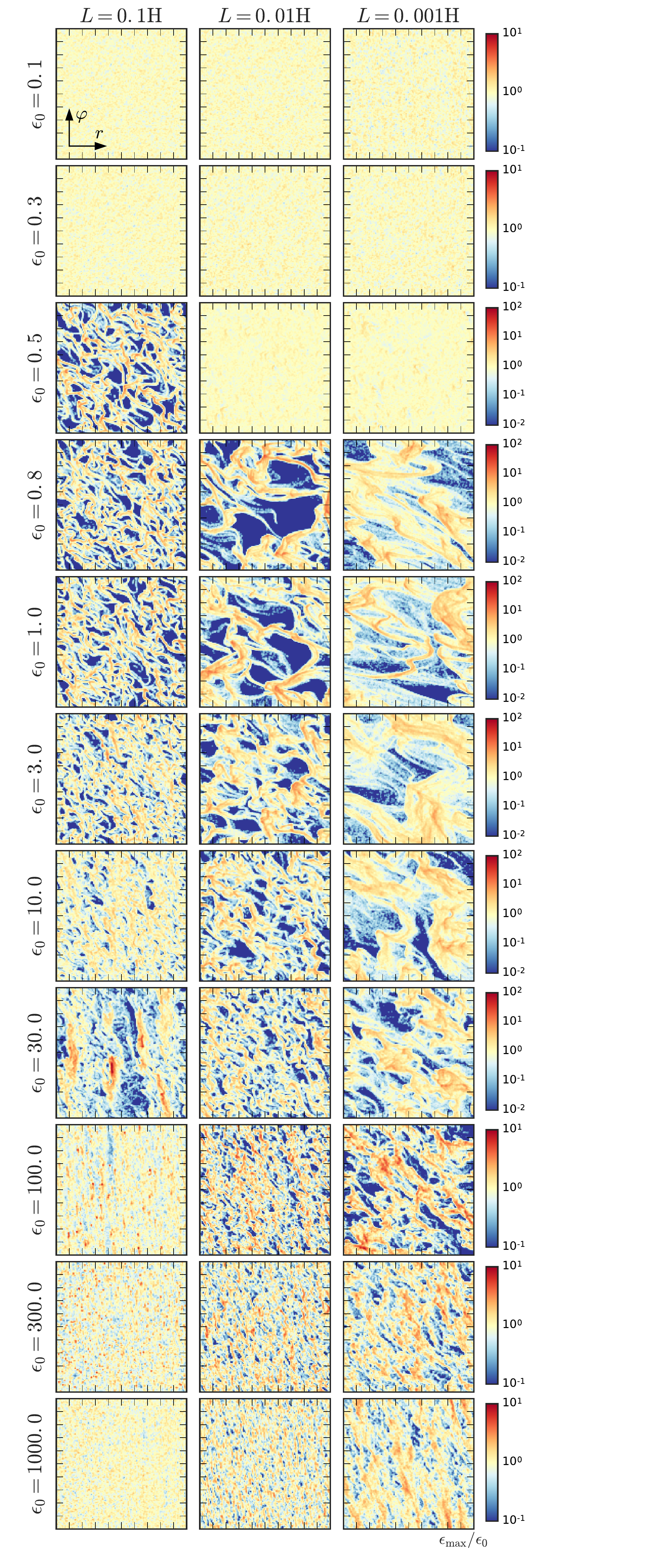}%
  {fig:01_xy_imggrid}%
  {$\stokes=0.1$ - $r$-$\varphi$ plane: Last snapshots of the dust-to-gas ratio normalized to $\einit$ (yellow). Over-densities are colored in red, particle voids in blue. For the $\epsilon$-range around unity, the aSI shows the same behavior on all scales, i.e., from left to right it appears like a zoom-in. In fact, the relevant modes are visually merely similar but identical in the ability to concentrate dust. For $\einit>30$ the aSI slowly vanishes on large scales, but remains active on small scales. For $\epsilon<1$ there is a sharp cut-off in aSI activity. All simulations have the same number of grid cells. Color mapping changes for high and low $\epsInit$.}%
  {.9}{{.2cm, .5cm, 4.cm, .2cm}}%
\fig{fig:01_xy_imggrid} shows the last snapshots of all simulations. The grid ticks mark a tenth of a box size $L$. This distance corresponds to the box size of the next smaller simulation, located to the right. Since the numerical resolution is $128^2$, the ticks of the smaller simulation mark approximately the grid resolution of the larger simulation.

The visible aSI-pattern is similar to the one of the SI, known from e.g., JY07, see following sections, though here in the $r$-$\varphi$ plane. If what is observed would be particle concentrations resulting out of initial random densities or velocities, it would lead to a non-length scale dependent pattern, as is seen in the upper two rows of \Fig{fig:01_xy_imggrid}.

Colored in red are the over-densities where dust is getting concentrated and in blue particle voids. The rows for $\epsilon=0.8$ till $10$ show an agreement in the wave pattern on all scales. Meaning, from left to right, one can not distinguish a smaller run from being just a zoom-in of the larger simulation by a factor of ten. This implies that no smaller wave modes become suddenly dominant, e.g., compare with $\stokes=0.01$ simulations in the low $\epsilon$ realm. For higher $\einit$ the aSI modes become smaller and one needs to go to small simulation grid sizes to resolve them.
The aSI does not die out, even at very high $\epsInit$, like $\eta=0$ would to, but becomes weaker with $\zeta_\mathrm{Nakagawa} \sim \eps^{-1}$. A global increase of $\epsilon$ has no effect on the global pressure gradient $\eta$. 
Local variations of $\eps$, especially in the non-linear (a)SI phase, can introduce an additional local pressure gradient which can lead to deviation from the mean rotation profile. This can be seen in the emerging zonal flows in \Sec{sec:01_xy_zonalFlow}. Still, in this situation we do not observe a decrease in aSI-activity, though dust-to-gas ratios increase by a great amount.
Additionally, \cite{Laibe2017} showed that a zonal flow with a locally vanishing pressure gradient does not prevent the SI from growing, as the second derivative of the pressure can also drive this instability.
\subsubsection{Particle diffusion - $\dx$}
\label{sec:01_xy_diff}
\begin{figure} %
	\centering %
	\hspace{0.6cm}\includegraphics[width=0.8\columnwidth]{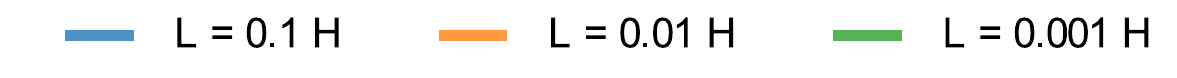}\\
	\includegraphics[width=\columnwidth, trim={0.2cm, 0.3cm, 0.2cm, 1.6cm}, clip]{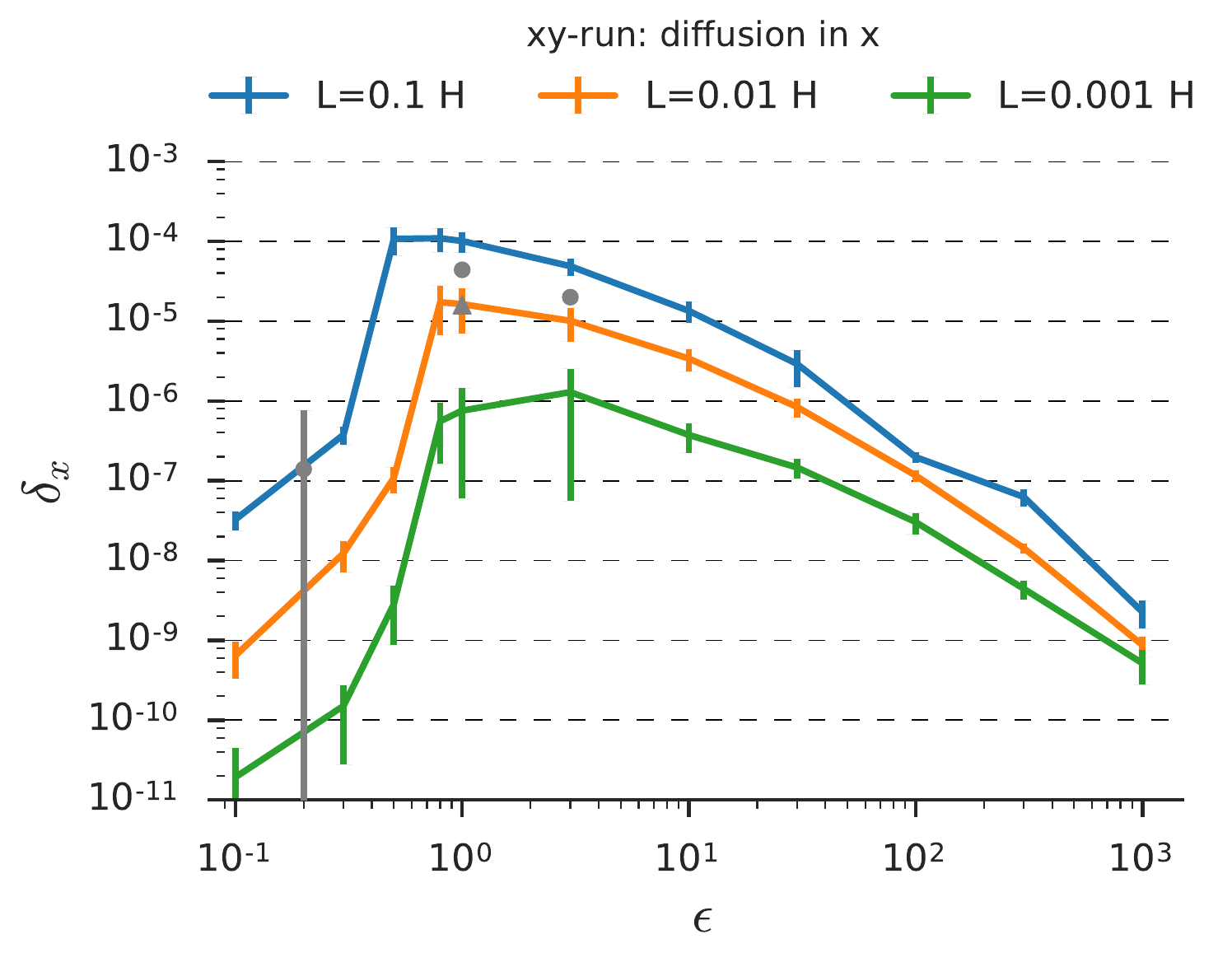}
	\caption{$\stokes=0.1$ - $r$-$\varphi$ plane: Radial particle diffusion $\dx$ estimated by treating the streaming instability as a random walk. When exceeding $\epsilon=30$, the aSI diffusivity shows a steeper drop on large scales (blue) than on smaller scales. The slope of the radial diffusion for this setup goes with $\eps^{-2}$ on the large scales and with $\eps^{-1}$ on smaller scales. For comparison plotted in grey circles (2D) and triangle (3D) are the diffusivities for $\stokes=0.1$ particles and $L\geq\SI{2}{\scaleheight}$, from JY07 Tab.~3. The found curve shape is in good agreement with the particle rms-velocity $\sigma$, shown in \Fig{fig:01_xy_dispersion}.} %
	\label{fig:01_xy_diffusion} %
\end{figure}%
Using \Eq{eq:diffusioninorbs} and \Eq{eq:diffusion}, the diffusivity $\dx$ is measured and plotted in \fig{fig:01_xy_diffusion} for each domain size over the initial dust-to-gas ratio. For comparison, radial diffusivity values from JY07 are plotted in grey. The plot shows a decrease in radial diffusivity with increasing particle load, and the slope goes with $\delta \sim \epsilon^{-1.0}$ up to $\eps^{-2.0}$, depending on simulation domain size. Other than the constant dust density fluctuations, the diffusivity drops steadily, not indicating a sudden aSI breakdown. %
The diffusion values found are similar to that of JY07 for 2-d radial vertical (grey dots) and also comparable to the diffusion from their 3-d simulation (grey triangle). All were found for the same particle Stokes number of $0.1$, but for larger simulation domain sizes of $L \geq \SI{2}{\scaleheight}$. Unfortunately, after JY07 no other study of SI measured particle diffusivities with which we could compare.
\subsubsection{Particle dispersion and drift - $\sigma$ and $\zeta$}
\label{sec:01_xy_dispNdrift}
\begin{figure} %
	\centering 
	\includegraphics[width=\columnwidth]{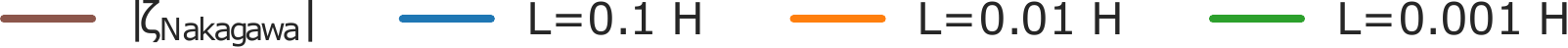}\\
	\begin{subfigure}
		[Global turbulent dispersion $\sg$ (lines) and local turbulent dispersion $\slo$ (contour).]
		{\includegraphics[width=\columnwidth, 
		trim={0.5cm, 0.3cm, 1.9cm, 1.6cm}, clip]
		{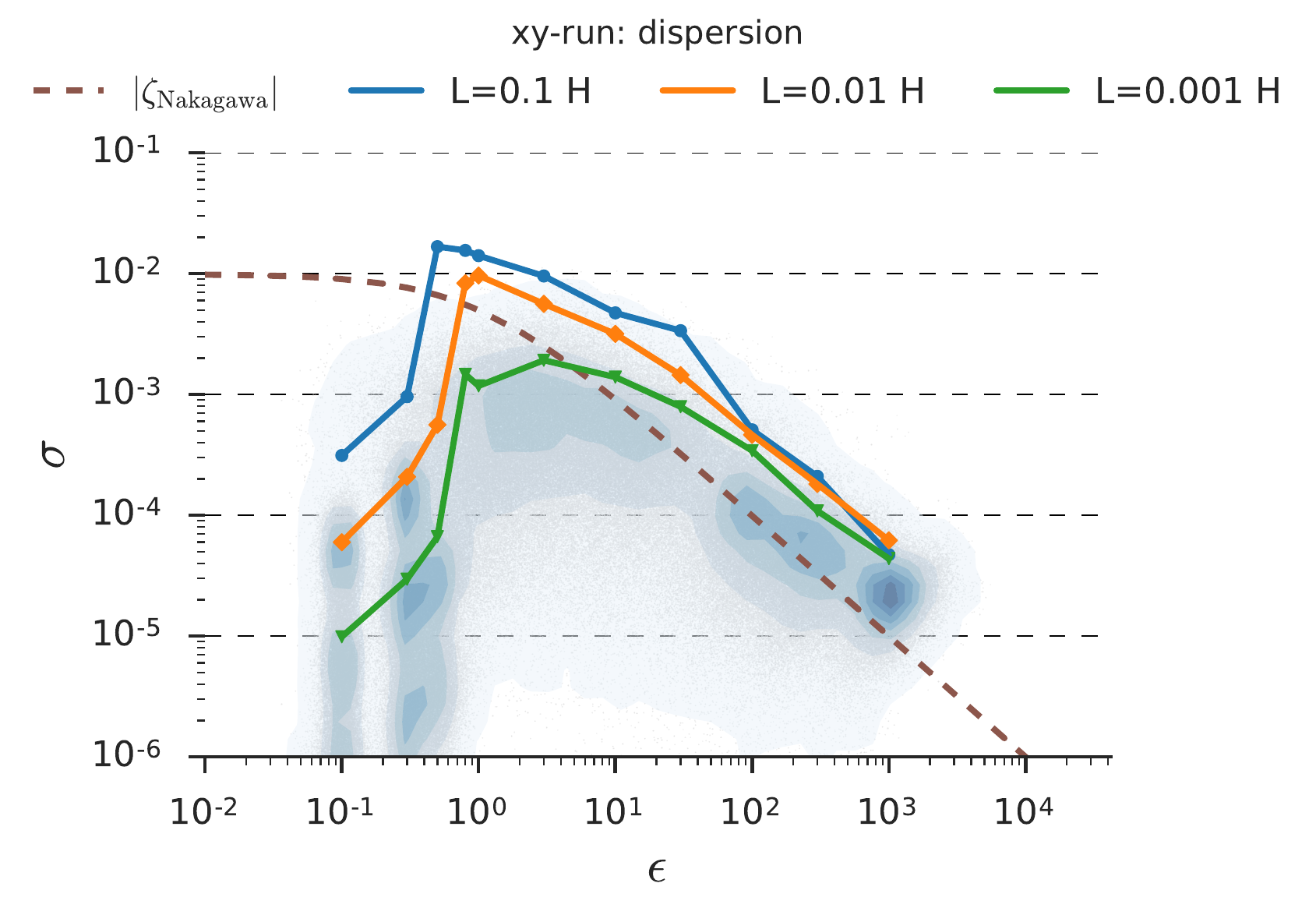}
	\label{fig:01_xy_dispersion}}
	\end{subfigure}\\
	\begin{subfigure}
		[Global drift $\zg$ (lines), local drift $\zlo$ (contour).]
		{\includegraphics[width=\columnwidth,
		trim={0.5cm, 0.3cm, 1.9cm, 1.6cm}, clip]
		{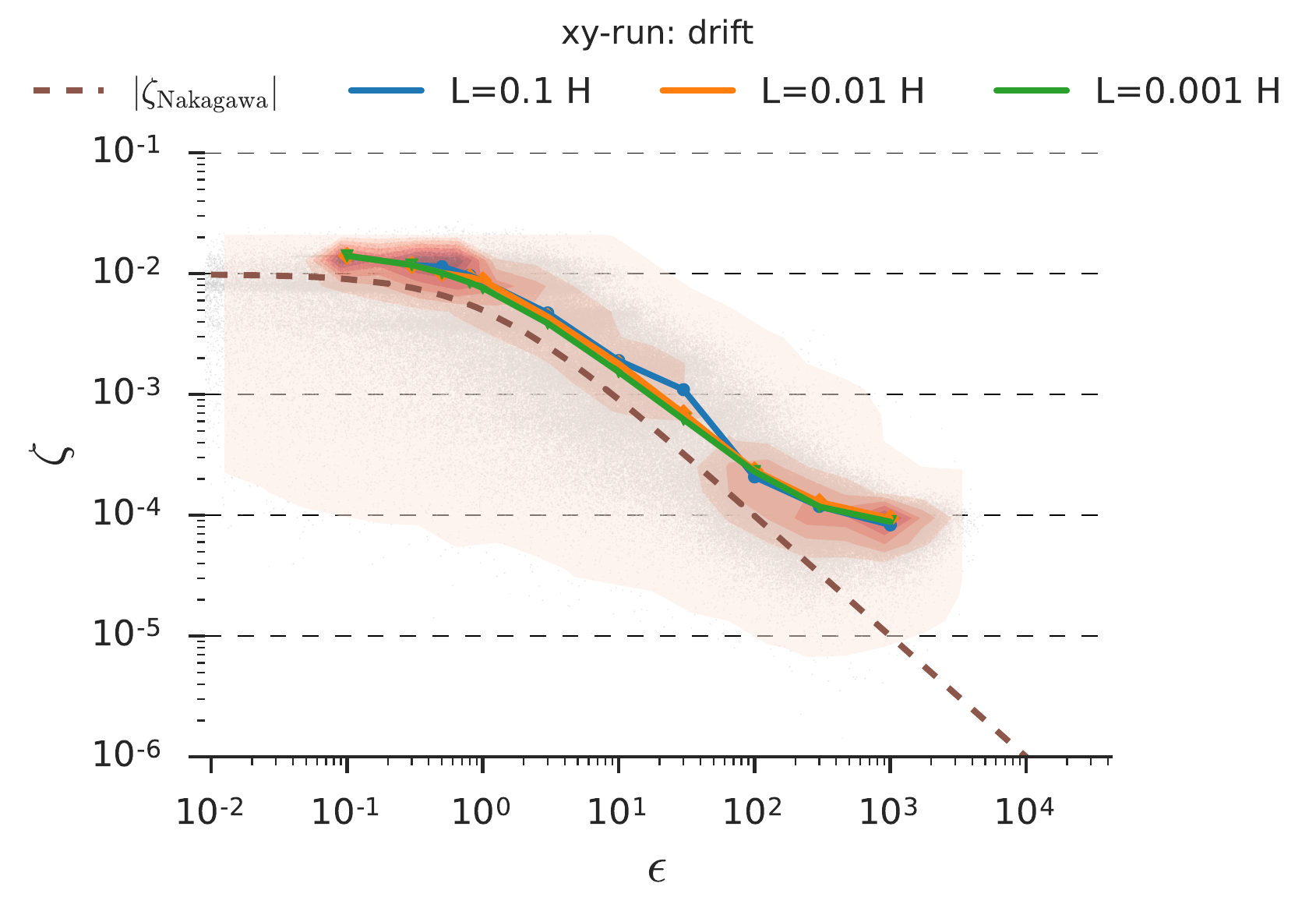}
	\label{fig:01_xy_drift}}
	\end{subfigure}
	\caption{$\stokes=0.1$ - $r$-$\varphi$ plane: Dispersion and drift for all simulations together. Blue, orange and green lines show the individual global values. The local, i.e., grid cell wise, values for all simulations combined are shown in shaded contours. As reference in dashed is shown the absolute magnitude of the Nakagawa drift speed from \Eq{eq:nakagawa_drift}. Global and local drift shows perfect agreement, whereas the local turbulent dispersion values are always well below the global values. The latter indicates that local particle groups move with similar velocity, but comparing two groups in distinct grid cells they move independently. The similarity in all $\zg$ indicates that particles and gas on all scales moves similar relative to each other, following Nagakawa drift prescription. But, clumps $\eps\geq10^2$ drift faster than actually predicted.} %
	\label{fig:01_xy_lc_drift_disp} %
\end{figure}%
In \fig{fig:01_xy_lc_drift_disp} global and local turbulent particle dispersion (\Eq{eq:dispersion}) and drift (\Eq{eq:drift}) are plotted. In the case of active turbulence by the (a)SI, the turbulent dispersion is a measure of the apparent turbulent velocity. \Fig{fig:01_xy_dispersion} compares global dispersion $\sg$ (colored lines) with local dispersion $\slo$ (shaded contours), were the local dispersion is calculated for each grid cell with two or more particles inside (grey dots), hence the scatter is large. 

The plot shows $\slo$ being on average much smaller than $\sg$, which is a result of the aSI having large extended modes, whereas grid cell wise the particles behave as a group, only slowly dispersing. For smaller $\epsilon_\local$-values the dispersion reduces as a result of the particle voids being the less turbulent regions (blue areas in \fig{fig:01_xy_imggrid}). For larger $\epsilon_\local$ the particles locally dominate with their momentum over the frictional influence of the gas, consequently particle groups stay longer together. In between the aSI is actively stirring the particles.

\label{sec:corrTime} %
Assuming a correlation between $\dx$ and $\sg$ via $\tauCorr$,  as stated in \Eq{eq:tauCorr} for aSI-active runs, we find $\tauCorr \approx 0.3 \Omega^{-1}$, though we here do not consider the run within the particle dominated regime, i.e., $L=\SI{e-1}{\scaleheight}$ and $\eps\geq30$, see \Sec{sec:01_xy_zonalFlow}.

\Fig{fig:01_xy_drift} shows the particle motion relatively to the gas. We find this drift $\zeta$ to be nearly identical on local and global scales and only marginally larger than the drift values of the steady state solution from \cite{Nakagawa1986}, see \Eq{eq:nakagawa_drift}. All particles were initially set to be in local Nakagawa drift equilibrium with the gas. That tells us that the aSI increases the particle drift speed by a factor of $\sim 2$, but still up to a factor of 100 times slower than without feedback.

We further find particles that group together at a very high dust-to-gas ratios of $\epsilon>100$, drifting one order of magnitude faster than what is predicted by \Eq{eq:drift}. 
We see in our simulations such particle heavy clouds radially drift inwards with significantly higher speed than the dust background does. This indicates a limit around $\epsilon=100$ on the validity of the Nakagawa equations.
\subsubsection{$\alpha$-value and Schmidt number}
\begin{figure} %
	\centering 
	\hspace{0.6cm}\includegraphics[width=.8\columnwidth]{legend_f.pdf}\\
	\begin{subfigure}
			[$\alpha$-value: turbulent gas viscosity]
			{\includegraphics[width=\columnwidth, 
			trim={0.2cm, 0.3cm, 0.2cm, 1.6cm}, clip]
			{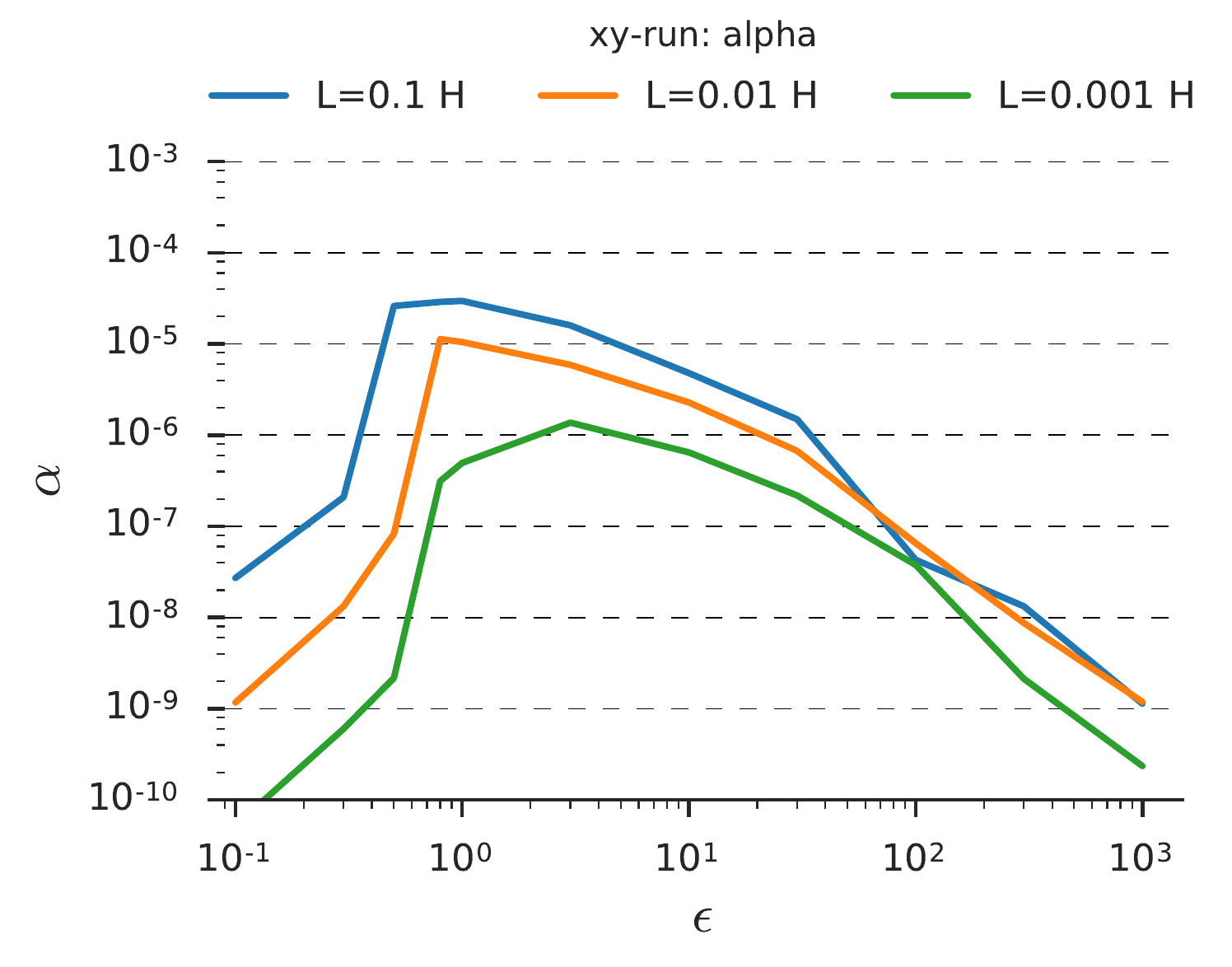}
	\label{fig:01_xy_alpha_paramplot}}
	\end{subfigure}\\
	\begin{subfigure}[Schmidt number]{\includegraphics[width=\columnwidth, trim={0.2cm, 0.3cm, 0.2cm, 1.6cm}, clip]{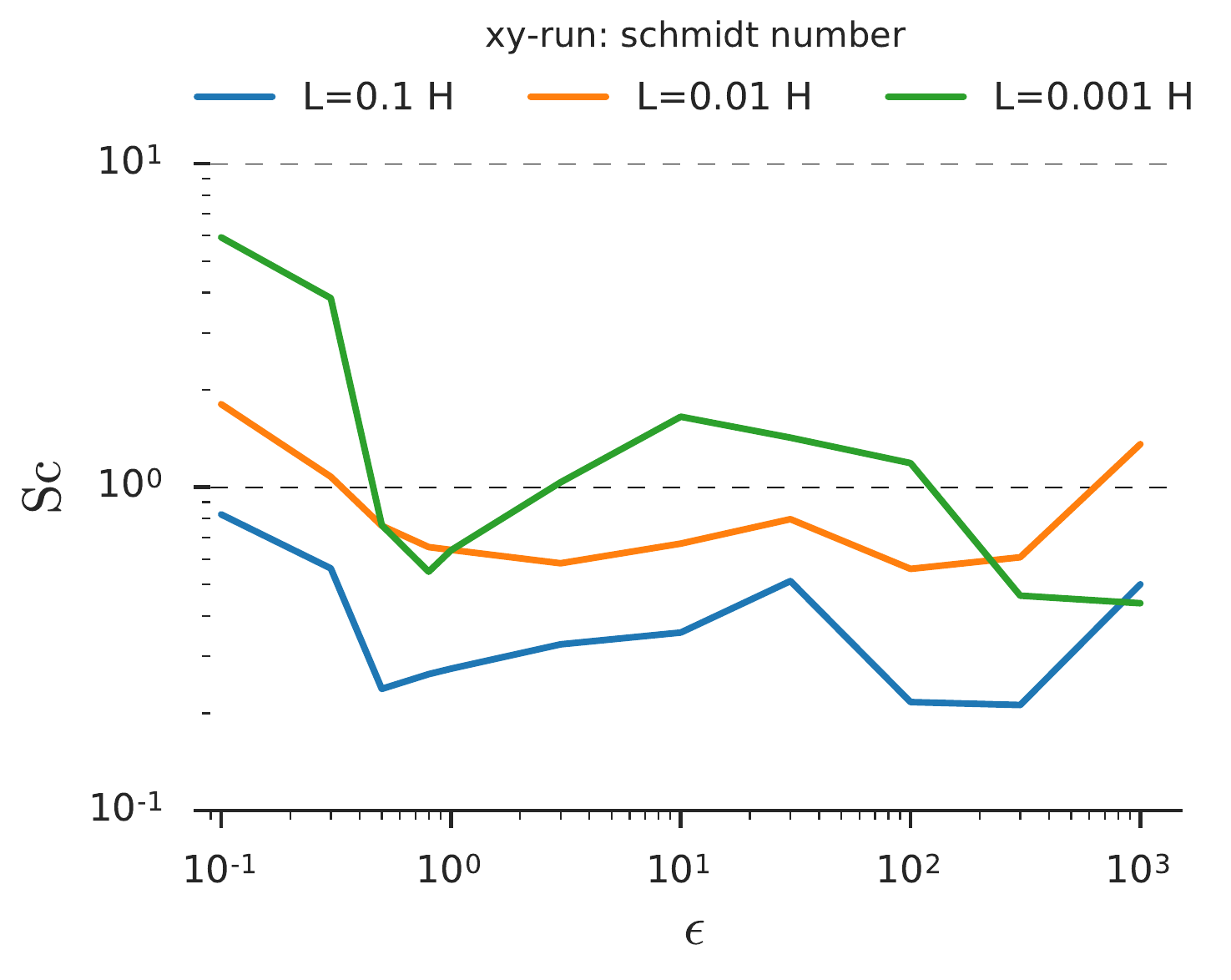}
	\label{fig:01_xy_schmidtnum}}
	\end{subfigure}
	\caption{$\stokes=0.1$ - $r$-$\varphi$ plane: As with most investigated quantities, the gas $\alpha$-turbulence shows a steep drop-off once the aSI is weaker at high dust-to-gas ratios. The Schmidt number in the case of active aSI is mostly below $\approx1$, showing that particle diffusion is in these cases stronger then the gas turbulence.} %
	\label{fig:01_xy_alpha_schmidt} %
\end{figure}%
\Fig{fig:01_xy_alpha_paramplot} shows the measured $\alpha$-values. Similar to the particle diffusion, the $\alpha$ gas turbulence shows a strong falloff with SI becoming inactive. We find $\alpha$ to drop towards higher $\epsilon$-values as strong as the diffusivity. 

One can see this better in the Schmidt number, plotted in \fig{fig:01_xy_schmidtnum}, which is the ratio of $\alpha$ transport against radial particle diffusion $\dx$, see \Eq{eq:schmidt}. This ratio shows a rather flat profile for the aSI-active range. We find the Schmidt number to depend on the size of the simulation domain. Generally, the particle turbulence is stronger or at least equality strong as the gas turbulence. Only on the smallest scales (green) the gas turbulence is slightly stronger within a larger fraction of the aSI-active range.

It may seem that $\alpha$ from the aSI is lower than the values known from MRI or the vertical shear instability. However, the $\alpha$-turbulence stemming from these large scale turbulence first needs to cascade down via the Kolmogorov cascade of gas turbulence, onto the considered scales of $L=0.1 H \dots 0.001 H$ or even $\de x = \n{e-3} \dots \n{e-5}$. Also, the dust load has to be taken into account when comparing, additionally weakening the turbulent strength, since for $\epsilon > 0$, the momentum of initially pure gas turbulence needs to pass over onto the dust-gas mixture. This is further discussed in \app{app:kolmogorov}.
\subsubsection{Special case: Zonal flows in ($L=\SI{0.1}{\scaleheight}$, $\epsInit=30$)}
\label{sec:01_xy_zonalFlow}
\FIGURE%
  {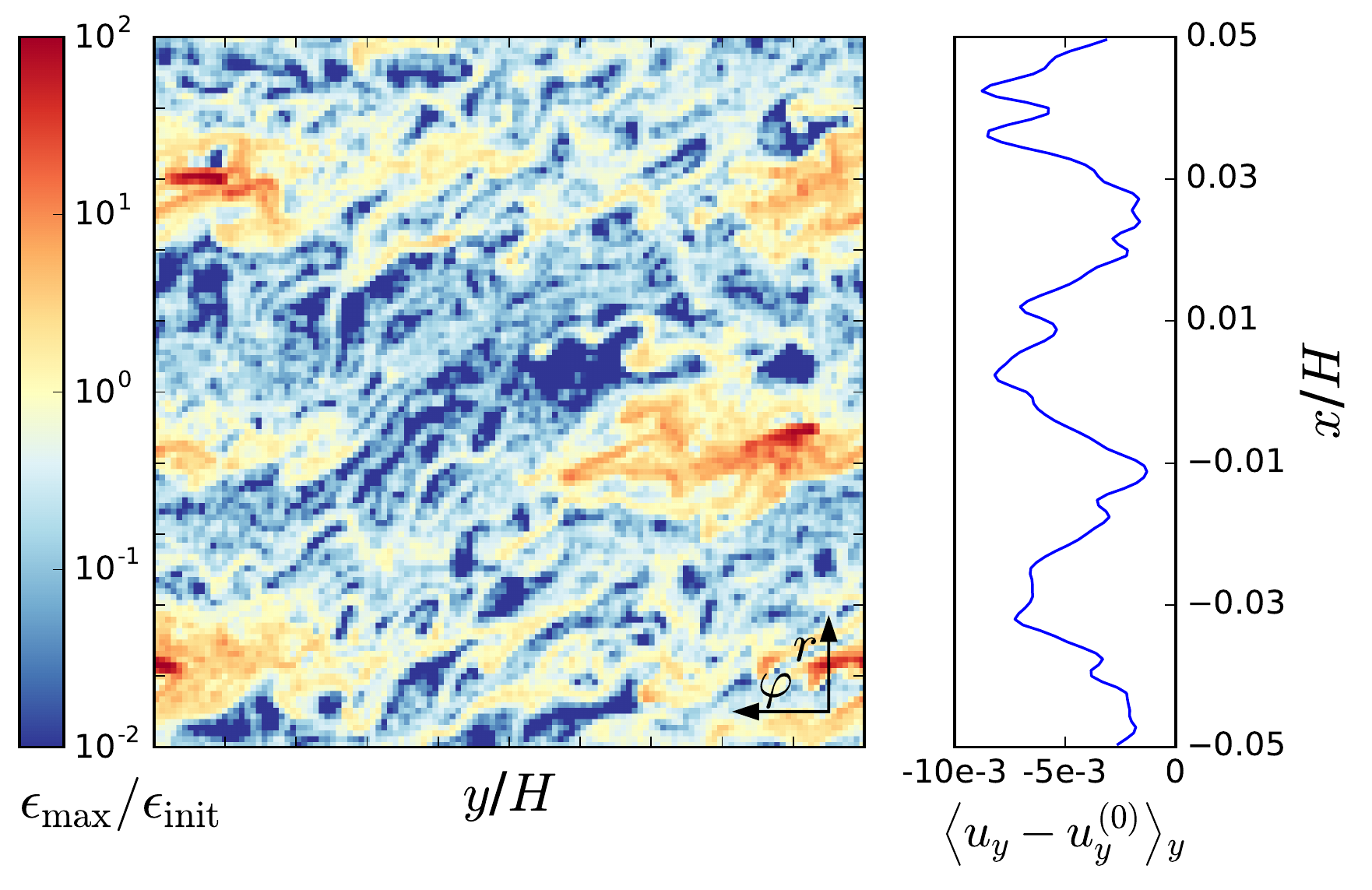}%
  {fig:01_xy_01e30ZF}%
  {$\stokes=0.1$ - $r$-$\varphi$ plane: The ($L=\SI{0.1}{\scaleheight}$, $\epsInit=30$) run is able to produce three zonal flows with strong particle clustering. Left shows the normalized particle density, right shows the underlying azimuthally averaged gas flow perturbation. The flows emerge after $t=30 \Torb$ and further concentrate particles up to $\eps=100$ and do not show up in other simulations. Snapshot is taken at $t=120\Torb$.}%
  {1}{{0, 0, 0, 0}}%
In this special case, we observed the aSI to saturate in a time between $20$ and $30$ orbits on a comparable similar dust density fluctuation level of $10$ as in all other aSI-active runs. After 30 orbits, the dust further concentrates reaching peak dust densities above $\emax/\einit=100$. This concentration happens in local non-axisymmetric particle heaps residing in zonal gas flows, see \fig{fig:01_xy_01e30ZF}. They are stable for the rest of the simulation duration and unaffected by the underlying aSI. %
These zonal flows are around $\SI{0.02}{\scaleheight}$ to $\SI{0.03}{\scaleheight}$ in width and are limited in azimuthal extend by no more than $\SI{0.08}{\scaleheight}$. Similar structures could not be observed in other runs for this parameter setup, probably because its radial wavenumber is larger than the largest wavenumber fitting into the next smaller simulation, but also for the cases with $\epsilon\geq 100$ we do not observe a similar phenomenon. One could argue that these bands would disappear if the spatial resolution is increased, therefore a resolution study for the whole parameter set is shown in \Sec{app:res_study}. Whether these zonal flows are similar to the ones observed in \cite{Carrera2015} needs to be shown in future work.

One might expect particle trapping in these elongated over-densities and hence a decrease in particle mobility and diffusivity, but averaged over the whole simulation domain, these band structures do not affect the global diffusivity to a significant amount. We find for $20 < t < 30 \torb$ a diffusivity of %
\[\dx= \left( {2.62 \pm 1.35} \right) \cdot \num{E-6},\] %
whereas in the case of the fully developed zonal flows %
\[\dx= \left( {3.71 \pm 2.14} \right) \cdot \num{E-6}.\] %
An additional reason might be that the mixing time of particles to get into and out of the heap is comparably short, because of aSI being fully active even within the over-densities.

The global particle drift $\zg$ in this case is lightly increased as the particle heaps have a significant higher dust-to-gas ratio and consequently radially drift faster, as seen in the dip in the blue line for $\epsInit=30$ in \Fig{fig:01_xy_drift}.
\subsection{$\stokes=0.1$ - $r$-$z$ plane}
\label{sec:01-xz}
In the $r$-$z$ plane, we observe the known non-linear SI but explore larger $\epsilon$ than usual. We find it to have surprisingly similar properties as the aSI in $r$-$\varphi$ from the previous section.
\subsubsection{Dust density fluctuations and growth rates}
\FIGURE%
  {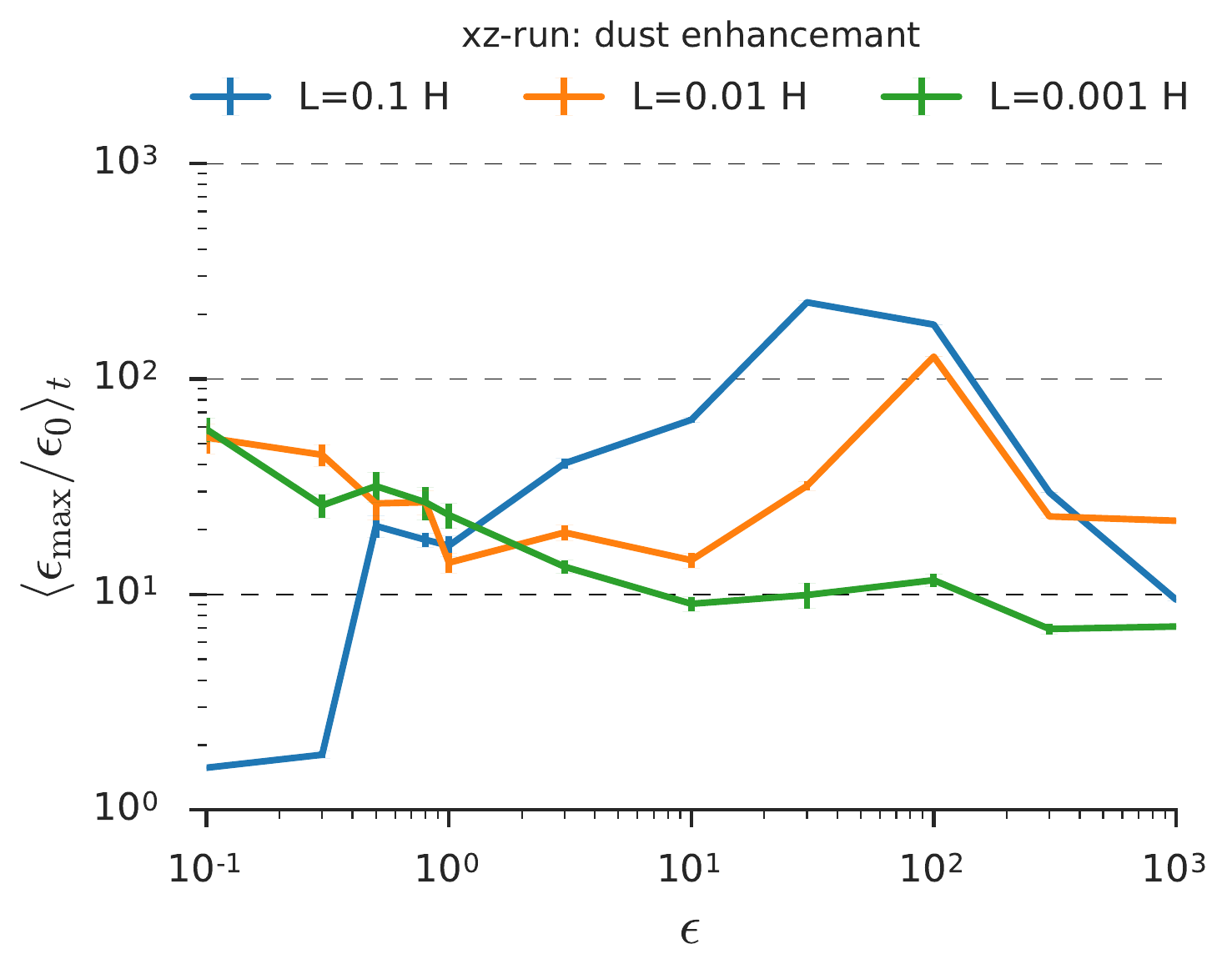}%
  {fig:01_xz_dust_enhancement}%
  {$\stokes=0.1$ - $r$-$z$ plane: All simulation with active SI manage to enhance their dust concentration locally well above $10$. Emerging vertical modes and horizontal band are capable to further enhance the dust-to-gas ratio to values up to $\approx200$.}%
  {1}{{.3cm, .3cm, .3cm, .7cm}}%
The achieved maximum dust-to-gas ratios in the case of pure SI are similar to the one of the aSI in $r$-$\varphi$, see \Fig{fig:01_xz_dust_enhancement}. But in many cases the simulations are dominated by horizontal or vertical modes, see \Fig{fig:01_xz_imggrid}. They strongly concentrate particles, similar to the zonal flows that were discussed in \Sec{sec:01_xy_zonalFlow}. In many cases the measured peak dust densities thus exceed the aSI values. These dominant horizontal bands appear in most of the investigated SI-active simulations and are a consequence of the chosen simulation domain sizes. One sees this, e.g., in the row for $\epsInit=3$ in \Fig{fig:01_xz_imggrid}, where there is a strong horizontal band in the smallest simulation (right), but a typical non-linear SI mode behavior in the next larger run (center) without a band-like structure. The largest of the three simulations (left) then shows the presence of a strong vertical mode that looks different if one goes to even larger domain sizes, see run BC in Fig.~3 in JY07, where there are many vertically aligned particle concentrations, but no clear single mode structure resides. %
One can follow this also up by looking at the green curve in \Fig{fig:01_xz_dust_enhancement}. It shows the emergence of a single mode dominated behavior when following this curve from large (right) so small (left) initial dust-to-gas ratios. In our setups, the many-mode turbulent SI is capable of enhancing the dust-to-gas ratio only up to the typical value of $\approx 10$. It is then the presence of single horizontal modes that induce stronger particle clumping. This is also true for the largest simulations (blue) where, towards higher dust-to-gas ratios, vertical modes appear that come along with high particle trapping therein, that reach up to values of $\epsMax/\epsInit \geq 100$.

The measured growth rates for the SI-active simulations we find to primarily depend on the simulation domain size. The growth rate only slightly varies with intial dust-to-gas ratio. For the SI-active simulations, the growth rates are on average: $s\left( L=\SI{0.1}{\scaleheight} \right) \approx \num{4e-1}\Omega$, $s\left( L=\SI{0.01}{\scaleheight} \right) \approx \num{8e-1}\Omega$ and $s\left( L=\SI{0.001}{\scaleheight} \right) \approx \num{1.5e0}\Omega$.
\subsubsection{End-state snapshots}
\FIGURE%
  {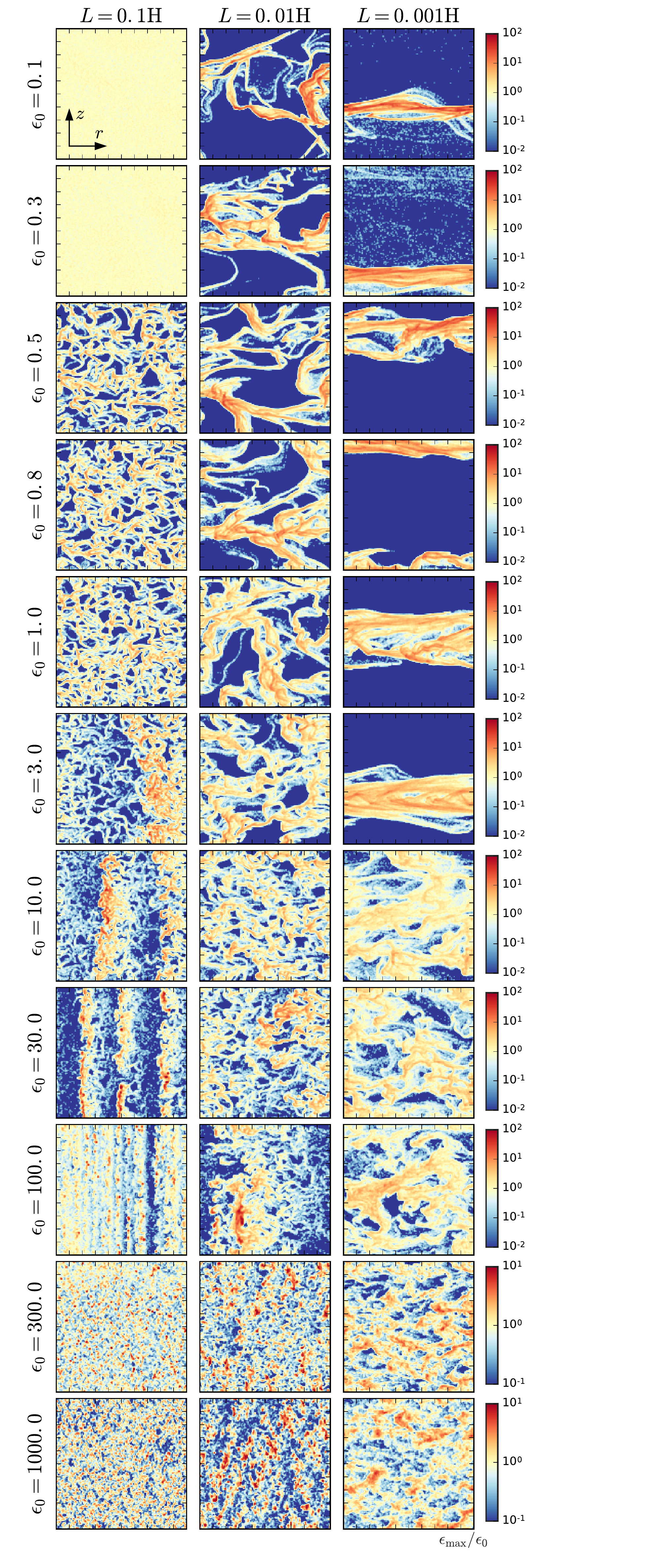}%
  {fig:01_xz_imggrid}%
  {$\stokes=0.1$ - $r$-$z$ plane: Last snapshots of the dust-to-gas ratio normalized to $\einit$ (yellow). Over-densities are colored in red, particle voids are colored in blue. There is no gravity or sedimentation active and the horizontal particle concentration at low $\eps$ are the result of gas drag interactions and is not the disk mid-plane. SI modes get radially shorter with increasing $\eps$, but radially larger with decreasing $\eps$. Consequently, vertical or horizontal bands appear that might be the incarnation of larger modes in too small simulation domains. Color mapping changes for high $\eps$.}%
  {.9}{{.2cm, .5cm, 4.cm, .2cm}}%
For moderate dust-to-gas ratios, the SI shows active modes that are very similar to the ones found in the $r$-$\varphi$ simulations, see \Fig{fig:01_xz_imggrid} and compare with \Fig{fig:01_xy_imggrid}, e.g., at ($L=\SI{0.1}{\scaleheight}$, $\epsInit=1$), ($L=\SI{0.01}{\scaleheight}$, $\epsilon=3$) or ($L=\SI{0.001}{\scaleheight}$, $\epsilon=10$). %
For $L=\SI{0.1}{\scaleheight}$ (left column) the simulations are dominated by horizontal modes (vertical bands) starting from $\epsilon\geq3$, with decreasing wavelength for increasing $\epsilon$, i.e., more vertical bands appear. Similar on the scale $L=\SI{0.01}{\scaleheight}$ (middle column) for runs with $\epsilon\geq30$. 

We also find single horizontal bands for $L=\SI{0.001}{\scaleheight}$ with dust-to-gas ratios $\epsilon\leq10$. In contrast to the horizontal modes on the large scales, here only a single band appears that is vertically more compact with smaller $\epsilon$ but no second band is present within our parameter range. These dominant horizontal bands do not show up in the next larger simulations with $L=\SI{0.01}{\scaleheight}$ (center), though particle concentrations increase here as well. This is indicates that the non-linear SI modes might more strongly concentrate particles at lower dust-to-gas ratio, than expected from the $L=0.1$ simulations.

With the ($L=\SI{0.01}{\scaleheight}$, $\epsilon=0.1$) and ($L=\SI{0.01}{\scaleheight}$, $\epsilon=0.3$) simulations we find SI activity in them to highly depend on the chosen value for the hyper-viscosity/-diffusivity. As discussed in \App{app:hyperhyper} this can lead to simulations where small modes are suppressed and subsequently large modes cannot grow as they lack initial perturbations of a significant amplitude. Going to higher resolutions or seeding-in dedicated SI modes might change the SI activity, too.
\subsubsection{Particle diffusion - $\dx$ and $\dz$}
\begin{figure} %
	\centering 
	\includegraphics[width=\columnwidth]{legend_f.pdf}\\
	\begin{subfigure}[Radial particle diffusion with $\dx\sim \epsilon^{-1.1} \dots \epsilon^{-1.5}$ slope.]{\includegraphics[width=\columnwidth, 
			trim={0.2cm, 0.3cm, 0.2cm, 1.6cm}, clip
			]{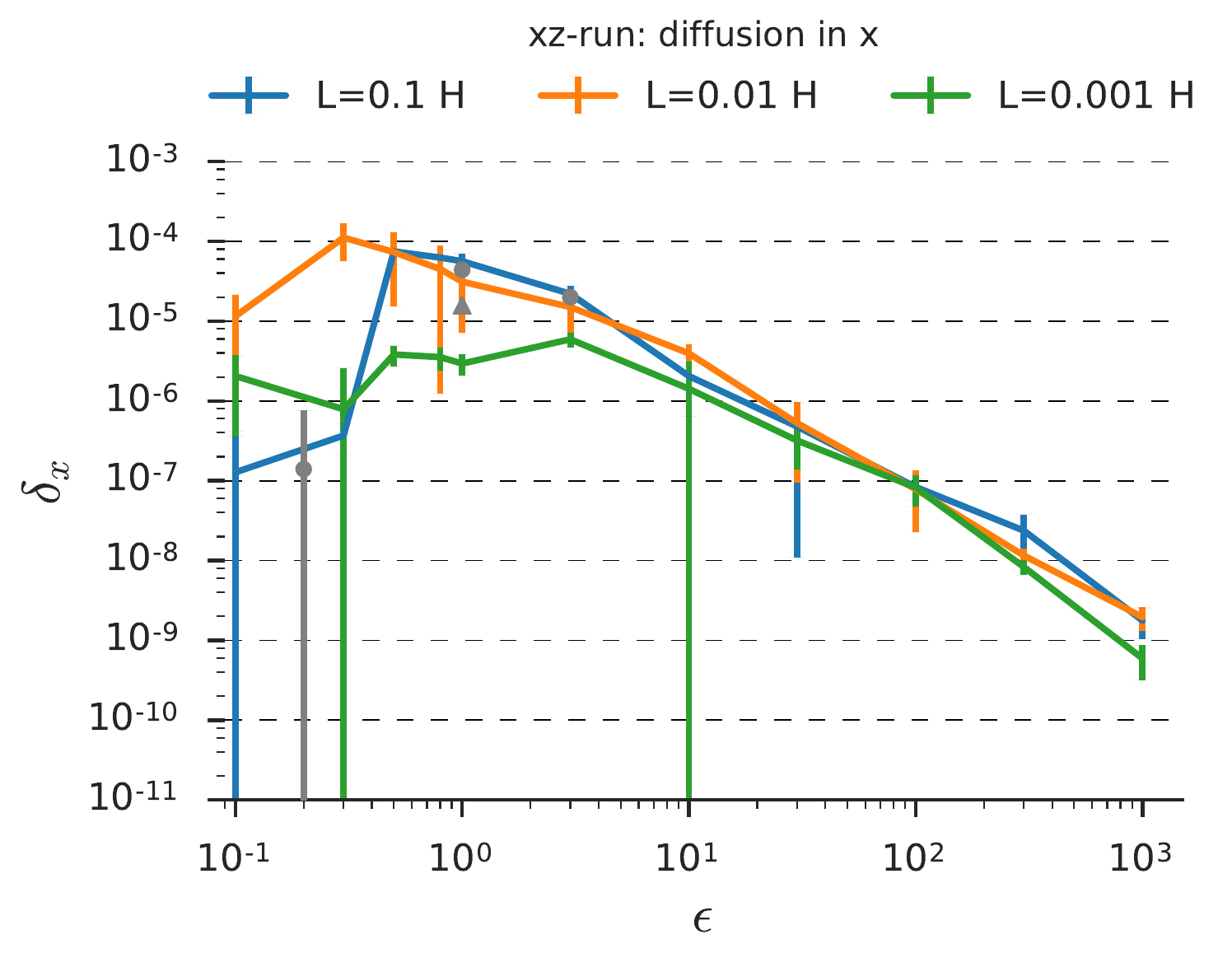}
	\label{fig:01_xz_diffusion_x}}
	\end{subfigure}\\
	\begin{subfigure}[Vertical particle diffusion with $\dz \sim \eps^{-0.3} \dots \eps^{-1.5}$ slope.]{\includegraphics[width=\columnwidth,
			trim={0.2cm, 0.3cm, 0.2cm, 1.6cm}, clip
			]{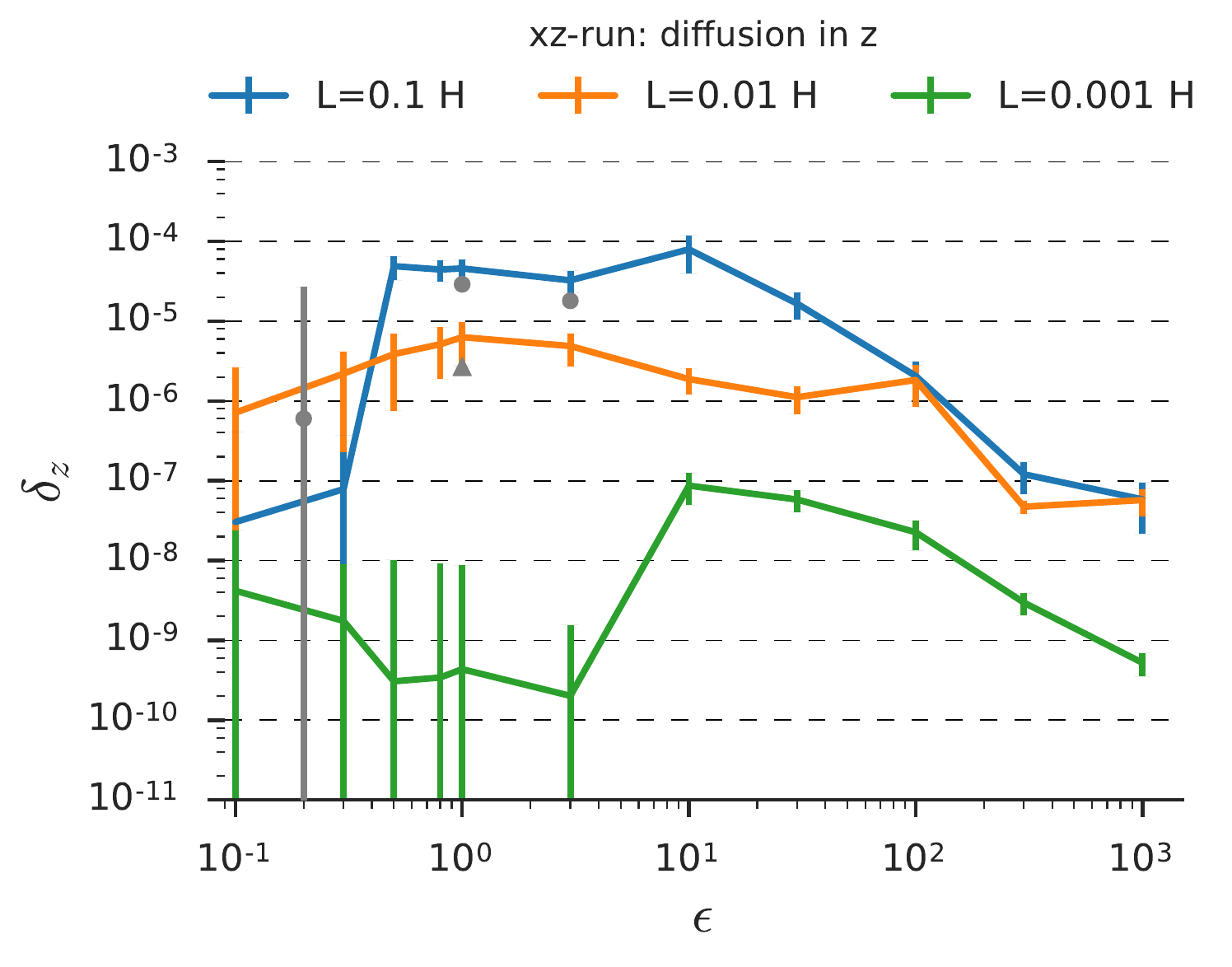}
	\label{fig:01_xz_diffusion_z}}
	\end{subfigure}
	\caption{$\stokes=0.1$ - $r$-$z$ plane: Radial (a) and vertical (b) particle diffusion estimated by treating particle movement as a random walk. For comparison plotted in grey circles (2D) and triangle (3D) are the diffusivities for $\stokes=0.1$ particles and $L\geq\SI{2}{\scaleheight}$, from JY07 Tab.~3.	 The presence of vertical modes (large boxes, high $\epsMax/\epsInit$) maintains vertical diffusion, but the presence of a strong horizontal band (small box, low $\epsMax/\epsInit$) shuts down the vertical diffusion but maintains radial diffusion. See \Fig{fig:01_xz_diffusion_ratio} for a comparison of $\dx/\dz$.} %
	\label{fig:01_xz_diffusion} %
\end{figure}%
\fig{fig:01_xz_diffusion} shows the particle diffusion for the radial $\dx$ and now also vertical direction $\dz$. For comparison plotted are radial and vertical diffusivity values from JY07, their values from 2-d radial-vertical simulations as grey dots and values from their 3-d simulations as grey triangles. In contrast to the aSI, here the radial diffusion is nearly identical on all scales, whereas the vertical diffusion changes in magnitude with simulation domain size, with the strongest diffusion on the largest scales.

For the radial diffusivity we measure a slope for larger $\epsilon$ of $\dx \sim \epsilon^{-1.1} \dots \epsilon^{-1.3}$, which is flatter then for the aSI (see  \Fig{fig:01_xy_diffusion}). In contrast, the vertical diffusion stays mostly on a fixed level for the SI-active range. Both plots show a decrease in diffusivity with increasing particle load. On the two larger scales (blue and orange) the presence of horizontal modes maintains the vertical diffusion strength whereas the radial diffusion seems to be mostly unaltered. The vertical bands in the small simulations (green) for low dust-to-gas ratios do the same with the $\dx$, but vertical diffusion is completely suppressed. Taking only the pure SI-active simulations, one finds that the radial diffusion is mostly stronger or at least as strong as the vertical diffusion. Again, the diffusion values found are similar to that of JY07.
\subsubsection{Particle dispersion and drift - $\sigma$ and $\zeta$}
\begin{figure} %
	\centering 
	\includegraphics[width=\columnwidth]{legend_nn.pdf}\\
	\begin{subfigure}[Global turbulent dispersion $\sg$ (lines) and local turbulent dispersion $\slo$ (contour), in blue the corresponding mean local dispersion values.]{\includegraphics[width=\columnwidth, 
			trim={0.5cm, 0.3cm, 1.9cm, 1.6cm}, clip
			]{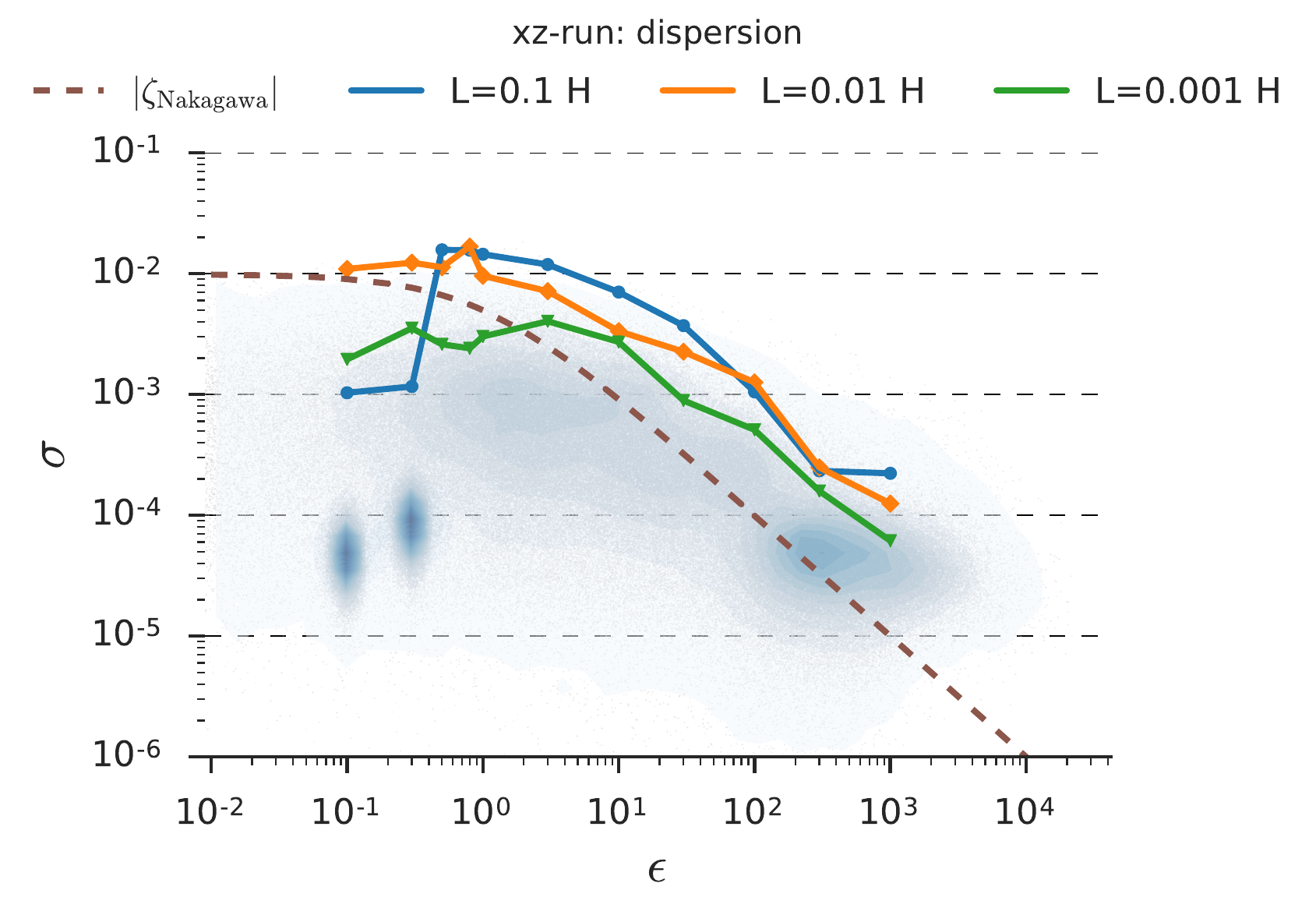}
	\label{fig:01_xz_dispersion}}
	\end{subfigure}\\
	\begin{subfigure}[Global drift $\zg$ (lines), local drift $\zlo$ (contour) and linear drift by \cite{Nakagawa1986} (dashed), in red the corresponding mean local drift values.]{\includegraphics[width=\columnwidth,
			trim={0.5cm, 0.3cm, 1.9cm, 1.6cm}, clip
			]{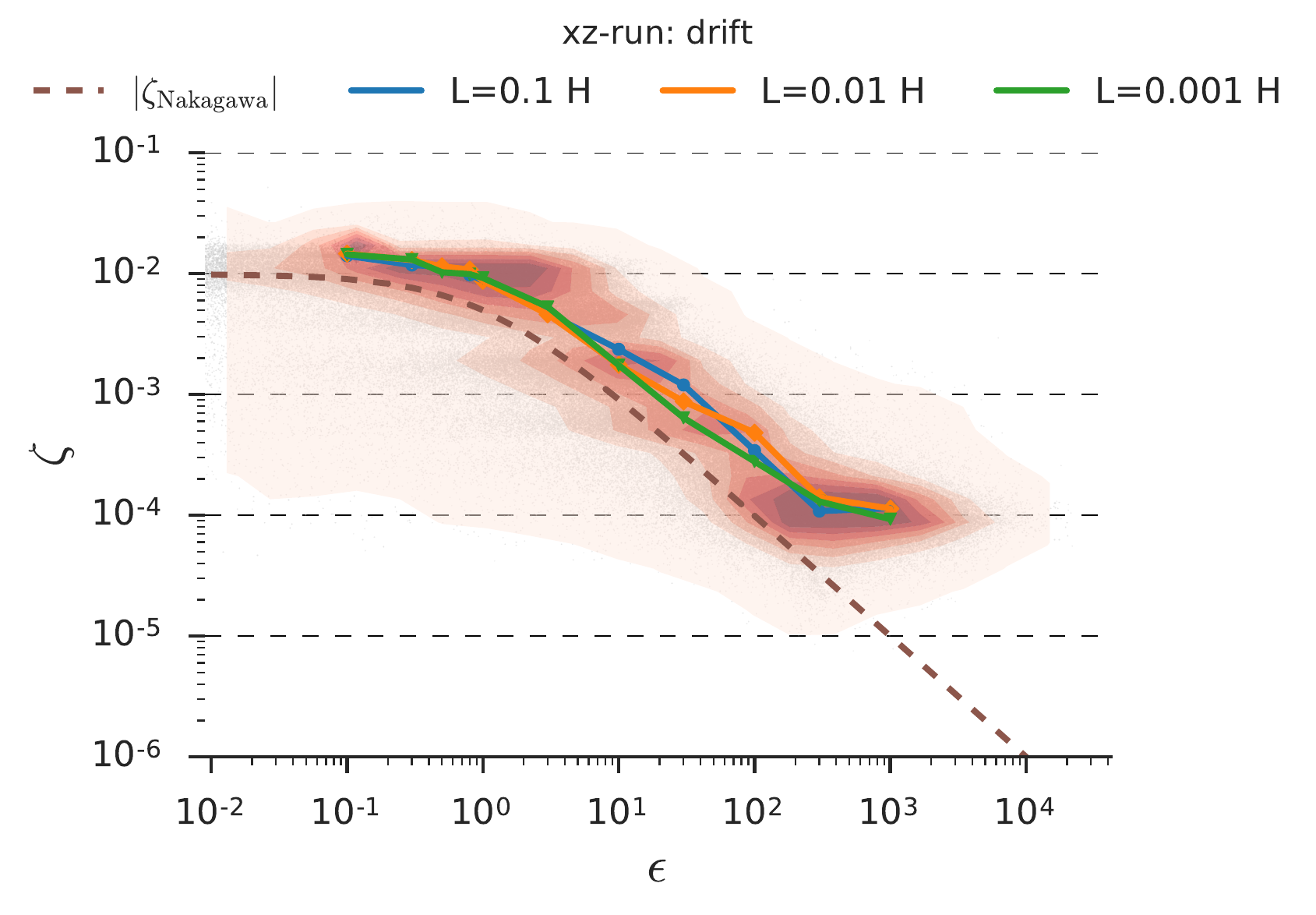}
	\label{fig:01_xz_drift}}
	\end{subfigure}
	\caption{$\stokes=0.1$ - $r$-$z$ plane: Dispersion and drift for all simulations together. Blue, orange and green lines show the individual global values. The local, i.e., grid wise, values for all simulations are shown combined in shaded contours. As reference in dashed is shown the absolute magnitude of the Nakagawa drift speed from \Eq{eq:nakagawa_drift}. Global and local drift shows perfect agreement, whereas the local dispersion values are always well below the global values. The latter indicates that local particle groups move with similar velocity, but comparing two groups in distinct grid cells they move independently. The similarity in all $\zg$ indicates that particles and gas on all scales moves similar relative to each other, following Nagakawa drift prescription. But, clumps with $\eps\geq10^2$ drift faster than predicted. What can be interpreted as their combined Stokes number as a particle group is higher than of the individual super particle.} %
	\label{fig:01_xz_lc_drift_disp} %
\end{figure}%
The measured turbulent particle dispersion and drift speed values are comparable to the ones from the $r$-$\varphi$ simulations. In simulations with non-active SI, the larger simulations with $\epsilon<1$ show a strong drop in particle dispersion but not as strong as in the $r$-$\varphi$ simulations. 

\label{sec:01_xz_corrTime} %
Linking $\dx$ with $\sg$ via $\tauCorr$, see \Eq{eq:tauCorr}, for SI-active simulations we find $\tauCorr \approx 0.3 \Omega^{-1}$. A similar value as found in \Sec{sec:corrTime} for the aSI. The correlation time is on average flat over $\eps$ and only increases, and more strongly varies, once horizontal or vertical modes emerge. For the cases with no active SI, the correlation time increases to $\tauCorr>1$. The same is true for the larger scales in the presence of radial modes.

The global particle drift $\zg$ again agrees well on the large and small scales. Slight increases in the drift speed can be found for the simulations with strong radial modes, as they produce zonal flows, e.g., ($L=\SI{0.1}{\scaleheight}$, $\epsInit=10$ - $\epsInit=100$) and ($L=\SI{0.01}{\scaleheight}$, $\epsilon=100$).
\subsubsection{$\alpha$-value and Schmidt number}
\begin{figure} %
	\centering 
	\hspace{0.6cm}\includegraphics[width=.8\columnwidth]{legend_f.pdf}\\
	\begin{subfigure}[$\alpha$-value: turbulent gas viscosity]{\includegraphics[width=\columnwidth, 
			trim={0.2cm, 0.3cm, 0.2cm, 1.6cm}, clip
			]{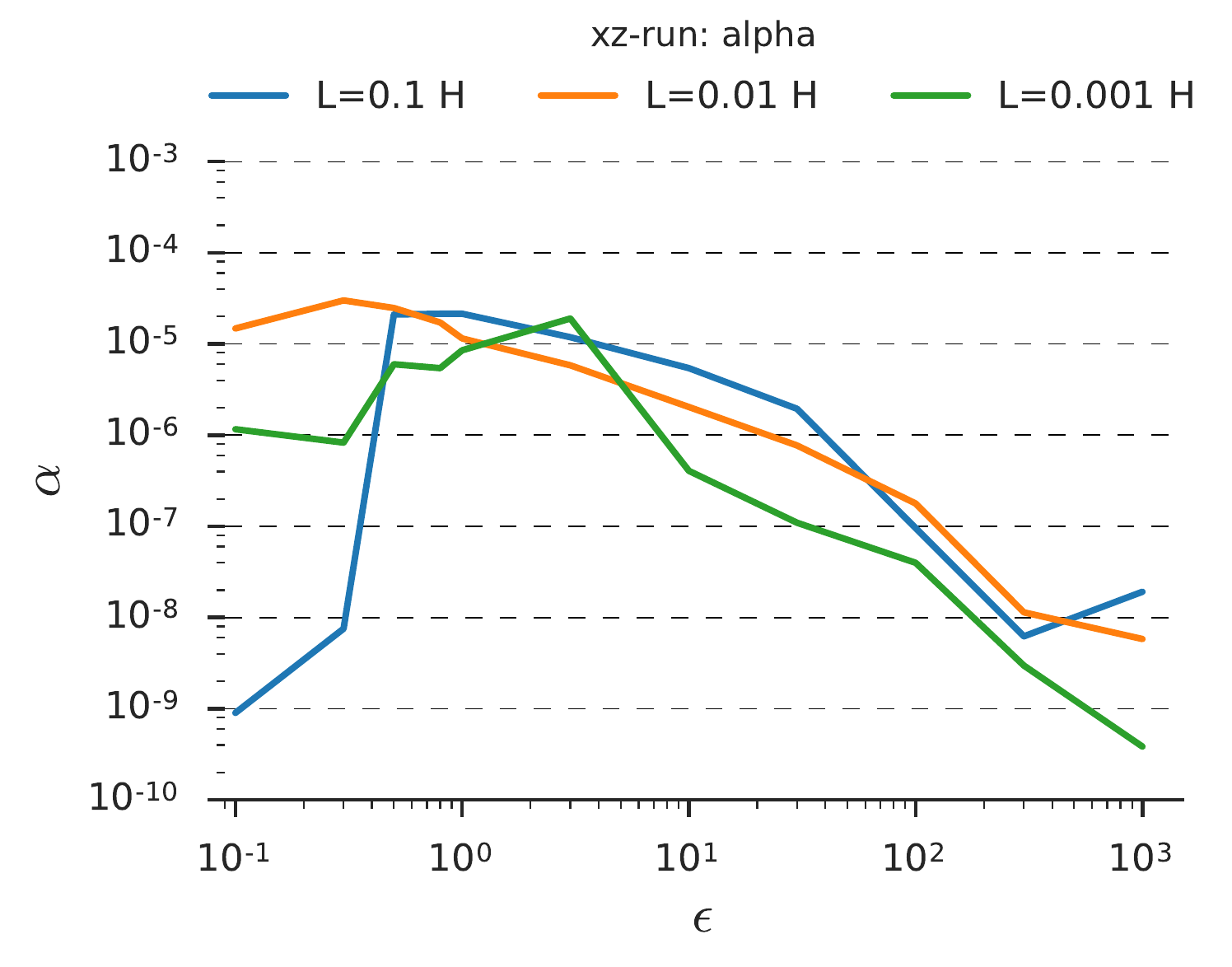}
	\label{fig:01_xz_alpha_paramplot}}
	\end{subfigure}\\
	\begin{subfigure}[Schmidt number]{\includegraphics[width=\columnwidth, trim={0.2cm, 0.3cm, 0.2cm, 1.6cm}, clip]{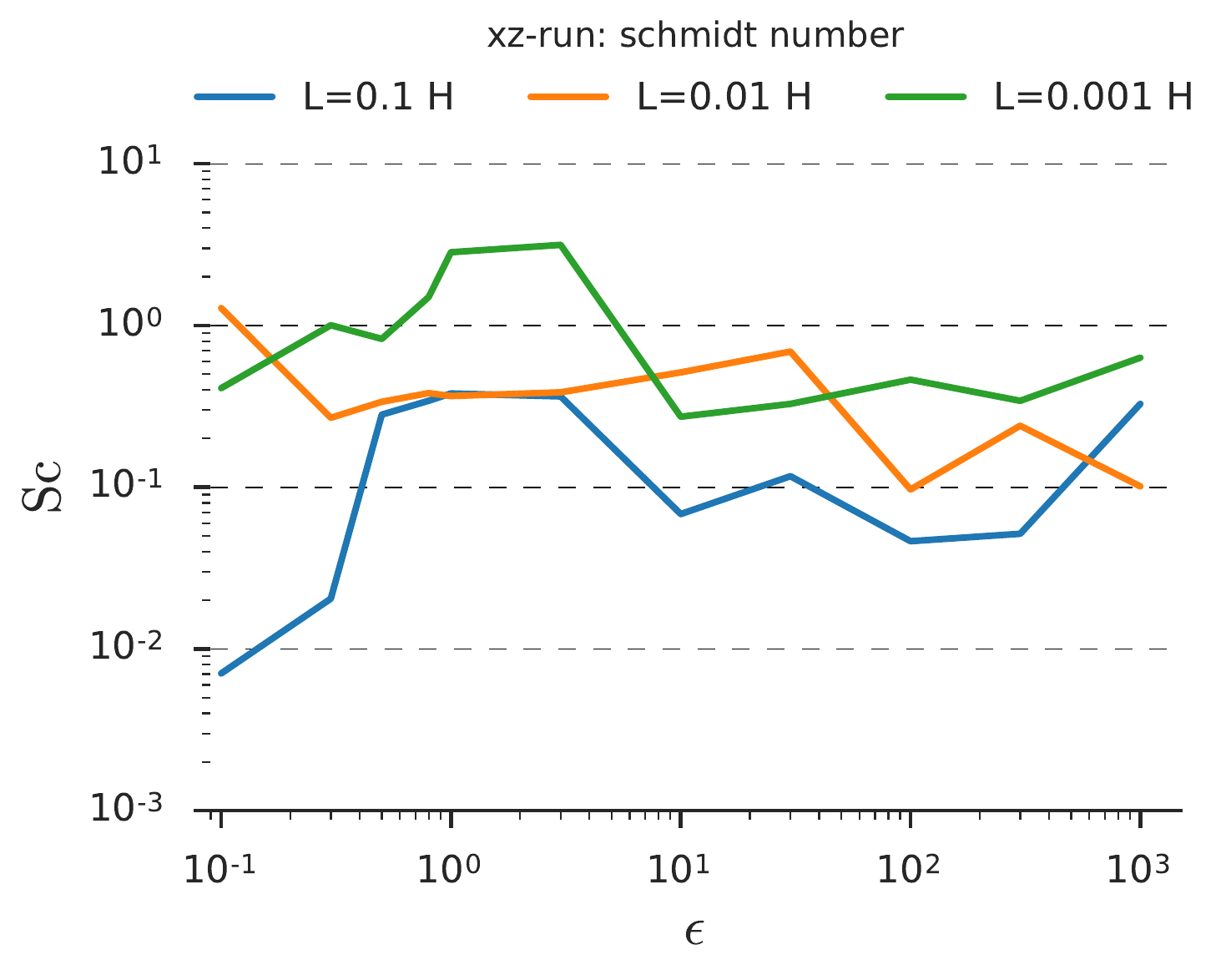}
	\label{fig:01_xz_schmidtnum}}
	\end{subfigure}
	\caption{$\stokes=0.1$ - $r$-$z$ plane: As most investigated quantities the gas $\alpha$-turbulence shows a steep drop-off once the SI is inactive at high dust-to-gas ratios. The Schmidt number in the case of active aSI is mostly below $\approx1$, showing that particle diffusion is stronger than the gas turbulence.} %
	\label{fig:01_xz_alpha_schmidt} %
\end{figure}%
\Fig{fig:01_xz_alpha_paramplot} shows the $\alpha$-values for the investigated parameter set. Similar to diffusion for particles, $\alpha$ is a measure of gas turbulence. It shows a strong falloff with SI becoming inactive for high $\epsInit$. We find the drop in $\alpha$ to higher $\epsilon$-values to be as strong as in the case of the particle diffusivity. 

One again sees this better in the Schmidt number, plotted in \fig{fig:01_xz_schmidtnum}, which is the ratio of $\alpha$-transport against particle diffusion, see \Eq{eq:schmidt}. This ratio shows a rather flat profile throughout the SI-active range. We find the Schmidt number to again depend on the simulation domain size. Mostly, the particle turbulence is stronger or at least equally strong as the particle turbulence. Only on the smallest scales (green) the gas turbulence is stronger within a larger fraction of the SI-active range, indicating that Schmidt number has a length scale dependency that leads to higher values on smaller scales. Comparing these runs with the $r$-$\varphi$ runs, one finds at the smallest scales and lowest dust-to-gas ratios the Schmidt number to decrease, whereas it increases in the $r$-$\varphi$ runs. The reason lies in the presence of vertical bands in the $r$-$z$ simulations that are strongly diffusing particles in radial direction. Though the gas flow is unaltered by these local bands, $\alpha$-turbulence stays low.
\section{Results of the Parameter Study II:\\$\stokes=0.01$ particles}
All studies on the pure streaming instability so far looked at $\stokes \geq 0.1$ particles, but \cite{Birnstiel2010} suggests that the dominant particles species might have $\stokes=0.01$. Thus, in this section we redo the parameter study from the previous section but with this smaller Stokes number.
\subsection{$\stokes=0.01$ - $r$-$\varphi$ plane}
\label{sec:001-xy}
This and the following section repeat the experiments from the previous two sections but for particles with one order of magnitude smaller Stokes number.
\subsubsection{Dust density fluctuations and growth rates}
\FIGURE
  {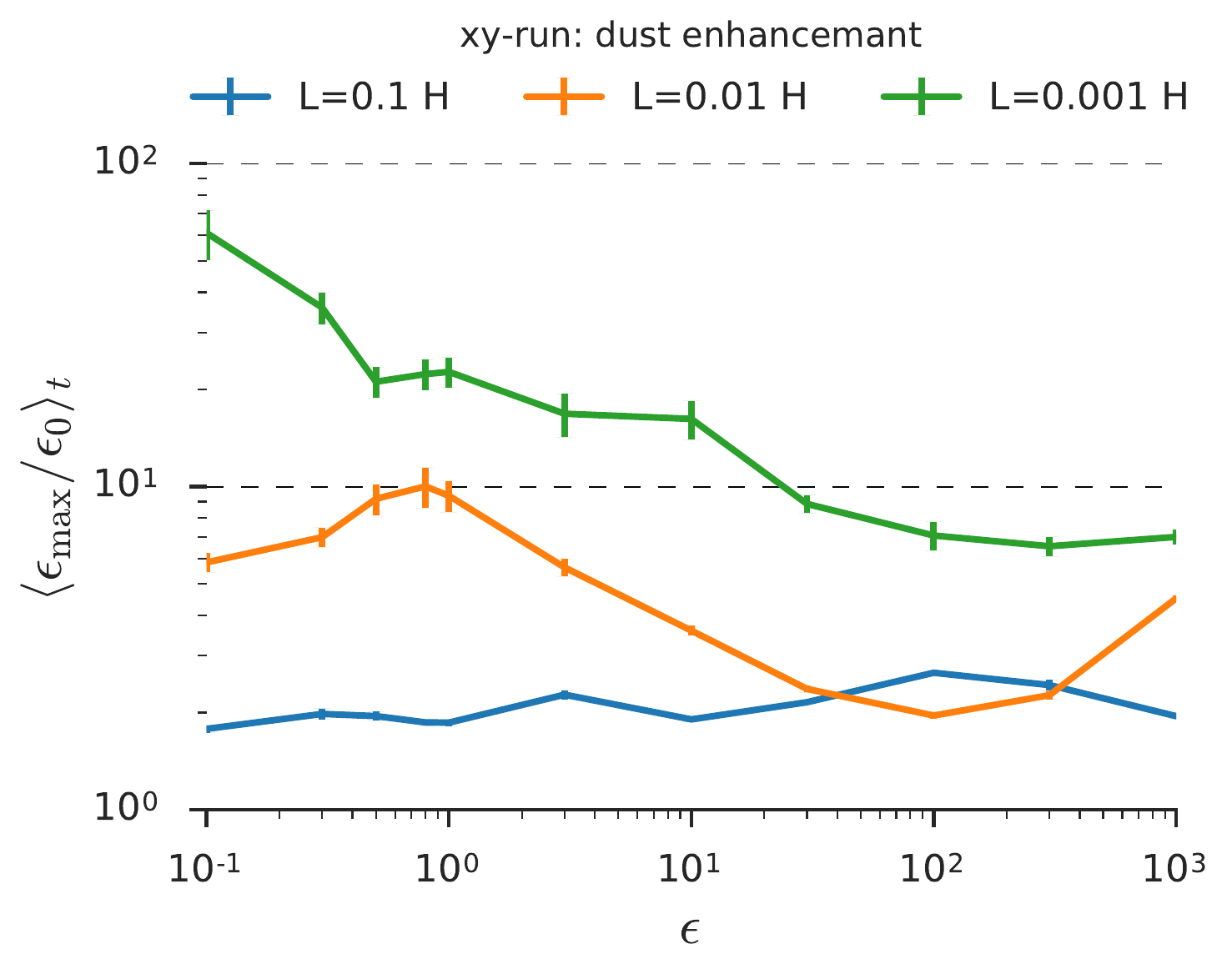}
  {fig:001_xy_dust_enhancement}
  {$\stokes=0.01$ - $r$-$\varphi$ plane: The aSI shows only active dust density fluctuations for rather low initial dust-to-gas ratios, even below $\epsInit=1$. On the intermediate scale (orange) the aSI is strongest at $\epsInit=1$ and peaks at the typical value of $\approx10$ as did the previous $\stokes=0.1$ runs. The aSI on the smallest scales is throughout the whole parameter space actively enhancing the dust-to-gas ratios up to values close and well above $10$. The strongest dust density fluctuations we observe for $\epsInit\approx\num{e-1}$.}
  {1}{{.3cm, .3cm, .3cm, .7cm}} 
For $\stokes=0.01$ particles we could not find any aSI on the largest scales (blue), see \Fig{fig:001_xy_dust_enhancement}. On the next smaller scale (orange) the aSI does appear up to a value of $\epsInit\approx10$. The smallest scale (green) surprisingly shows aSI activity throughout the whole parameter space, especially including simulations with $\epsInit\le 1$. In all the aSI-active runs, we do not find any zonal flows, as we did for $\stokes=0.1$ particles. It is also surprising to find a very high ability of the aSI to concentrate dust on the smallest scales up to values of $\epsMax/\epsInit\approx 60$ at the lowest initial dust-to-gas ratio. In contrast, the aSI on the intermediate scales peaks at $\epsInit=1$ with a value of $\epsMax/\epsInit\approx 10$.

The measured growth rates for the aSI-active simulations we find to depend on the simulation domain size. With initial dust-to-gas ratio the growth rate only slightly varies and for the aSI-active simulations are on average: $s\left( L=\SI{0.01}{\scaleheight} \right) \approx \num{5e-1}\Omega$ and $s\left( L=\SI{0.001}{\scaleheight} \right) \approx \num{4.0}\Omega$.
\subsubsection{End-state snapshots}
\FIGURE
  {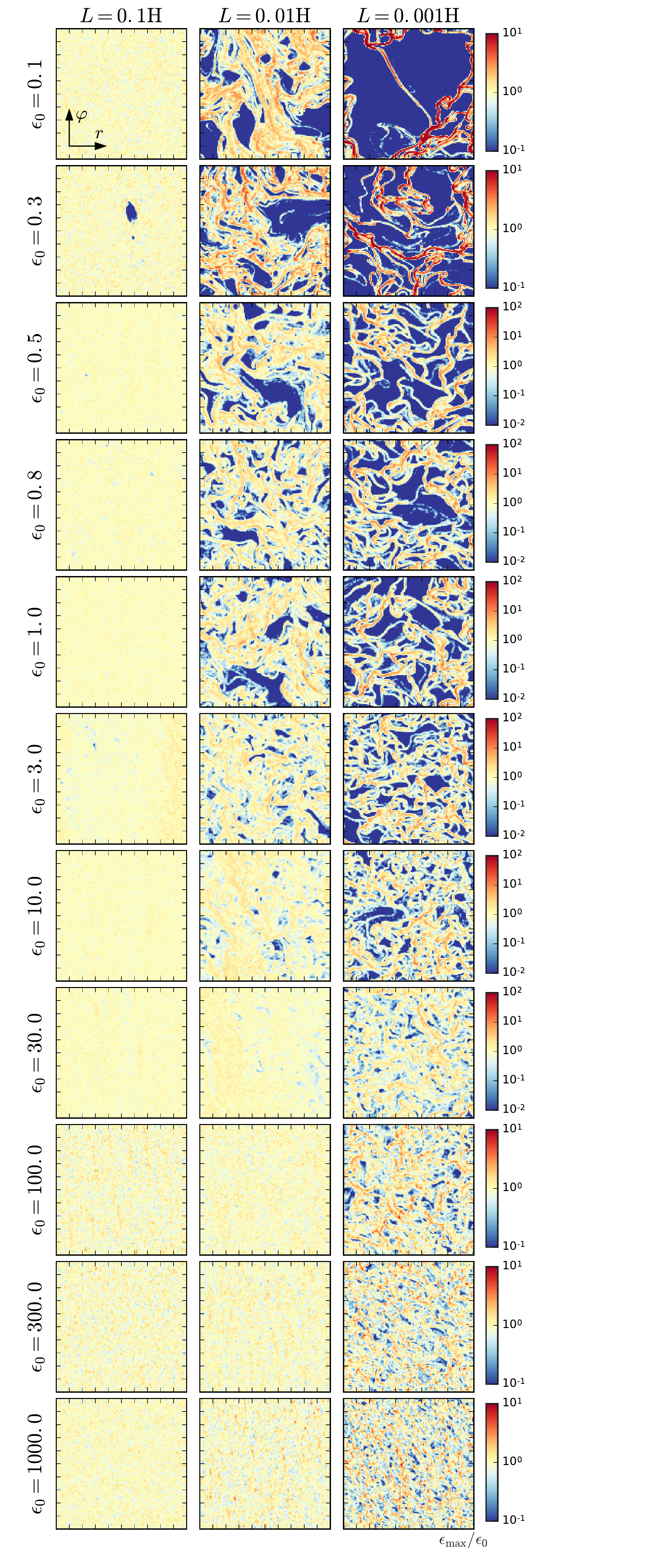}
  {fig:001_xy_imggrid}
  {$\stokes=0.01$ - $r$-$\varphi$ plane: Last snapshots of the dust-to-gas ratio normalized to $\einit$ (yellow). Over-densities are colored in red, particle voids in blue. Besides the transient void in ($L=\SI{0.1}{\scaleheight}$, $\epsilon=0.3$) we do not find any signs of aSI at $L=\SI{0.1}{\scaleheight}$. Though the parameter space for smaller simulations is fully populated with the aSI. We do not find any zonal flows emerging. Color mapping changes for high and low $\eps$.}
  {.9}{{.2cm, .5cm, 4.cm, .2cm}} 
We do not find any aSI-activity on the largest scale, see \Fig{fig:001_xy_imggrid}, for ($L=\SI{0.1}{\scaleheight}$, $\epsInit=0.3$) we find small transient voids appearing. A surprise is the ability of aSI to form at very low dust-to-gas ratios and the scale independence from the domain size of the active modes that appear in our simulations. We do not find any zonal flows or band structures appearing in all runs, as we found for $\stokes=0.1$. Though, for ($L=\SI{0.001}{\scaleheight}$, $\epsInit=0.1$) the aSI mode is extremely strong in its ability to concentrate dust, see \Fig{fig:001_xy_dust_enhancement}, which does not show up in the next larger simulation.
\subsubsection{Particle diffusion - $\dx$}
\begin{figure} %
	\centering %
	\hspace{0.6cm}\includegraphics[width=0.8\columnwidth]{legend_f.pdf}\\
	\includegraphics[width=\columnwidth, trim={0.2cm, 0.3cm, 0.2cm, 1.6cm}, clip]{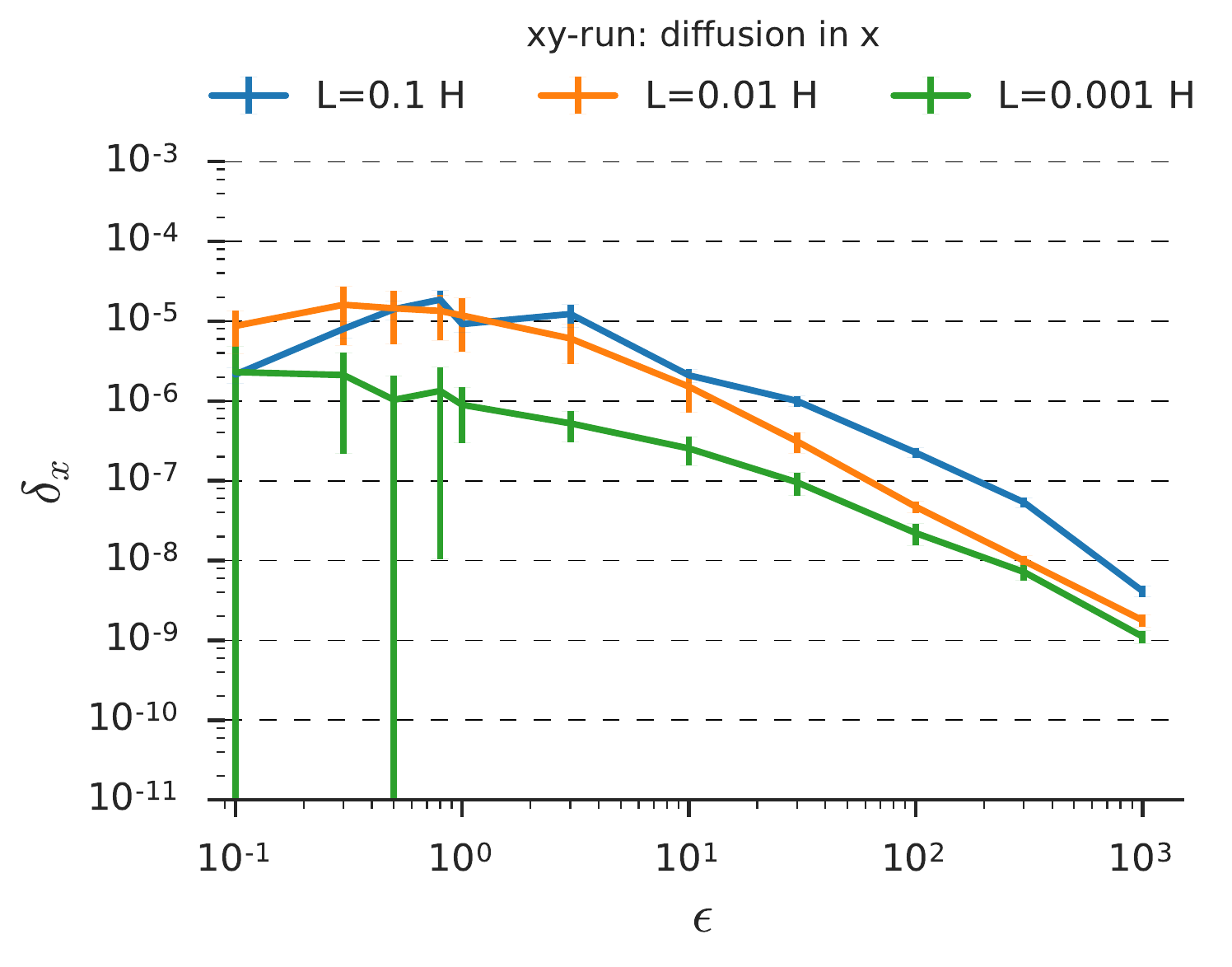}
	\caption{$\stokes=0.01$ - $r$-$\varphi$ plane: Radial particle diffusion $\dx$ estimated by treating the streaming instability as a random walk. The slope of the radial diffusion for this setup goes with $\eps^{-1.5}$ on intermediate scales (orange) and with $\eps^{-1.0}$ on small scales (green). Diffusion values on the largest scales (blue) are not from the aSI, see \Fig{fig:001_xy_imggrid}, but from lasting gas turbulence that is induced from the initialization of the particles in drag force equilibrium.}%
	\label{fig:001_xy_diffusion} %
\end{figure}%
The radial particle diffusion for $\stokes=0.01$ is surprisingly similar in its magnitude with the ones we find for $\stokes=0.1$ in \Sec{sec:01_xy_diff}. Though, the diffusion for $\stokes=0.01$ particles resides on the intermediate and small scales and is compared to the diffusion for $\stokes=0.1$ shifted towards lower $\epsInit$ values. At $\eps\leq1$ it is even stronger than for $\stokes=0.1$ particles. The measured slopes fit $\dx \sim \epsilon^{-1.0} \dots \epsilon^{-1.5}$.

For $L=\SI{0.1}{\scaleheight}$ we do measure diffusion, but that is from the initialization of the particles in drag force equilibrium with the gas. We find this value to be reduced with increasing number of particles in the simulation, i.e., once initial perturbations in dust density get reduced. The found diffusion, rms-values and $\alpha$-turbulence come from buoyancy effects of under- and over dense grid cells and are thus not (a)SI but results of a initial non-equilibrium setup. The snapshots in \Fig{fig:001_xy_imggrid} show no evidence for aSI and the correlation time in this case is $\tauCorr=1$ that shows the diffusion comes from the particles its rms-velocity.
\subsubsection{Particle dispersion and drift - $\sigma$ and $\zeta$}
\begin{figure} %
	\centering 
	\includegraphics[width=\columnwidth]{legend_nn.pdf}\\
	\begin{subfigure}[Global turbulent dispersion $\sg$ (lines) and local turbulent dispersion $\slo$ (contour).]{\includegraphics[width=\columnwidth, 
			trim={0.5cm, 0.3cm, 1.9cm, 1.6cm}, clip
			]{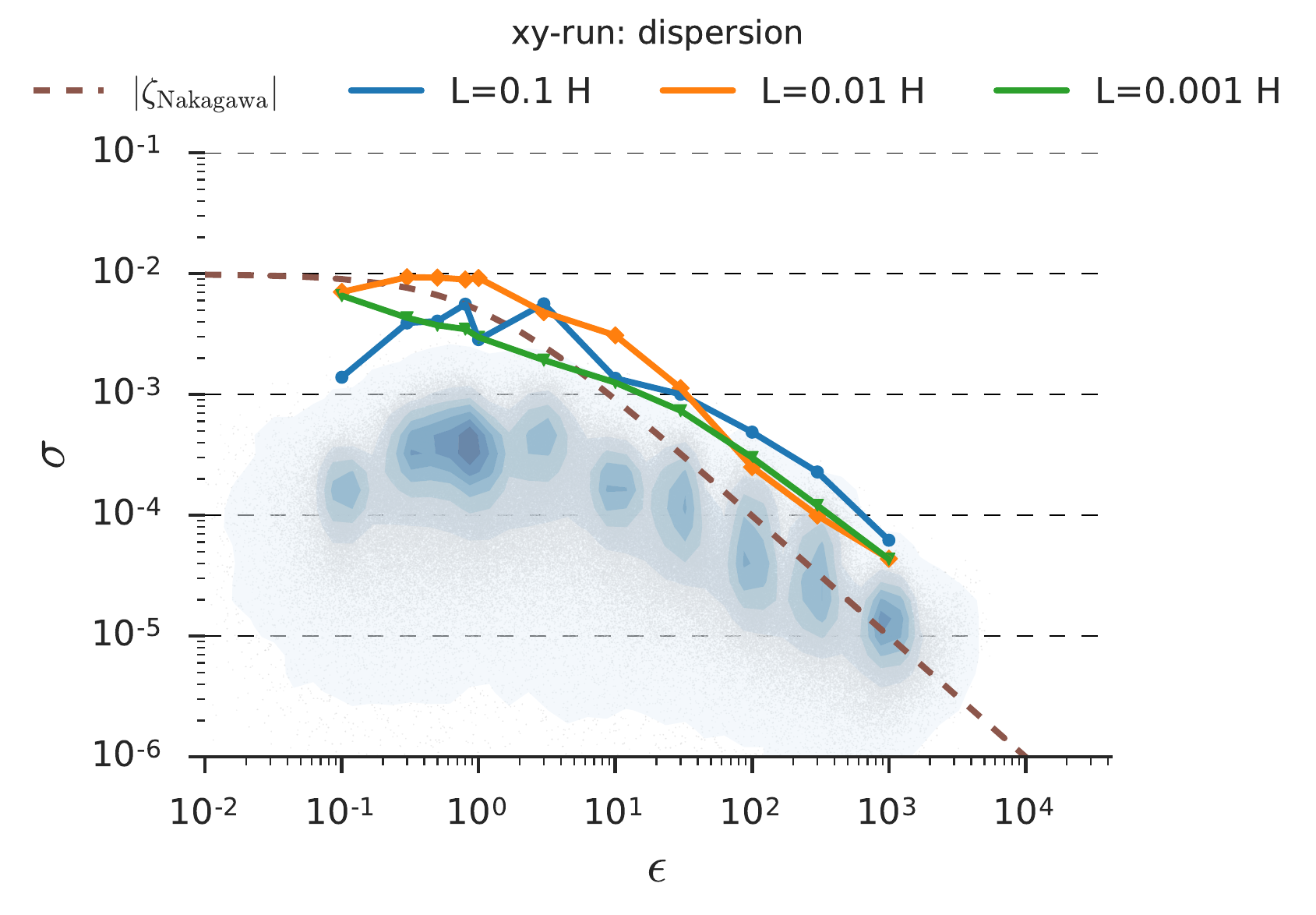}
	\label{fig:001_xy_dispersion}}
	\end{subfigure}\\
	\begin{subfigure}[Global drift $\zg$ (lines), local drift $\zlo$ (contour).]{\includegraphics[width=\columnwidth,
			trim={0.5cm, 0.3cm, 1.9cm, 1.6cm}, clip
			]{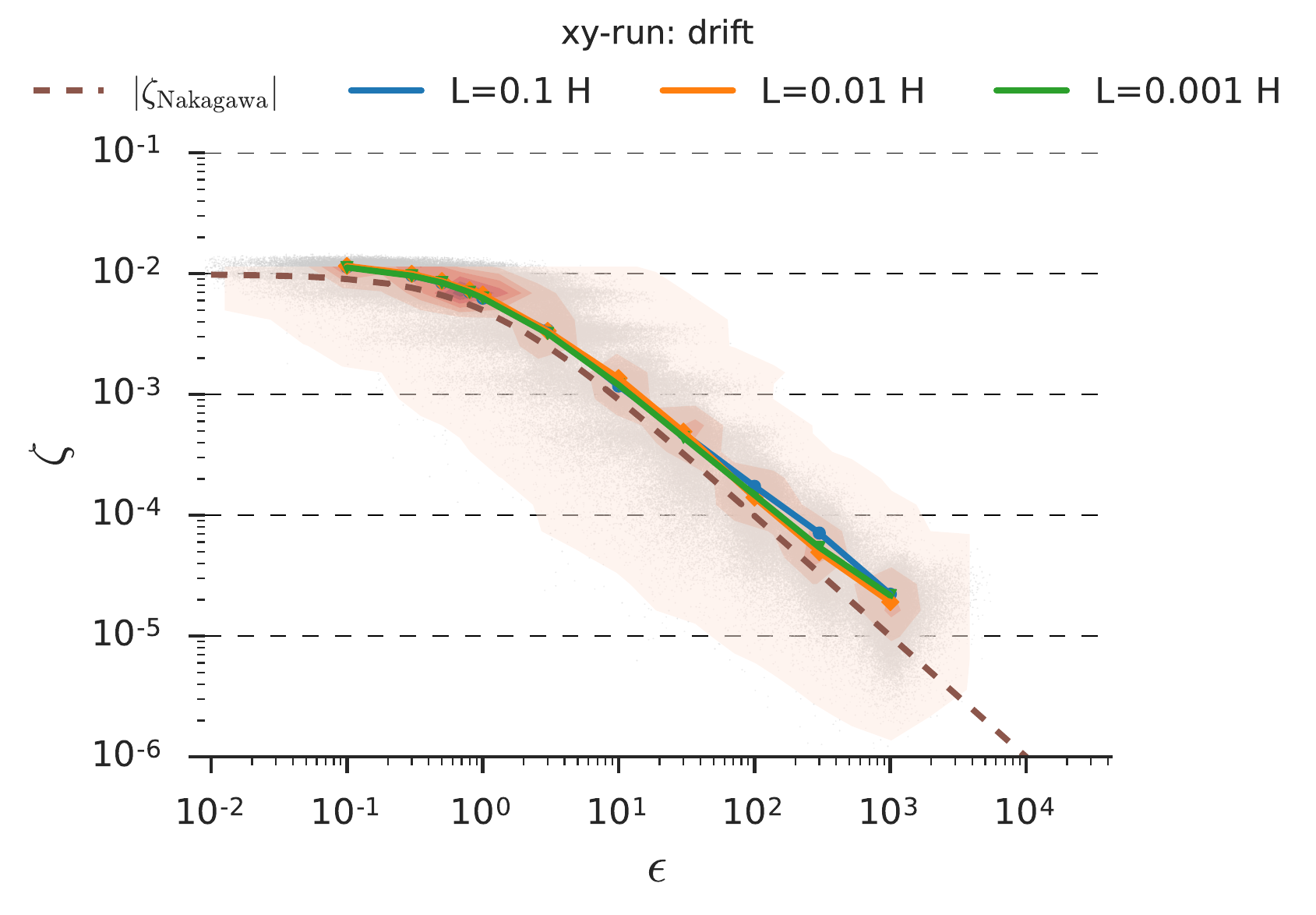}
	\label{fig:001_xy_drift}}
	\end{subfigure}
	\caption{$\stokes=0.01$ - $r$-$\varphi$ plane: Dispersion and drift for all simulations together. Blue, orange and green lines show the individual global values. The local, i.e., grid wise, values for all simulations are shown combined in shaded contours. As reference in dashed is shown the absolute magnitude of the Nakagawa drift speed from \Eq{eq:nakagawa_drift}. Global and local drift again shown perfect agreement, as in the runs with $\stokes=0.1$, but now closer to the Nakagawa solution. The local dispersion values are again always well below the global values. But, in constrast to the $\stokes=0.1$-runs now for $\epsInit\leq1$ the aSI-active simulations have dispersion velocities that are closer to the Nagakawa drift, as they remain aSI-active.} %
	\label{fig:001_xy_lc_drift_disp} %
\end{figure}%
The turbulent particle dispersion is shown in \Fig{fig:001_xy_dispersion}. In its magnitude the dispersion is again following the slope of the Nakagawa solution for particle drift (dashed line), but now, since aSI is active also for $\epsInit\leq1$ the global dispersion values for the intermediate and small simulations continues to follow the Nakagawa solution also at these values, whereas for $\stokes=0.1$ we observed a knee at around $\epsilon=1$.

The particle drift in \Fig{fig:001_xy_drift} again shows a perfect agreement between local and global drift values. For this Stokes number, we find the drift values to be even closer to the predicted value from \Eq{eq:nakagawa_drift}. In the scatter of the grey dots that represent $\zlo$, we also see that the drift velocity depends on the mean dust-to-gas ratio, which is equal to $\epsInit$, and does not follow the expected value from the Nakagawa drift solution.

The correlation time from \Eq{eq:tauCorr} for $\stokes=0.01$ particles on the smallest and the intermediate scales is very flat, around $\tauCorr \approx 0.1 \Omega^{-1} \dots 0.2 \Omega^{-1}$. Once the aSI is dead it rises onto $\tauCorr = 1 \Omega^{-1}$, the level where the correlation time for the $L=\SI{0.1}{\scaleheight}$ runs is. This indicates that there is no aSI activity on the largest scale but only dispersion and diffusion from random particle movement.
\subsubsection{$\alpha$-value and Schmidt number}
\begin{figure} %
	\centering 
	\hspace{0.6cm}\includegraphics[width=.8\columnwidth]{legend_f.pdf}\\
	\begin{subfigure}[$\alpha$-value: turbulent gas viscosity]{\includegraphics[width=\columnwidth, 
			trim={0.2cm, 0.3cm, 0.2cm, 1.6cm}, clip
			]{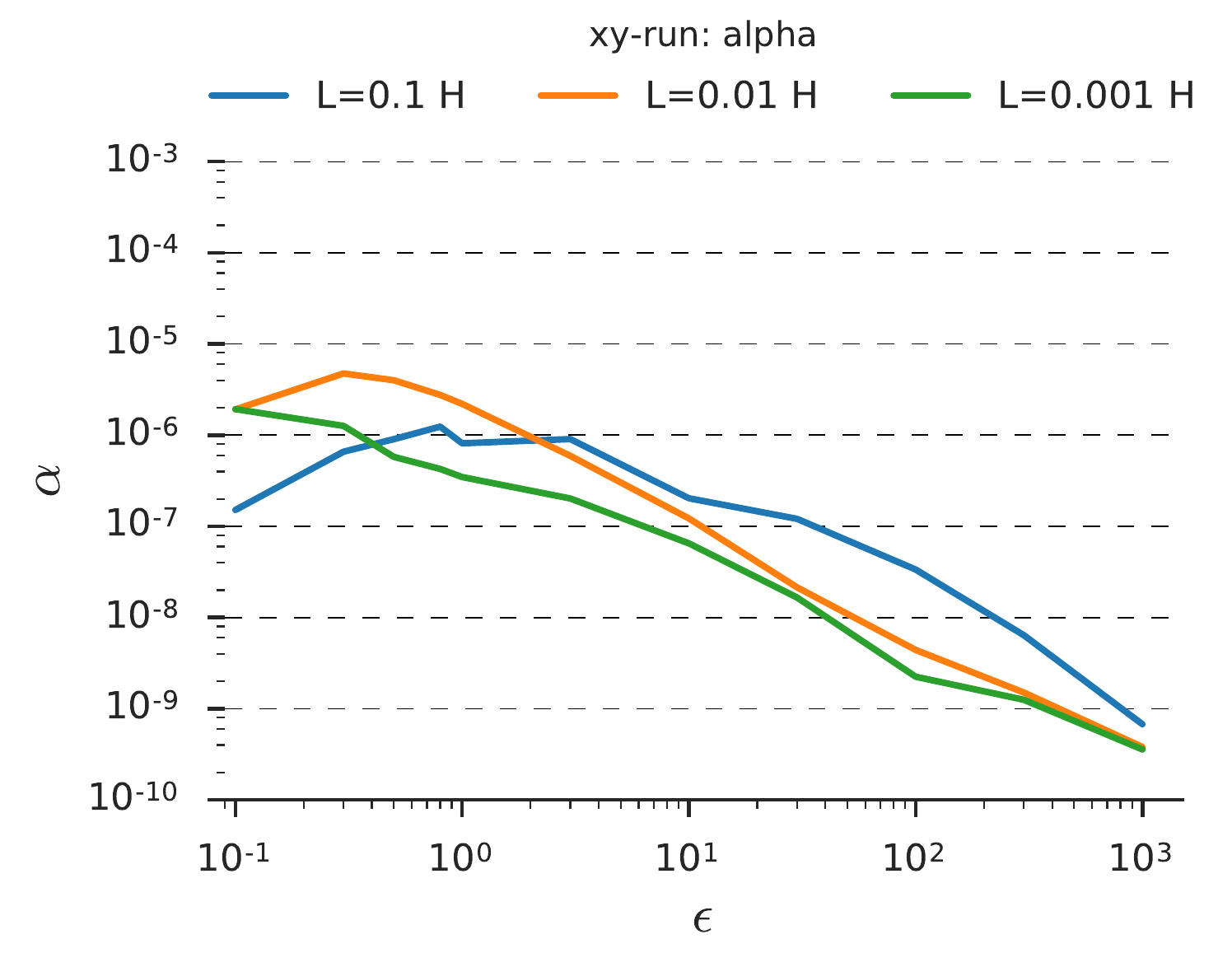}
	\label{fig:001_xy_alpha_paramplot}}
	\end{subfigure}\\
	\begin{subfigure}[Schmidt number]{\includegraphics[width=\columnwidth, trim={0.2cm, 0.3cm, 0.2cm, 1.6cm}, clip]{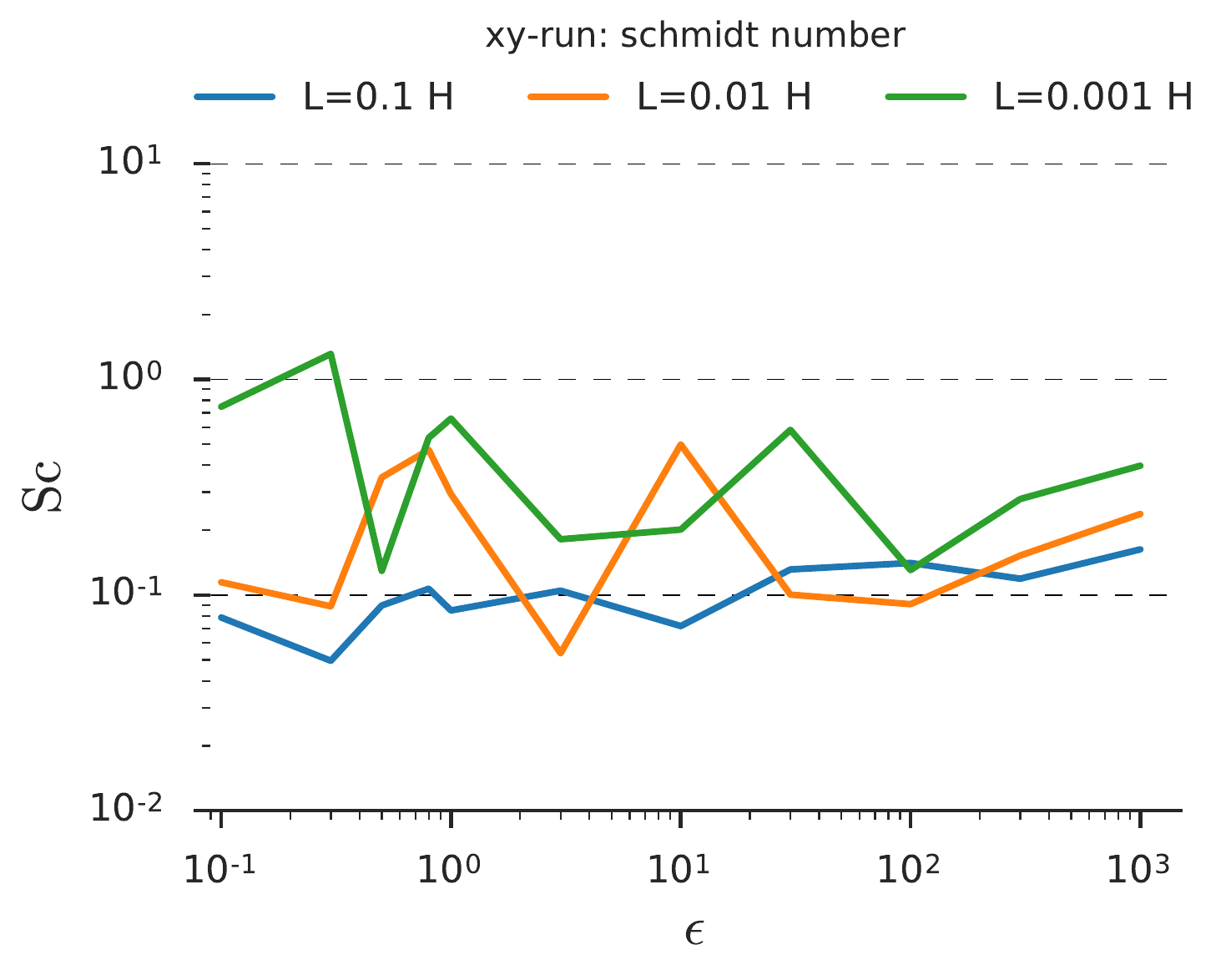}
	\label{fig:001_xy_schmidtnum}}
	\end{subfigure}
	\caption{$\stokes=0.01$ - $r$-$\varphi$ plane: The gas $\alpha$-turbulence shows again a steep drop-off once the aSI is weaker for high dust-to-gas ratios. Since aSI is for $\stokes=0.01$ active throughout the whole parameter space, this results in a completely constant Schmidt number mostly below $\approx1$, showing that the particle diffusion is stronger than the gas turbulence.} %
	\label{fig:001_xy_alpha_schmidt} %
\end{figure}%
\Fig{fig:001_xy_alpha_paramplot} shows the measured level of $\alpha$-turbulence. We find it to be just a bit lower to the case of $\stokes=0.1$, see \Fig{fig:01_xy_alpha_paramplot}. The difference here is that again aSI and thus gas turbulence is active for $\epsInit\leq1$.
\subsection{$\stokes=0.01$ - $r$-$z$ plane}
\label{sec:001-xz}
This section repeats the experiments on the streaming instability $r$-$z$ from \Sec{sec:01-xz} but for particle with one order of magnitude smaller Stokes number.
\subsubsection{Dust density fluctuations and growth rates}
\FIGURE
  {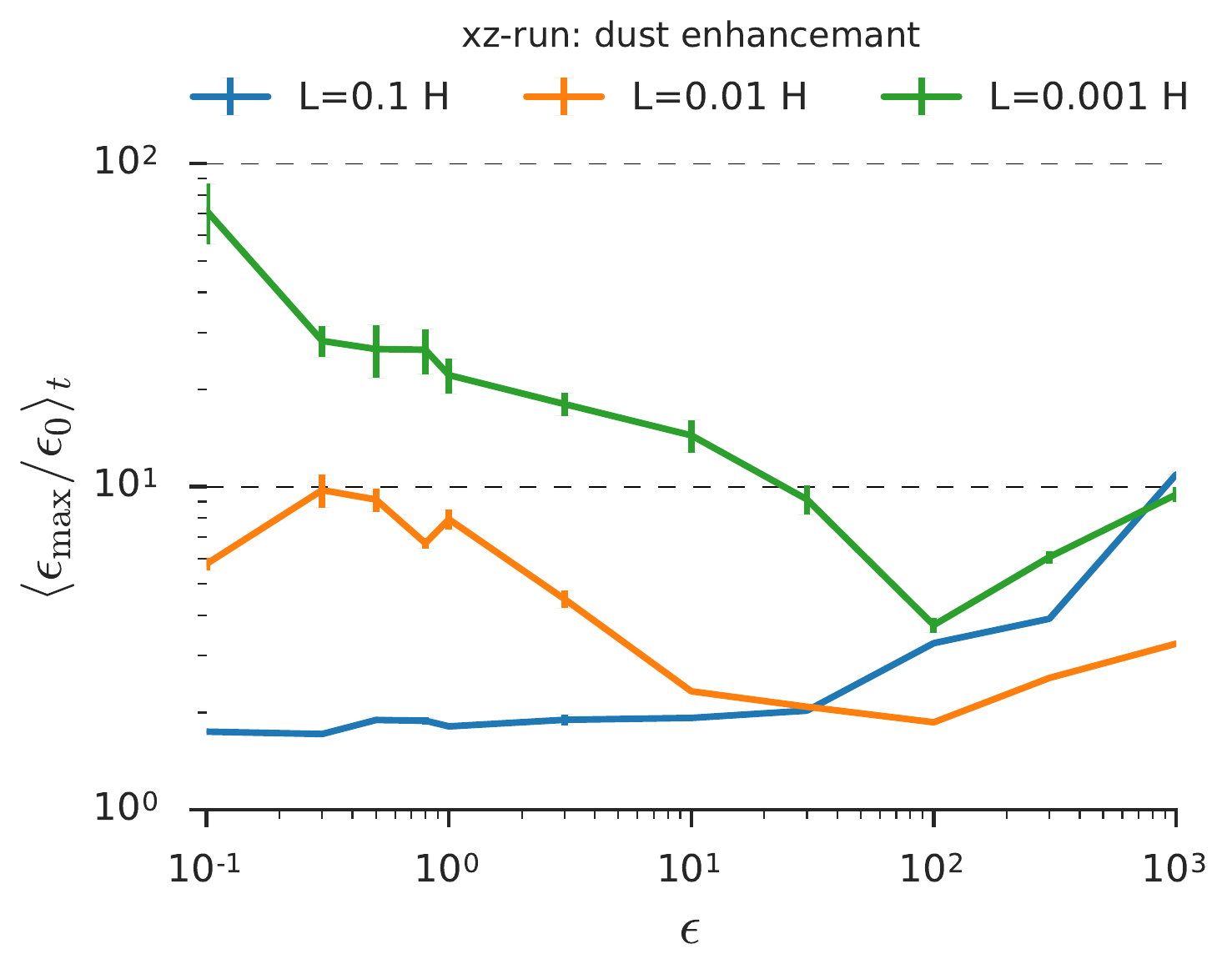}
  {fig:001_xz_dust_enhancement}
  {$\stokes=0.01$ - $r$-$z$ plane: The SI shows only active dust density fluctuations for rather low dust-to-gas ratios, even below $\epsInit=1$. On the intermediate scale (orange) the SI is the strongest around $\epsInit=1$ and peaks at a value of $\approx 0.5$, which is a bit lower than in the previous $r$-$\varphi$ runs. The SI on the smallest scales is throughout the whole parameter space actively enhancing the dust-to-gas ratios up to values close and well above $10$. The strongest dust density fluctuations we observe for $\epsInit\approx\num{e-1}$. That values are very similar to the findings in \Fig{fig:001_xy_dust_enhancement}.}
  {1}{{.3cm, .3cm, .3cm, .7cm}} 
The maximum dust density fluctuation values are plotted in \Fig{fig:001_xz_dust_enhancement}. They are very similar to the ones in $r$-$\varphi$ in the previous section. On the largest scale (blue) we do not find any signs of SSI. This finding is not in contrast to the finding of \textit{SI-activity} in \cite{Carrera2015} (see Fig.~4 therein), since they investigate gravity assisted particle clumping that can only occur if turbulent diffusivity by the SI is weak. Moreover, their simulations with smaller $\eta$ but same domain size correspond to our intermediate sized simulations that show SI presence and thus the stalled formation of bands in \cite{Carrera2015}. This results in a more turbulent picture, and can be explained by an increased turbulent particle diffusion. %
For the intermediate scale (orange) the peak in the dust density fluctuations is shifted a bit towards lower initial dust-to-gas ratios to $\eps=0.5$, where for the $r$-$\varphi$ runs it was at $\eps=1$. On the smallest scales (green) the parameter space is covered with SI up to a value of $\epsInit=100$. Comparing the slope with the ones from \Fig{fig:001_xy_dust_enhancement} for the $r$-$\varphi$ runs, the drop in the ability to enhance dust is slightly steeper for the $r$-$z$ runs, i.e., for ($L=\SI{0.001}{\scaleheight}$, $\epsInit=100$) the SI is active in the $r$-$\varphi$ case, but seems rather inactive in the $r$-$z$ case.

The measured growth rates for the SI-active simulations we find to depend on the simulation domain size. With initial dust-to-gas ratio the growth rate only slightly varies and for the SI-active simulations are on average: $s\left( L=\SI{0.01}{\scaleheight} \right) \approx \num{5e-1}\Omega$ and $s\left( L=\SI{0.001}{\scaleheight} \right) \approx \num{3.0}\Omega$.
\subsubsection{End-state snapshots}
\FIGURE
  {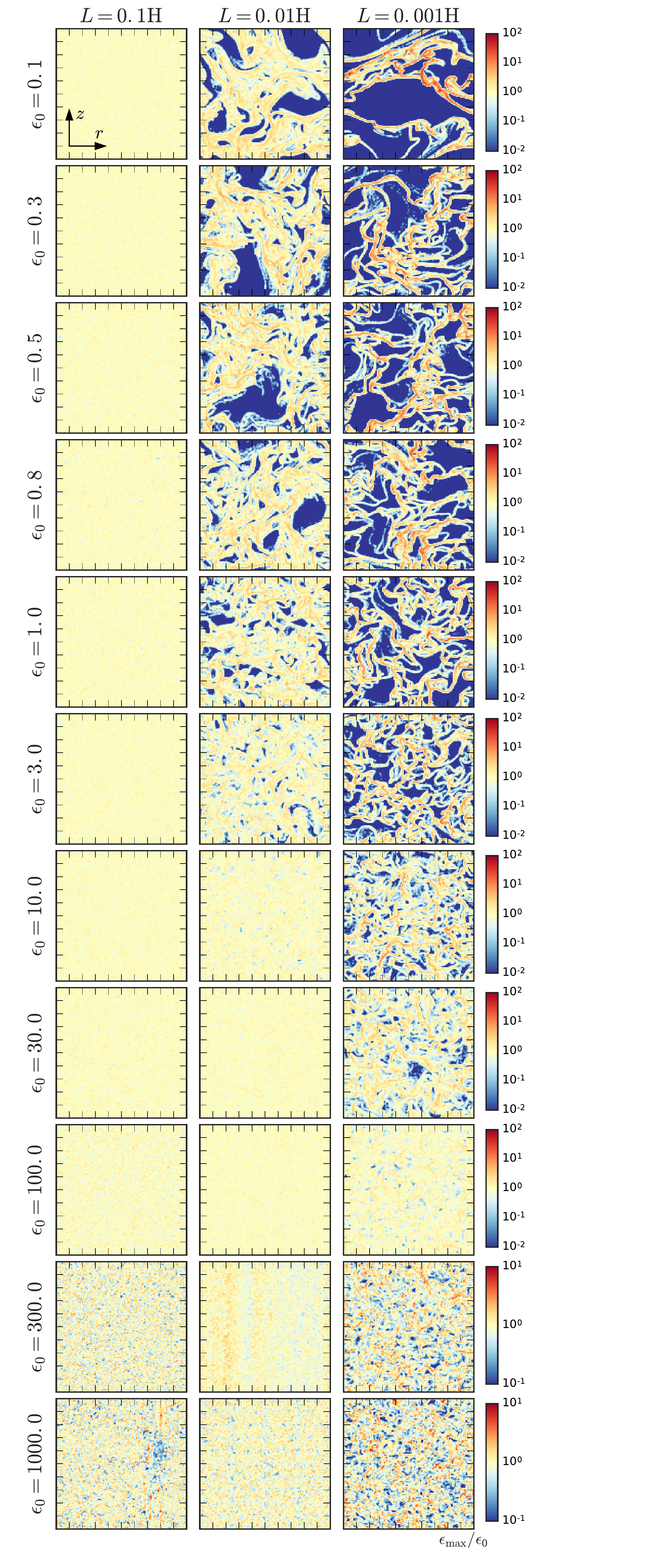}
  {fig:001_xz_imggrid}
  {$\stokes=0.01$ - $r$-$z$ plane: Last snapshots of the dust-to-gas ratio normalized to $\einit$ (yellow). Over-densities are colored in red, particle voids in blue. We do not find any signs of SI at $L=\SI{0.1}{\scaleheight}$. Though the parameter space for smaller simulations is fully populated with the SI. We do not find any zonal flow emerging. Color mapping changes for high and low $\eps$.}
  {.9}{{.2cm, .5cm, 4.cm, .2cm}} 
The last snapshots of all simulations in \Fig{fig:001_xz_imggrid} are also very similar to the previous case of aSI. The only difference is the strength of the SI pattern is weaker for high dust-to-gas ratios compared with the corresponding aSI run.
\subsubsection{Particle diffusion - $\dx$ and $\dz$}
\begin{figure} %
	\centering 
	\includegraphics[width=\columnwidth]{legend_f.pdf}\\
	\begin{subfigure}[Radial particle diffusion with $\dx\sim \epsilon^{-2.0}$ slope.]{\includegraphics[width=\columnwidth, 
			trim={0.2cm, 0.3cm, 0.2cm, 1.6cm}, clip
			]{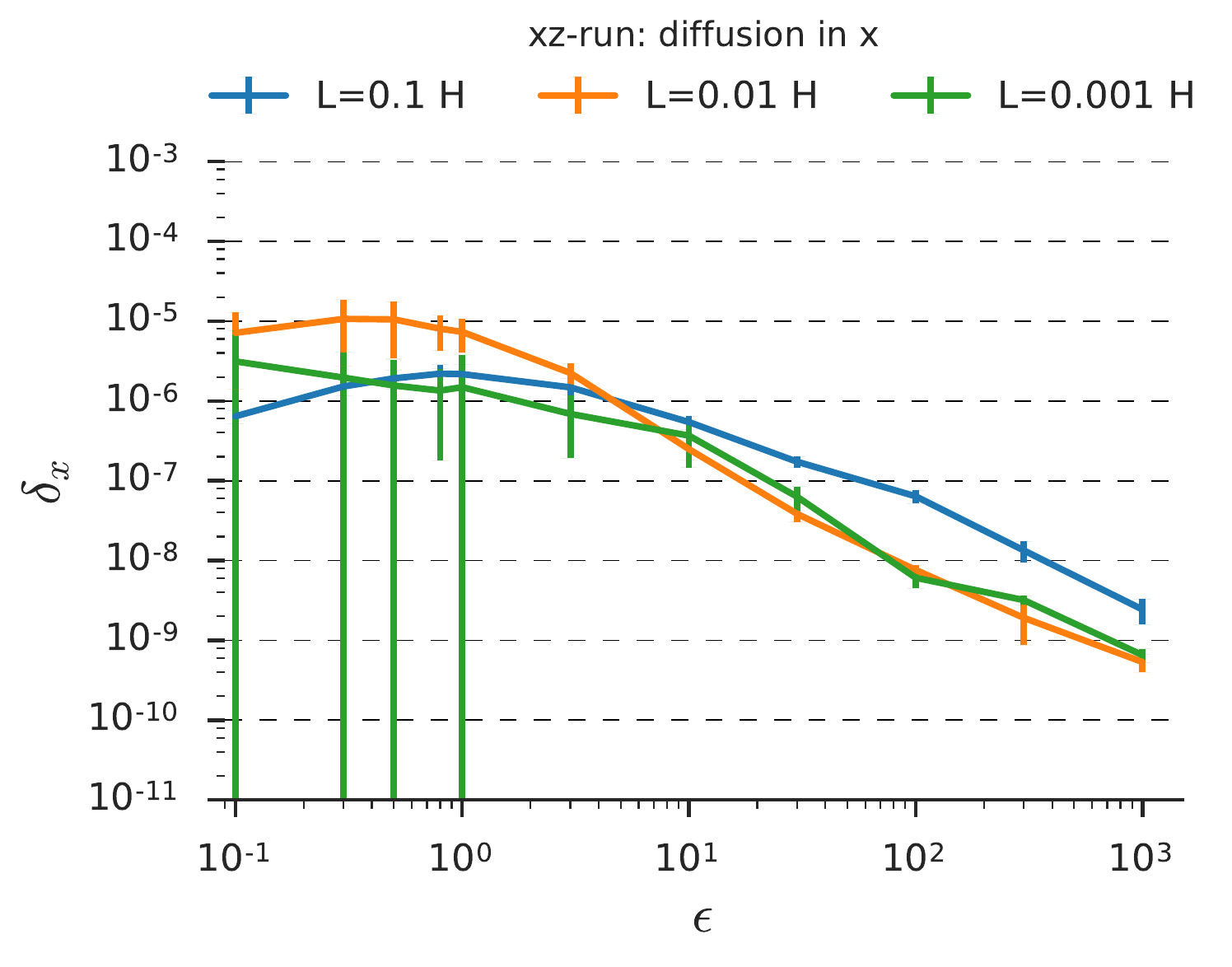}
	\label{fig:001_xz_diffusion_x}}
	\end{subfigure}\\
	\begin{subfigure}[Vertical particle diffusion with $\dx\sim \epsilon^{-0.5}$ slope.]{\includegraphics[width=\columnwidth,
			trim={0.2cm, 0.3cm, 0.2cm, 1.6cm}, clip
			]{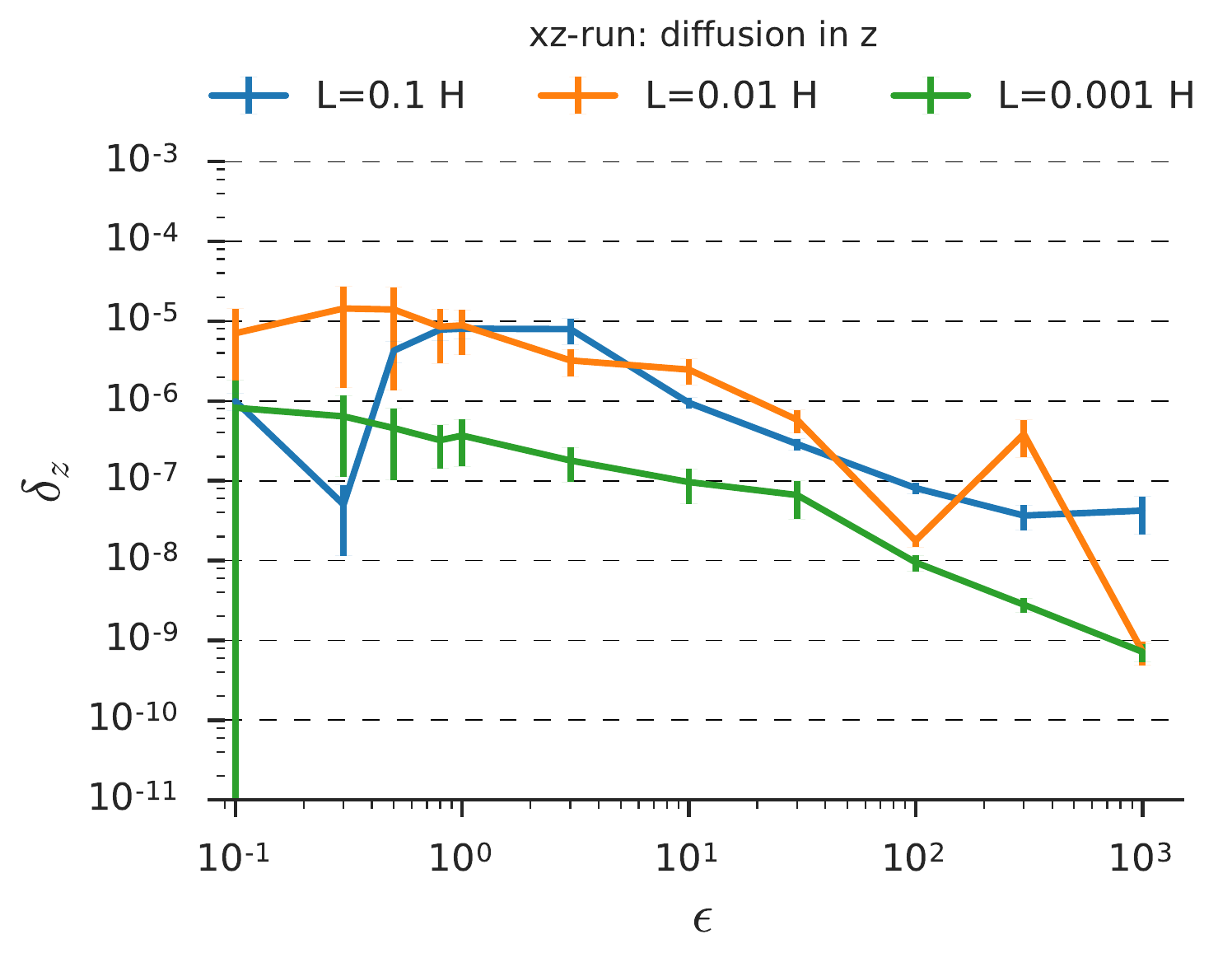}
	\label{fig:001_xz_diffusion_z}}
	\end{subfigure}
	\caption{$\stokes=0.1$ - $r$-$z$ plane: Radial (a) and vertical (b) particle diffusion estimated by treating particle movement as a random walk. Diffusion values on the largest scales (blue) are not from the SI, see \Fig{fig:001_xy_imggrid}, but from the initial particle density noise and resulting rms-velocity introduced by the drag force equilibrium between dust and gas.} %
	\label{fig:001_xz_diffusion} %
\end{figure}%
The radial particle diffusion is very similar to the ones in the $r$-$\varphi$ case. But, we find the diffusion to be a factor 2 stronger on the smallest scale (green) and SI to decay in $r$-$z$ slightly faster for higher $\epsInit$. The measured slope of this decay is $\dx\sim \epsilon^{-2.0}$. The values for the $L=\SI{0.1}{\scaleheight}$ runs is not due to SI, as can be seen in \Fig{fig:001_xz_imggrid}, but again from the initial state.

The diffusion in the vertical direction is as strong or even stronger as in the case of $\stokes=0.1$ particles. In the case where for $\stokes=0.1$ horizontal bands emerged on the smallest scales, here the SI shows no such features and the turbulent diffusion remains active. The slope is $\dz \sim \epsilon^{-0.5}$ towards larger $\epsInit$.
\subsubsection{Particle dispersion and drift - $\sigma$ and $\zeta$}
\begin{figure} %
	\centering 
	\includegraphics[width=\columnwidth]{legend_nn.pdf}\\
	\begin{subfigure}[Global turbulent dispersion $\sg$ (lines) and local turbulent dispersion $\slo$ (contour).]{\includegraphics[width=\columnwidth, 
			trim={0.5cm, 0.3cm, 1.9cm, 1.6cm}, clip
			]{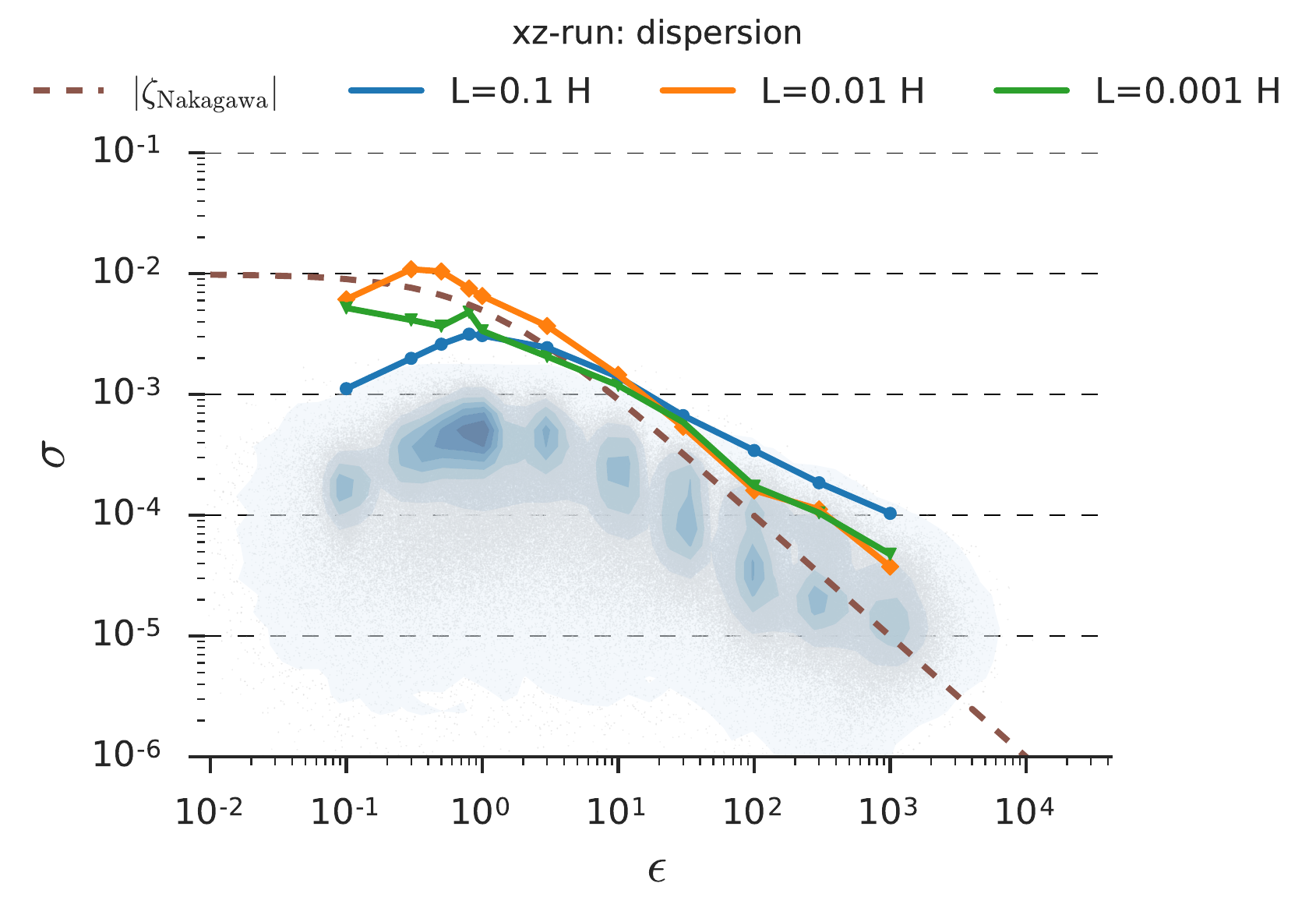}
	\label{fig:001_xz_dispersion}}
	\end{subfigure}\\
	\begin{subfigure}[Global drift $\zg$ (lines), local drift $\zlo$ (contour).]{\includegraphics[width=\columnwidth,
			trim={0.5cm, 0.3cm, 1.9cm, 1.6cm}, clip
			]{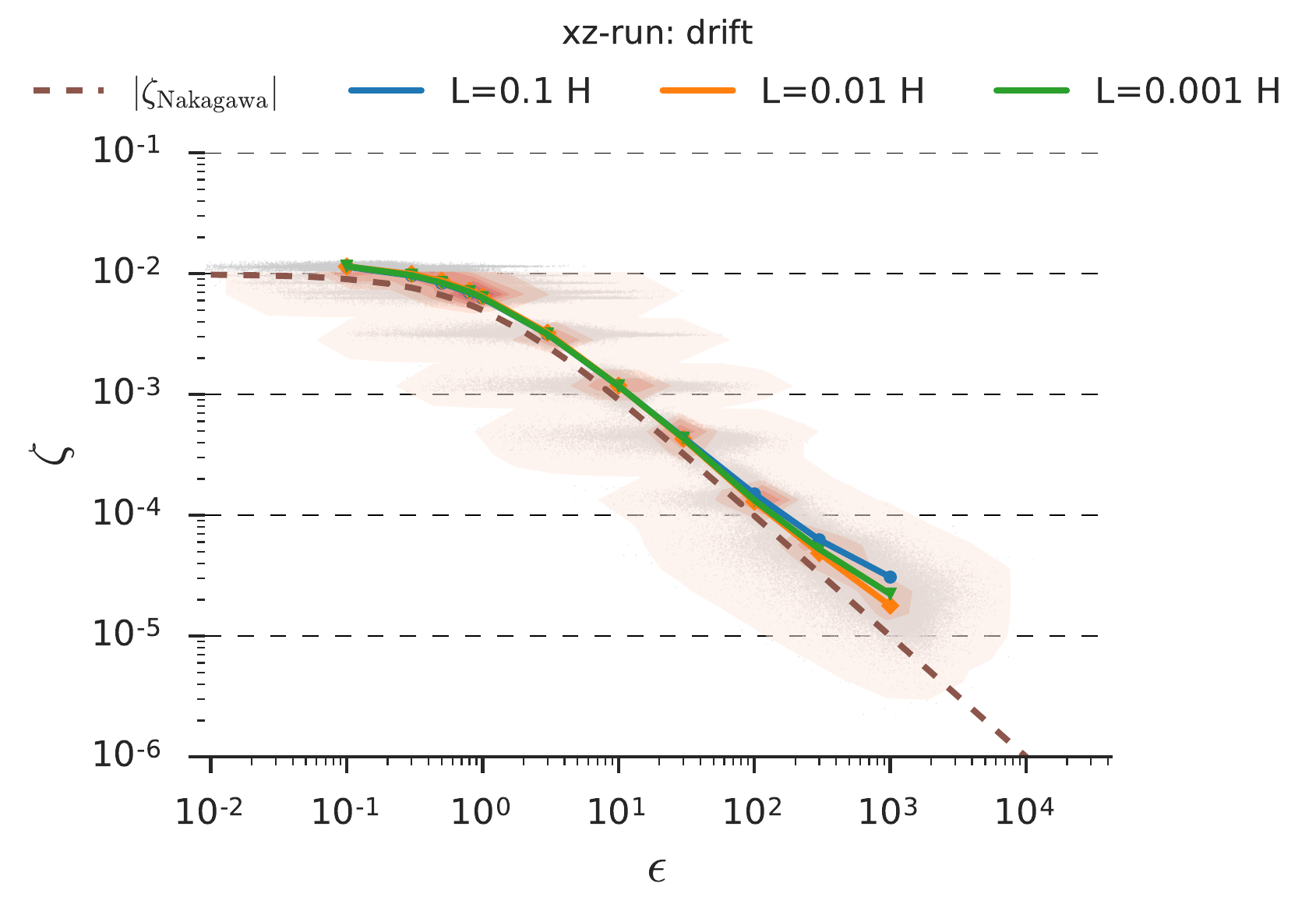}
	\label{fig:001_xz_drift}}
	\end{subfigure}
	\caption{$\stokes=0.01$ - $r$-$z$ plane: Dispersion and drift for all simulations together. Blue, orange and green lines show the individual global values. The local, i.e., grid wise, values for all simulations are shown combined in shaded contours. As reference in dashed is shown the absolute magnitude of the Nakagawa drift speed from \Eq{eq:nakagawa_drift}. Global and local drift again shown perfect agreement, as in the runs with $\stokes=0.1$, but now closer to the Nakagawa solution. The local dispersion values are again always well below the global values. But, in constrast to the $\stokes=0.1$-runs now for $\epsInit\leq1$ the SI-active simulations have dispersion velocities that are closer to the Nagakawa drift.} %
	\label{fig:001_xz_lc_drift_disp} %
\end{figure}%
The turbulent particle dispersion, shown in \Fig{fig:001_xz_dispersion}, is almost identical to the dispersion we found for the $r$-$\varphi$ case.

\Fig{fig:001_xz_drift} shows the particle drift. Again the situation is very similar to the $r$-$\varphi$ case.
The correlation time from \Eq{eq:tauCorr} is consequently also similar to the case of aSI for $\stokes=0.01$ particles. On the smallest and the intermediate scales $\tauCorr$ is flat, around $\tauCorr \approx 0.1 \Omega^{-1} \dots 0.2 \Omega^{-1}$. It also rises once the aSI is dead and reaches a value of $\tauCorr\approx 1\Omega^{-1}$.
\subsubsection{$\alpha$-value and Schmidt number}
\begin{figure} %
	\centering 
	\hspace{0.6cm}\includegraphics[width=.8\columnwidth]{legend_f.pdf}\\
	\begin{subfigure}[$\alpha$-value: turbulent gas viscosity]{\includegraphics[width=\columnwidth, 
			trim={0.2cm, 0.3cm, 0.2cm, 1.6cm}, clip
			]{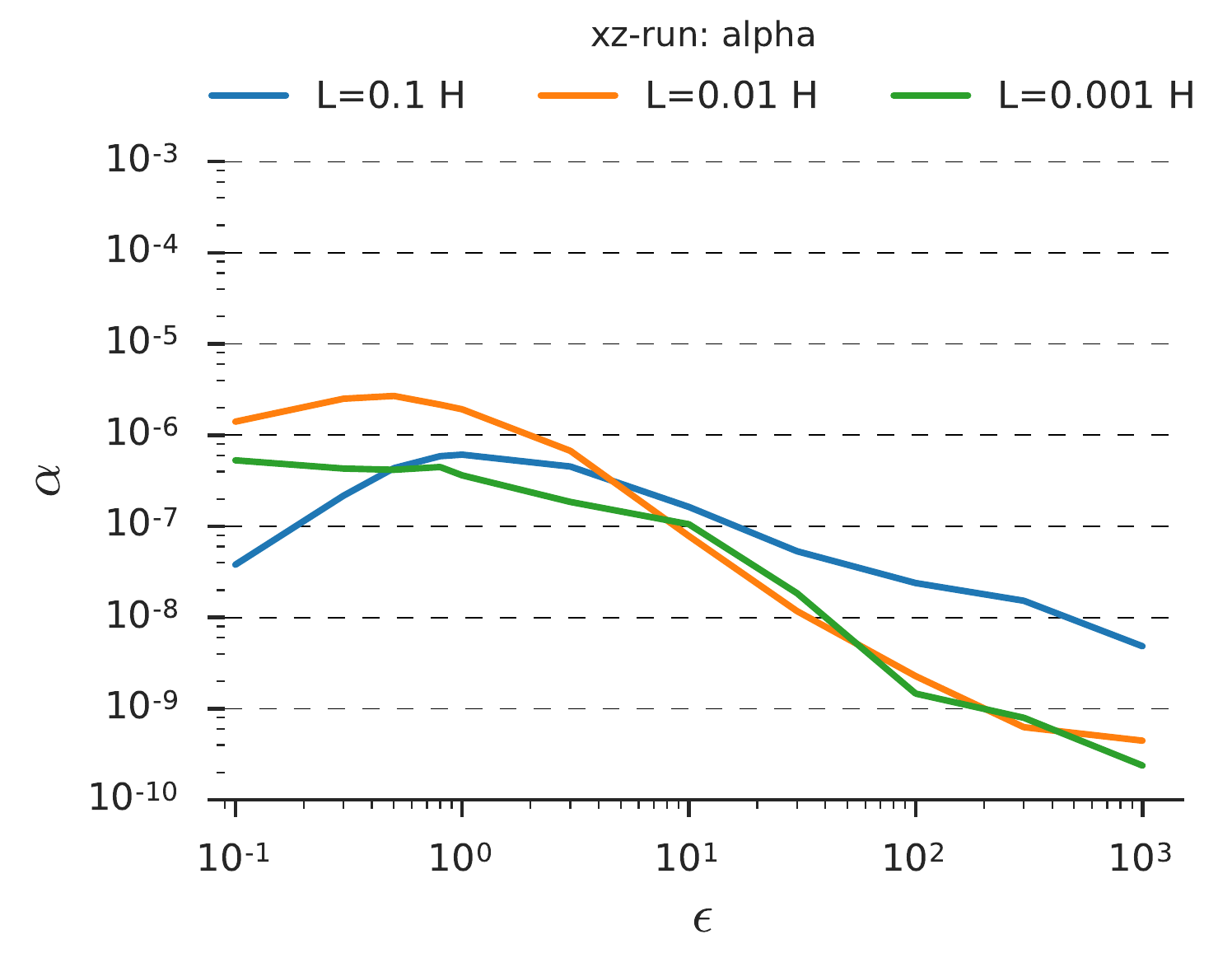}
	\label{fig:001_xz_alpha_paramplot}}
	\end{subfigure}\\
	\begin{subfigure}[Schmidt number]{\includegraphics[width=\columnwidth, trim={0.2cm, 0.3cm, 0.2cm, 1.6cm}, clip]{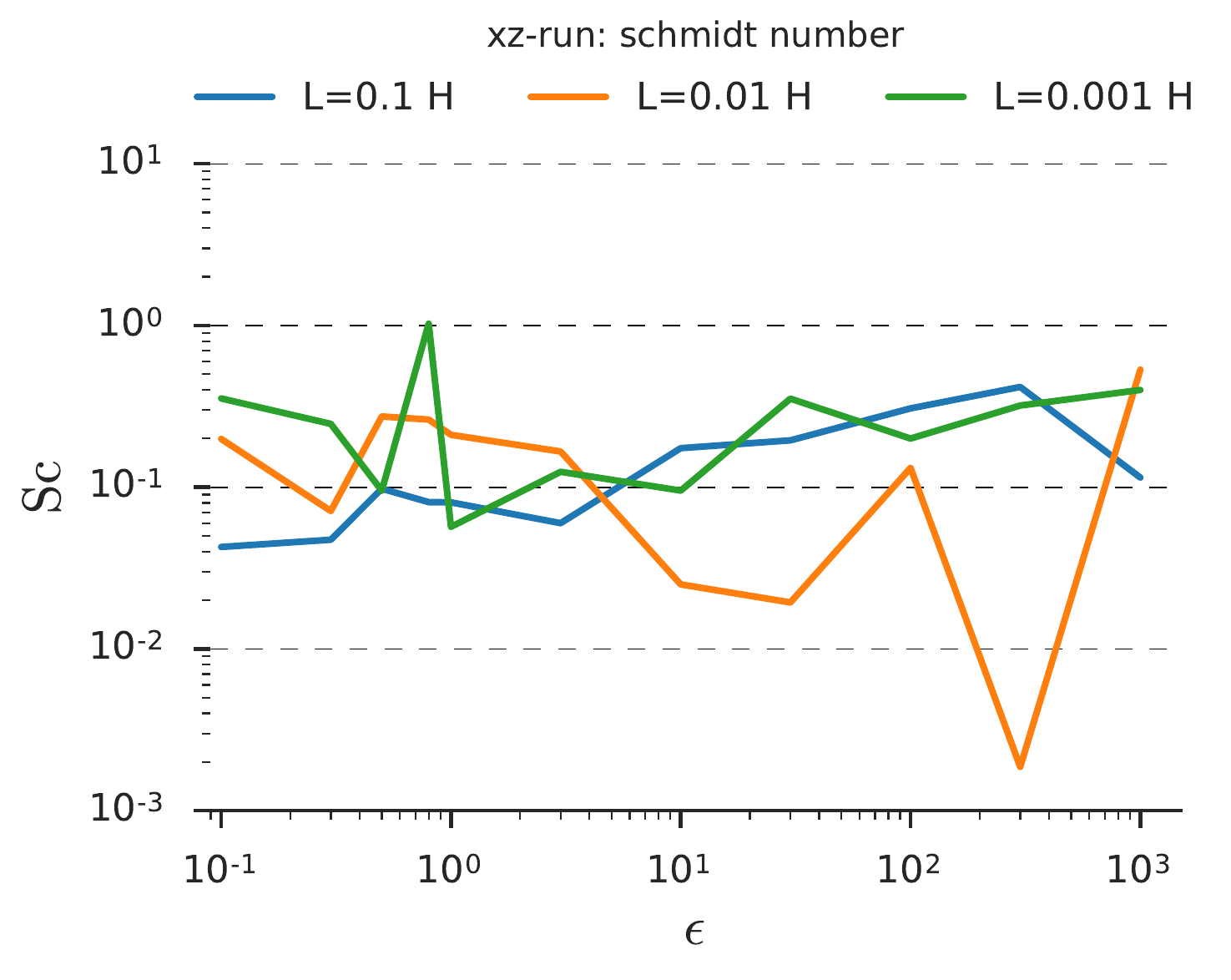}
	\label{fig:001_xz_schmidtnum}}
	\end{subfigure}
	\caption{$\stokes=0.01$ - $r$-$z$ plane: The gas $\alpha$-turbulence shows a steep drop-off once the SI is weaker for high dust-to-gas ratios. Since SI is for $\stokes=0.01$ active throughout the whole parameter space, this results in a completely constant Schmidt number mostly below $\approx1$, again showing that particle diffusion is stronger than the gas turbulent transport.} %
	\label{fig:001_xz_alpha_schmidt} %
\end{figure}%
\Fig{fig:001_xz_alpha_paramplot} shows the measured level of $\alpha$-turbulence. We find it to be just a bit lower to the case of $\stokes=0.1$, see \Fig{fig:01_xz_alpha_paramplot}. The difference is again that the SI and thus gas turbulence is active for $\epsInit\leq1$.
\section{Resolution study on aSI at $L=\SI{0.1}{\scaleheight}$ with $\stokes=0.1$ particles}
\label{app:res_study}
Here we study the resolution dependency of the aSI at the $L=0.1$ length scale. We set up simulations as in \Sec{sec:01-xy} but limit ourself to the ones with $\epsInit \geq 1$. We use grid resolutions of $N_x=N_y=256$ and $1260$, hyper-viscosity and -diffusivity are altered to resolve the very small length scales. The simulation end-states of this study are shown in \Fig{fig:res_study_simgrid} and compared with the corresponding original simulations from \Sec{sec:01-xy}.

We again measure the maximum dust density fluctuation normalized to the initial dust-to-gas ratio and now compare it with the values from the $N=128$ simulations, see \Fig{fig:res_study_dust_enhancement_comp}. Due to the higher resolution, the azimuthal streaming instability is able to develop smaller modes that we find to be active even within the larger aSI modes and emerging zonal flows. We find this further concentration of the dust increasing the maximum dust-to-gas ratio but only by a factor of $\sim 2$ to $3$. We observe especially the zonal flows from \Sec{sec:01_xy_zonalFlow} to get further refined, though dust is not significantly stronger concentrated. For $\epsInit=100$ we now find clear azimuthal band structures that have a larger wavenumber than the zonal flows in $\eps=30$ and mark a transition to a realm where the radial aSI wavelengths get very small and are hardly resolved. What we find is that the dust density fluctuations at very large dust-to-gas ratios still increases with resolution and three times higher dust-to-gas ratios are found than in the $N=128$ stimulations.

\Fig{fig:res_study_diffusion_comp} compares the measured $\dx$ with the values from the $128$ runs. We find mostly similar values, only for $\eps=1000$ the simulation with $N=256$ doubles the diffusion value. Overall, we conclude that our $N=128$ simulations were not hampered by its coarser resolution.
\FIGURE%
  {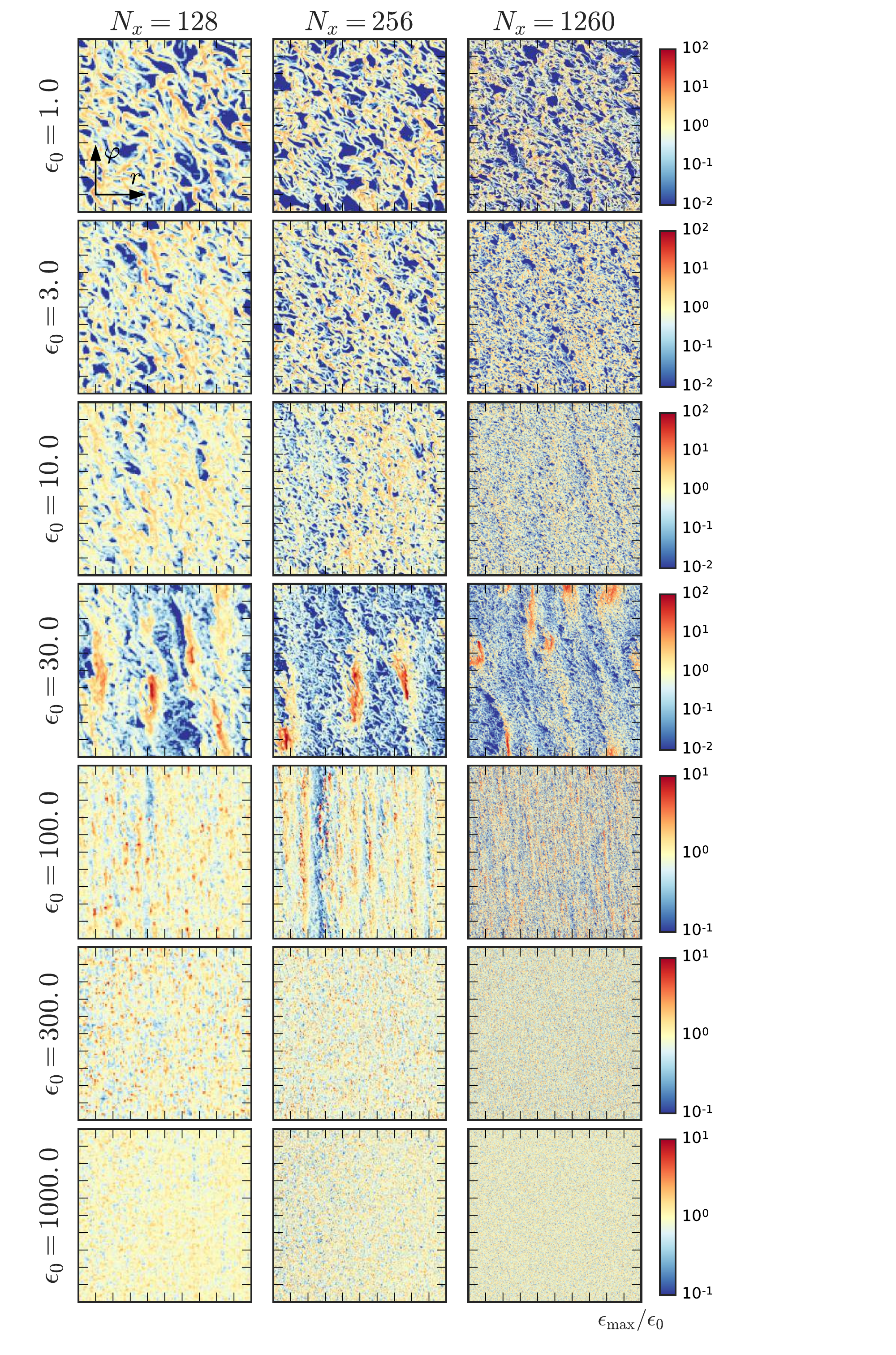}%
  {fig:res_study_simgrid}%
  {$\stokes=0.1$ - $r$-$\varphi$ plane - $L=\SI{0.1}{\scaleheight}$: Last snapshots of the dust-to-gas ratio normalized to $\einit$ (yellow). Over-densities are colored in red, particle voids in blue. The aSI mode pattern is more refined with increasing numerical resolution. The three zonal flows from \Sec{sec:01_xy_zonalFlow} for $\epsInit=30$ also emerge at the doubled resolution. At $N=1260$ we find seven zonal flows with high particle concentration that are smaller in both azimuthal and radial direction than in the runs with coarser resolution. The emerging particle over-densities show aSI activity inside and signs of erosion, i.e., particle flows in radial outward direction  at the azimuthal back of the heap.}%
  {.9}{{.2cm, .5cm, 4.cm, .2cm}}%
\FIGURE%
  {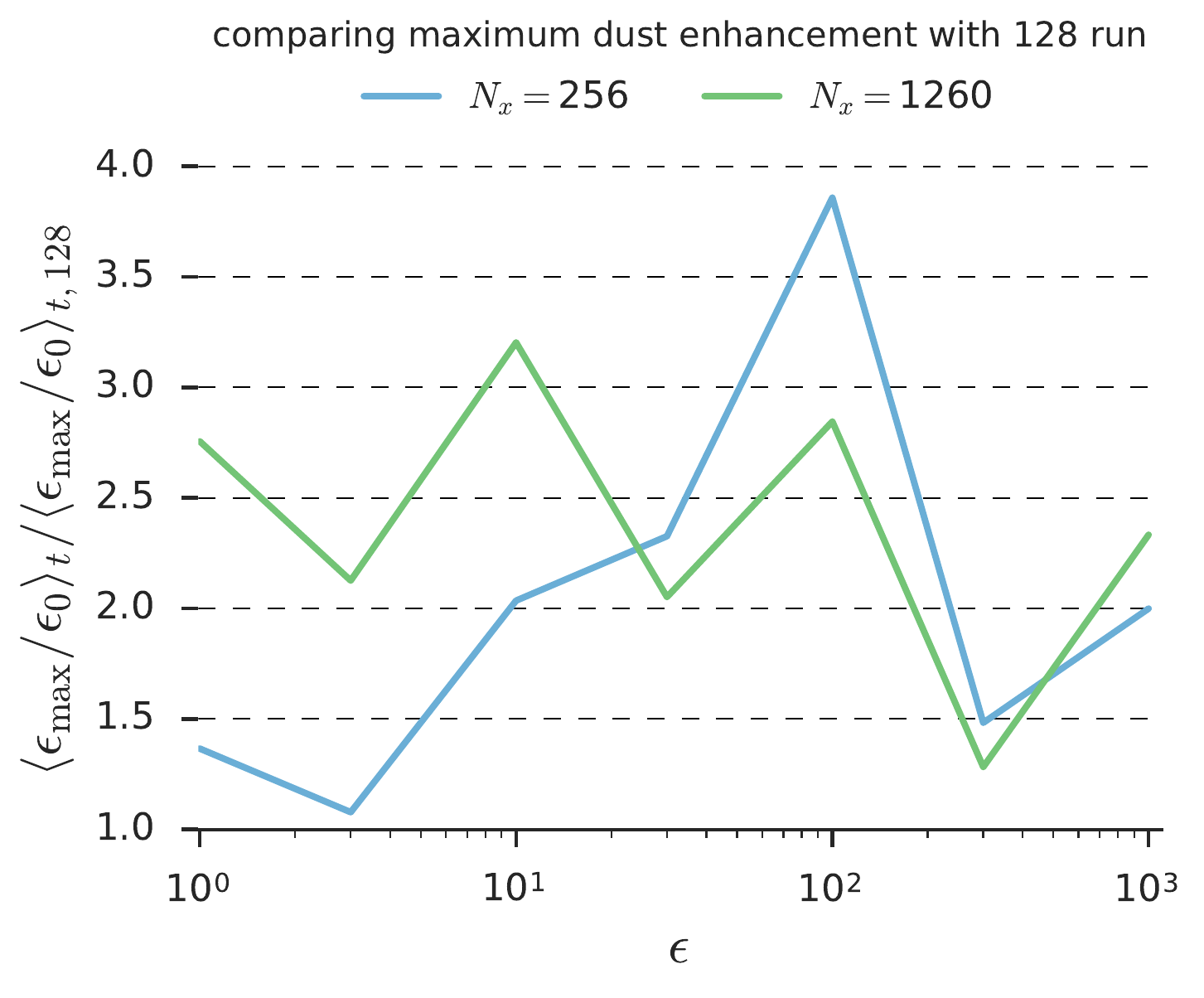}%
  {fig:res_study_dust_enhancement_comp}%
  {$\stokes=0.1$ - $r$-$\varphi$ plane: Comparing dust density fluctuation values of the resolution study with the values from the original $N=128$ simulations. The aSI is capable to increase maximum dust-to-gas ratios up to a factor of 4 stronger, when resolution is increased. Not all the simulations for $N=1260$ where for a long time in saturation, but the highest values we find are not indicating that this high resolution is strongly increasing the maximum dust-to-gas ratio.}%
  {1}{{.2cm, .3cm, .3cm, .7cm}}%
\FIGURE%
  {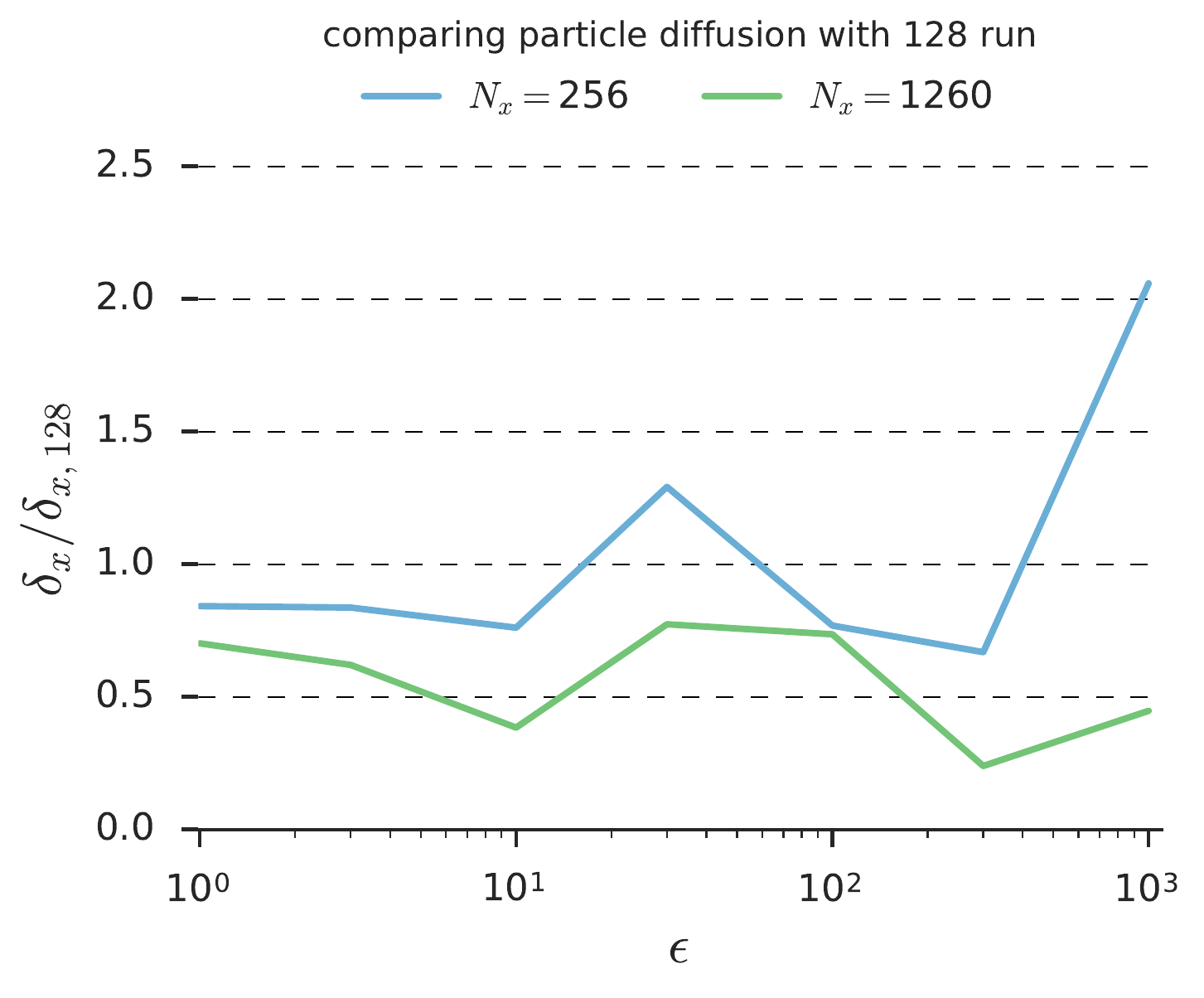}%
  {fig:res_study_diffusion_comp}%
  {$\stokes=0.1$ - $r$-$\varphi$ plane: Comparing radial particle diffusion $\dx$ of the resolution study with the values from the original $N=128$ simulations. We can find a decrease in the diffusivity with increasing numerical resolution by a factor of $0.5$ at maximum.}%
  {1}{{.2cm, .3cm, .3cm, .7cm}}%
%
%
%
\section{Discussion and Summary}
\label{sec:discussion}
The SI as described in YG04 is an instability acting only in combination with radial with vertical directions and also \cite{Squire2017a} did not change that picture. We presented in this paper unstratified simulations for $\stokes=0.1$ and $0.01$ particles in a 2-d/2.5-d approach that well resolves either $r$-$\varphi$ or $r$-$z$, but limits the vertical/azimuthal direction to a single grid cell, i.e., vertical/azimuthal modes are suppressed. In agreement with literature, we  find in our $r$-$z$ simulations the SI to be active, however, we go to initial dust-to-gas ratios of up to $\einit=1000$ which has not been covered in previous works. 

$\bullet$ \textit{Azimuthal streaming instability:} %
Moreover, when going to $r$-$\varphi$ simulations, we find a very similar instability appearing that shares the characteristics of the SI to enhance dust concentrations locally and provide particle diffusion. Both at almost identical strengths and with similar mode patterns plus similar growth rates, see \Tab{tab:growthrates}.
\begin{table}[b]
\resizebox{\columnwidth}{!}{%
\begin{tabular}{rcccccc}
	\toprule
	               &   $L=0.1$    &   $L=0.1$    &   $L=0.01$   &   $L=0.01$   &  $L=0.001$  &  $L=0.001$  \\
	               &      SI      &     aSI      &      SI      &     aSI      &     SI      &     aSI     \\ \midrule
	 $\stokes=0.1$ & $\num{4e-1}$ & $\num{3e-1}$ & $\num{8e-1}$ & $\num{1e0}$  & $\num{1.5}$ & $\num{2e0}$ \\
	$\stokes=0.01$ &      -       &      -       & $\num{5e-1}$ & $\num{5e-1}$ & $\num{3.0}$ & $\num{4.0}$ \\ \bottomrule
\end{tabular} 
}
\caption{Mean growth rates for SI and aSI simulations in comparison. They tend to not significantly depend on dust-to-gas ratio and be similar not only for SI and aSI, but also for $\stokes=0.1$ and $0.01$ particles. Growth rates in units of $\Omega$.}
\label{tab:growthrates}
\end{table}
We refer to this instability as the azimuthal streaming instability (aSI), but suggest calling it streaming instability nevertheless, for simplicity. So far no dispersion relation for this aSI has been solved and detailed growth rates are unknown. With this in mind, we propose the SI in 3-d simulations can only be understood if it is seen as a combination of SI and aSI, that future work will have to account for. We suggest an approximation of the YG04 findings for the case of growth time-scale being shorter than the shearing time-scale, i.e. $\tau_{\rm shear} = 2/3 \Omega^{-1} \leq s^{-1}$, which is a factor of 2 away from our smallest measured growth rates of $s = \num{3e-1}$.

$\bullet$ \textit{Diffusion on the smallest scales:} %
Comparing the two different Stokes numbers the (a)SI on a fixed length scale, the (a)SI with the larger particles grows faster, which is consistent with usual expectations for fastest growth at $\stokes\approx1$. We further confirm YG04 findings on the SI growth rate to be the largest on small length scales. A consequence for simulations that deal with SI and planetesimal formation is that the SI might seem dead on large scales, but remains active on these very small scales, providing particle turbulence that might alter the outcome of particle cloud collapse. For an increasing dust-to-gas ratio, we find a decrease in particle diffusivity following a $\delta \sim \epsilon^{-1} \dots \epsilon^{-2}$ slope. Moreover, we observe aSI and SI to be active even for values above $\einit > 100$, though the strength in dust density fluctuation as well as particle diffusion drops. We find for our specific simulation setups that the SI is capable of local particle density fluctuations up to a factor of $\sim10$ to $40$, depending on numerical resolution. The diffusion could be the reason why no further particle concentration occurs in our simulations. For the aSI we find with increasing numerical resolution an decrease in particle diffusion by a factor of $<2$ and an increase in maximum particle density of $<3$, see \Sec{app:res_study}.

Comparing the aSI and SI at two different Stokes numbers but fixed dust-to-gas ratio, the largest active modes we observe to get smaller with smaller Stokes number. For example, for $\stokes=0.01$ we do not find much of (a)SI activity on the scales of $L=\SI{0.1}{\scaleheight}$, on which scales the (a)SI for $\stokes=0.1$ is actually at its strongest.
In our setups for $\stokes=0.1$, we further find vertical mode structures and zonal flows emerging when going to high dust-to-gas ratios that both seem to have the same wave numbers. They are limited to our largest domain size for $r$-$\varphi$ simulations, but for $r$-$z$ simulations also appear in the $L=\SI{0.01}{\scaleheight}$ simulations and in general are more prominent in $r$-$z$ setups. These bands and zonal flows are themselves fully turbulent. This suggest that SI remains active inside of these structures, explaining the significant particle diffusion value measured in the respective direction of the structure.
For dust-to-gas ratios around and below unity, the (a)SI is active down to $\epsInit \approx 0.5$ and especially for $\stokes=0.01$ down to $ \epsInit \approx 0.1$ and probably even lower.

\FIGURE%
  {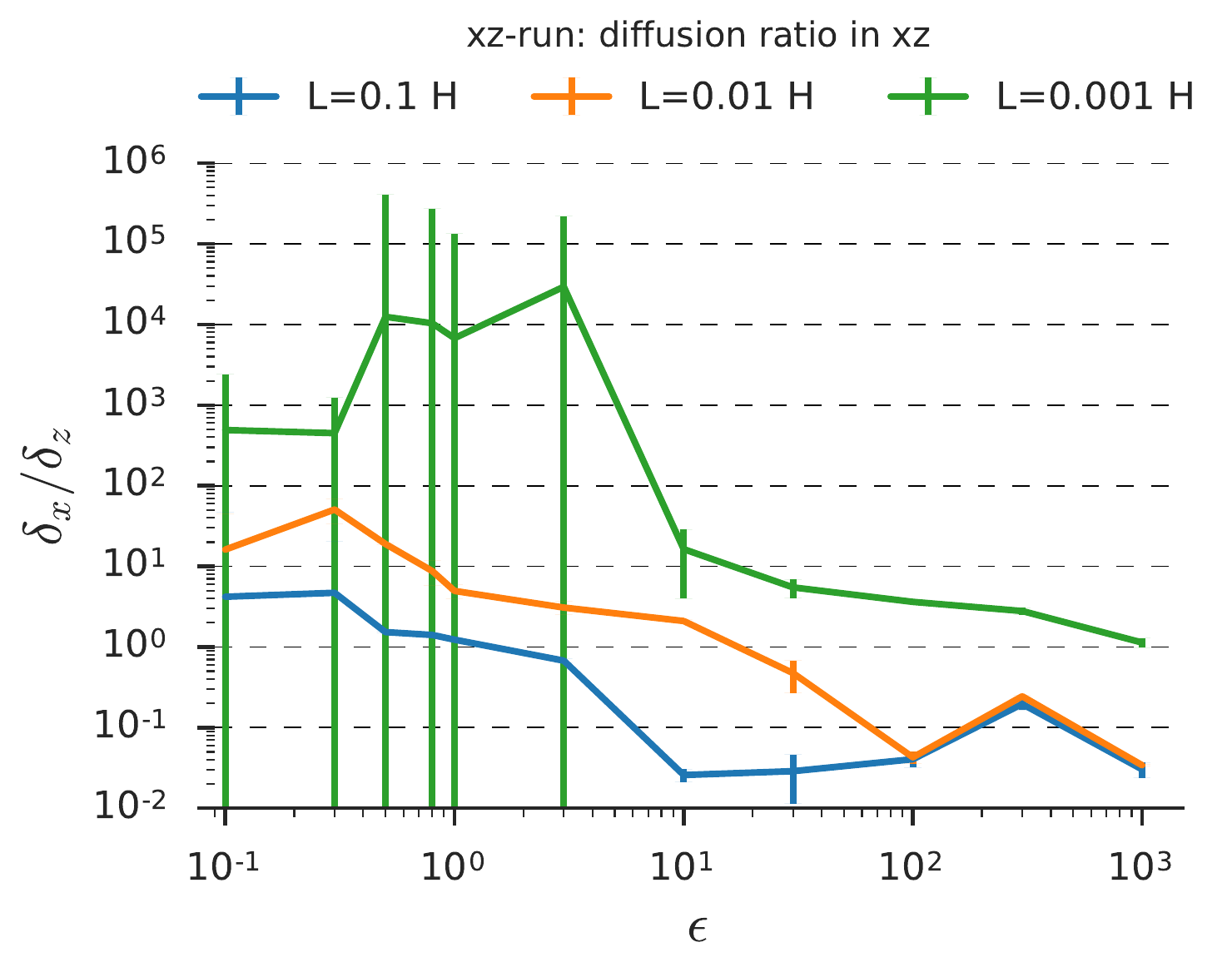}%
  {fig:01_xz_diffusion_ratio}%
  {Comparing horizontal with vertical particle diffusion of the SI for the case of $\stokes=0.1$ particles. Mostly the radial diffusion is the strongest. Only when bands structures appear in the simulations, compare with \Fig{fig:01_xz_imggrid}, the corresponding direction gets the preferred diffusion direction.}
  {1}{{.3cm, .3cm, .3cm, .7cm}}%
We find the vertical diffusion to be mostly lower than the radial diffusion, see \Fig{fig:01_xz_diffusion_ratio} for a comparison for the $x$-$z$ SI simulations with $\stokes=0.1$. Once vertical modes or horizontal bands appear, the strongest diffusion is in the direction of the corresponding particle concentrating structure.
Comparing turbulent gas transport $\alpha$ with particle diffusivity $\delta$, Schmidt values are of $\num{e-1} < \Sc < \num{e0}$. For smaller Stokes number the Schmidt number is slightly lower. Hence, one cannot assume $\delta \approx \alpha$ for the case of active (a)SI. When trying to link particle rms-velocities to their diffusion values, we find an average correlation time of $\tauCorr\approx0.3$ for all (a)SI-active runs. Non-(a)SI-active simulations typically show $\tauCorr=1.0$ and where (a)SI-activity is decaying, values are in-between these two values.

$\bullet$ \textit{Zonal flows:} %
For $\stokes=0.1$ we further find single horizontal bands to appear in the $r$-$z$ setups at these low dust-to-gas ratios, but they are limited to the smallest simulations and are not present in one order of magnitude larger simulations. 
In these cases, the SI wavelength in the intermediate simulation is much larger than the domain size of the smaller simulation. This suggests that the appearance of band structures and zonal flows could be an incarnation of larger (a)SI modes into the small simulations. Consequently, the large modes are limited down to the appearance of single direction modes. 
Whether this has an implication on the appearance of SI in nature, needs further investigations. Under the right circumstances they might assist in the formation of rings as observed recently with ALMA in PPDs, such as the HL Tau system.
Especially, since PPDs in reality are stratified by vertical gravity and hence are in a gravo-turbulent state of the SI, see \cite{Carrera2015} and \cite{Bai2010}. The (shear-)periodicity of our simulations might also enhance modes that would normally not grow in isolated particle clouds of finite size.

The often discussed 'traffic jam effect' that is emerging from the Nakagawa drift, where high dust-to-gas ratios bring radial particle drift to a halt, we could observe up to $\eps=100$. Above that, the particle drift speed deviates from the Nakagawa solution towards slightly higher drift speeds, but still lower than in the limit of single particles, i.e. $\eps\approx0$. We question how much of this is a result of the enforced gas pressure gradient, since it is the energy source of the (a)SI and 'unlimited' in our simulations. This must not be true for a PPD, where high dust-to-gas ratios might alter the gas pressure gradient on comparable large scales. Hence, the picture of the SI as a particle concentrating process might be limited by the fact that altering the gas pressure gradient does consequently reduce its strength in concentrating particles. This needs to be further investigated in future work.
Another finding is that the SI can already be active on very small scales for very low $\epsilon$-values for particles with $\stokes=0.01$, needing also further investigations.

$\bullet$ \textit{Summary:} In this work, we showed that the SI can also exist in the $r$-$\varphi$ plane, where vertical modes are suppressed. This SI in $r$-$\varphi$ we call the azimuthal streaming instability. It is found to operate with very similar properties as the SI, providing a 2-d test-bed for numerical experiments on planetesimal formation that include both shear and SI but remain low in computational cost. Still, how strongly the (a)SI deviates in 2-d from its 3-d equivalent needs to be studied in future work. We showed that even the small (a)SI modes actively contribute to particle diffusion, acting on the relevant scales of planetesimal formation and need to be resolved in particle cloud collapse simulations.
%
%
%
%
\section*{Acknowledgments}
{\small We are indebted to Andrew Youdin, Hans Baehr, Christian Lenz and Anders Johansen for many fruitful discussions.
This research has been supported by the Studienstiftung des deutschen Volkes, the Deutsche Forschungsgemeinschaft Schwerpunktprogramm (DFG SPP) 1385 "The first ten million years of the Solar System" under contract KL 1469/4-(1-3) "Gravoturbulente Planetesimal Entstehung im fr\"uhen Sonnensystem" and by (DFG SPP) 1833 "Building a Habitable Earth" under contract KL 1469/13-1 "Der Ursprung des Baumaterials der Erde: Woher stammen die Planetesimale und die Pebbles? Numerische Modellierung der Akkretionsphase der Erde." This research was supported by the Munich Institute for Astro- and Particle Physics (MIAPP) of the DFG cluster of excellence "Origin and Structure of the Universe and in part  at KITP Santa Barbara by the National Science Foundation under Grant No. NSF PHY11-25915. 
The authors gratefully acknowledge the Gauss Centre for Supercomputing (GCS) for providing computing time for a GCS Large-Scale Project (additional time through the John von Neumann Institute for Computing (NIC)) on the GCS share of the supercomputer JUQUEEN at J\"ulich Supercomputing Centre (JSC). GCS is the alliance of the three national supercomputing centres HLRS (Universit\"at Stuttgart), JSC (Forschungszentrum J\"ulich), and LRZ (Bayerische Akademie der Wissenschaften), funded by the German Federal Ministry of Education and Research (BMBF) and the German State Ministries for Research of Baden-W\"urttemberg (MWK), Bayern (StMWFK) and Nordrhein-Westfalen (MIWF). Additional simulations were performed on the THEO and ISAAC cluster owned by the MPIA and the HYDRA and DRACO clusters of the Max-Planck-Society, both hosted at the Max-Planck Computing and Data Facility in Garching (Germany). }
\appendix
%
%
\section{Influence of Hyper-Viscosity and -Diffusivity on the (a)SI}
\label{app:hyperhyper}
\FIGURE%
  {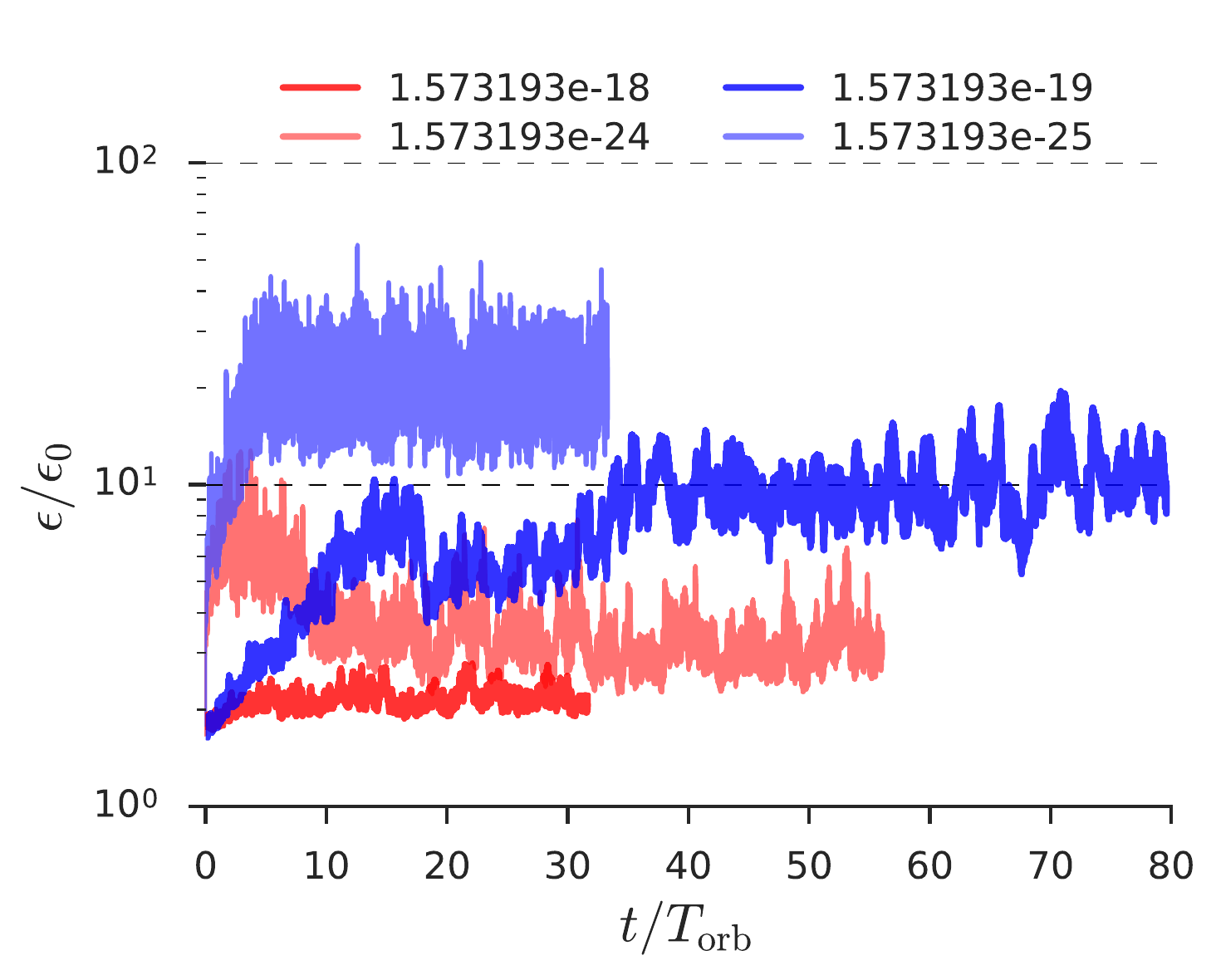}%
  {fig:hyper_diff_visc}%
  {Timeseries of maximum dust-to-gas ratio normalized to the initial dust-to-gas ratio for two cases of ($L=\SI{e-1}{\scaleheight}$, $r$-$\varphi$, $\epsInit=\num{100}$) and ($L=\SI{e-2}{\scaleheight}$, $r$-$\varphi$, $\epsInit=\num{0.5}$). The choice of hyper-viscosity and -diffusivity strength can suppress the aSI since small modes grow the fastest.}%
  {1}{{.3cm, .3cm, .3cm, .7cm}}%
We operate the whole parameter study with a resolution dependent but fixed hyper-viscosity and -diffusivity value of $\num{1.573e-19}$, $\num{1.573e-24}$ and $\num{1.573e-29}$ for the three simulation domain sizes respectively. We found these values to not be the lowest stable value for all simulations and by decreasing this value many of the (a)SI-inactive simulations could be populated with (a)SI. Still, we were interested in having the same value of hyper-viscosity and -diffusivity for all simulations of our parameter study to have them consistent. To comprehend this additional parameter we performed additional simulations with $\stokes=0.1$ particles.

The first run is ($L=\SI{e-1}{\scaleheight}$, $r$-$\varphi$, $\epsInit=\num{100}$) where we increased the hyper-diffusivity and -viscosity by one order of magnitude. The result is $\left< \epsMax/\epsInit \right>_t \approx 2$ instead of $10$, compare \Fig{fig:01_xy_dust_enhancement}. The particle diffusivity in this case dropped from $\dx = \num{1.93e-07}$ down to $\num{4.84e-08}$. %

The second run is ($L=\SI{e-2}{\scaleheight}$, $r$-$\varphi$, $\epsInit=\num{0.5}$) with one order of magnitude lower hyper-viscosity and -diffusivity. The previously aSI-dead simulation now was populated with aSI and showed $\left< \epsMax/\epsInit \right>_t \approx 20$ instead of $3$.  The particle diffusivity increased from $\dx = \num{1.09e-07}$ to $\num{1.42e-5}$. %

From the point of numerical resolution all of these simulations are able to resolve the (a)SI. But fastest growth happens on small wavelengths that can get suppressed by hyper-viscosity and -diffusion. This can prevent growth of larger modes as well, since the initial perturbation is not strong enough. We see this also in the time evolution of our simulations. First small modes grow, it takes the large modes many small-mode growth rates to develop, too. \pencilCode{} users we recommend using the mesh based hyper-diffusion and -viscosity as presented in \cite{Yang2012}.
\section{Correlation length}
\label{sec:01_xz_corrL} %
\begin{figure} %
	\centering %
	\hspace{0.6cm}\includegraphics[width=0.8\columnwidth]{legend_f.pdf}\\
	\includegraphics[width=\columnwidth, 
		trim={0.2cm, 0.3cm, 0.2cm, 1.6cm}, clip
		]{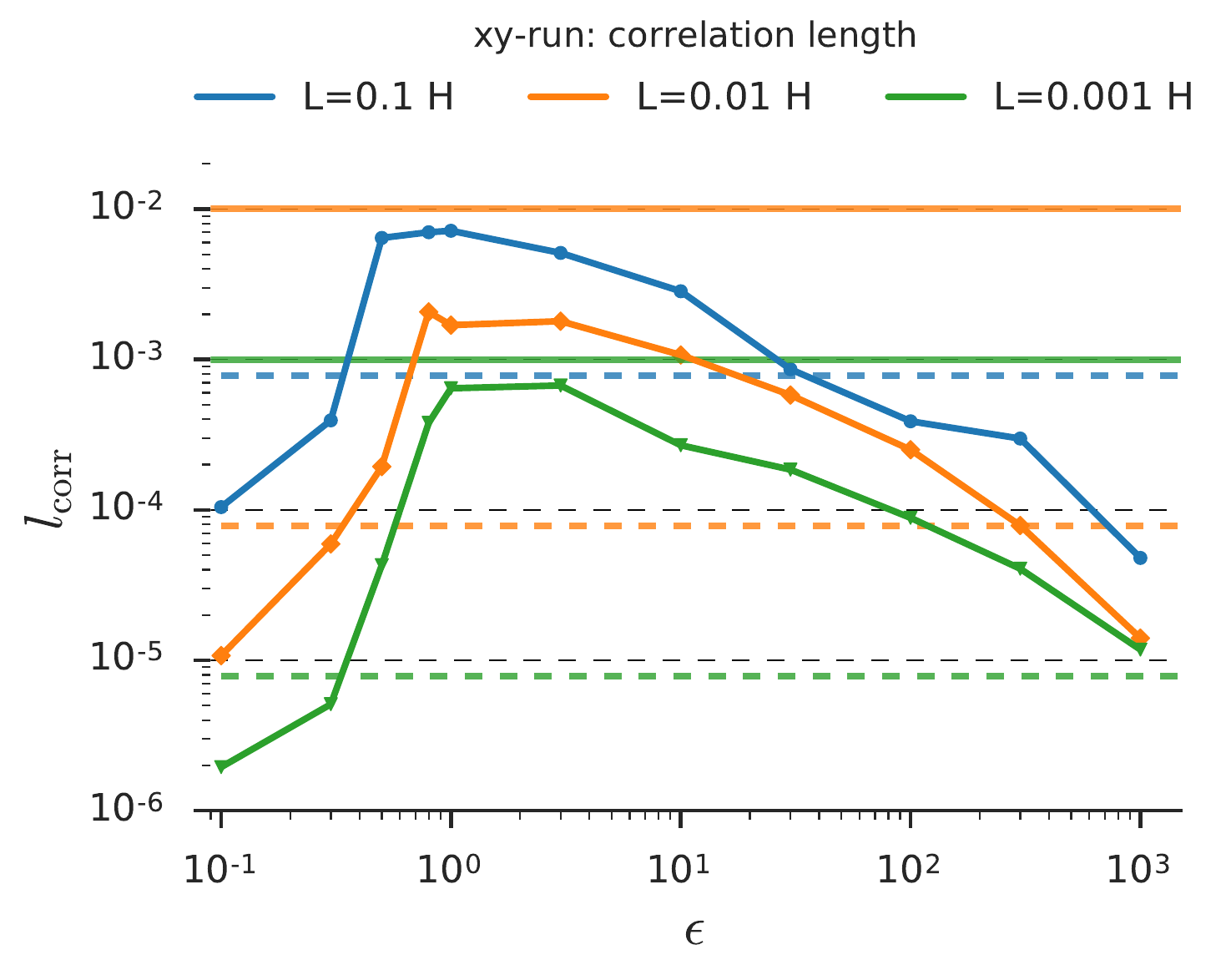}
	\caption{$\stokes=0.1$ - $r$-$\varphi$ plane: Correlation length compared with corrosponding $\de x$~(dashed) and $L_x$ (solid lines). The correlation length gives an indication whether characteristic modes are resolved since its a measure for the turbulent eddy size. \label{fig:01_xy_correlationlength}}
\end{figure}%
If is possible that our simulations do not resolve the dominant scales of the turbulent eddies. We thus compute the correlation length
\begin{equation}
\label{eq:lcorr}
\lCorr = \frac{\delta}{\sigma}\, ,
\end{equation}
which is a measure of the turbulent eddy size. If $\lCorr$ is getting smaller than $\de x$ they are not resolved by the simulation grid and eddies larger than the domain size $L$ should not be present as they do not fit inside the simulation domain. It can also be seen as a 'poor mans Fourier analysis', giving us the most prominent eddies size. \Fig{fig:01_xy_correlationlength} shows this as an example for the aSI simulations with $\stokes=0.1$ in the $r$-$\varphi$ plane. $\lCorr$ is always smaller than the domain size and only below the grid scale for the cases of vanishing SI. Results from simulations with $\lcorr < \de x$ should thus be trusted with reservations, see list of $\lcorr$ for all simulations in \Tab{tab:r-phi-runs} and \Tab{tab:r-z-runs}.
\section{Comparison with Turbulent Cascade}
\label{app:kolmogorov} %
\begin{figure} %
	\centering 
	\includegraphics[width=\columnwidth]{legend_f.pdf}\\
	\begin{subfigure}[$\stokes=0.1$ - $r$-$z$ plane]{\includegraphics[width=\columnwidth,
			trim={0.2cm, 0.3cm, 0.2cm, 1.6cm}, clip
			]{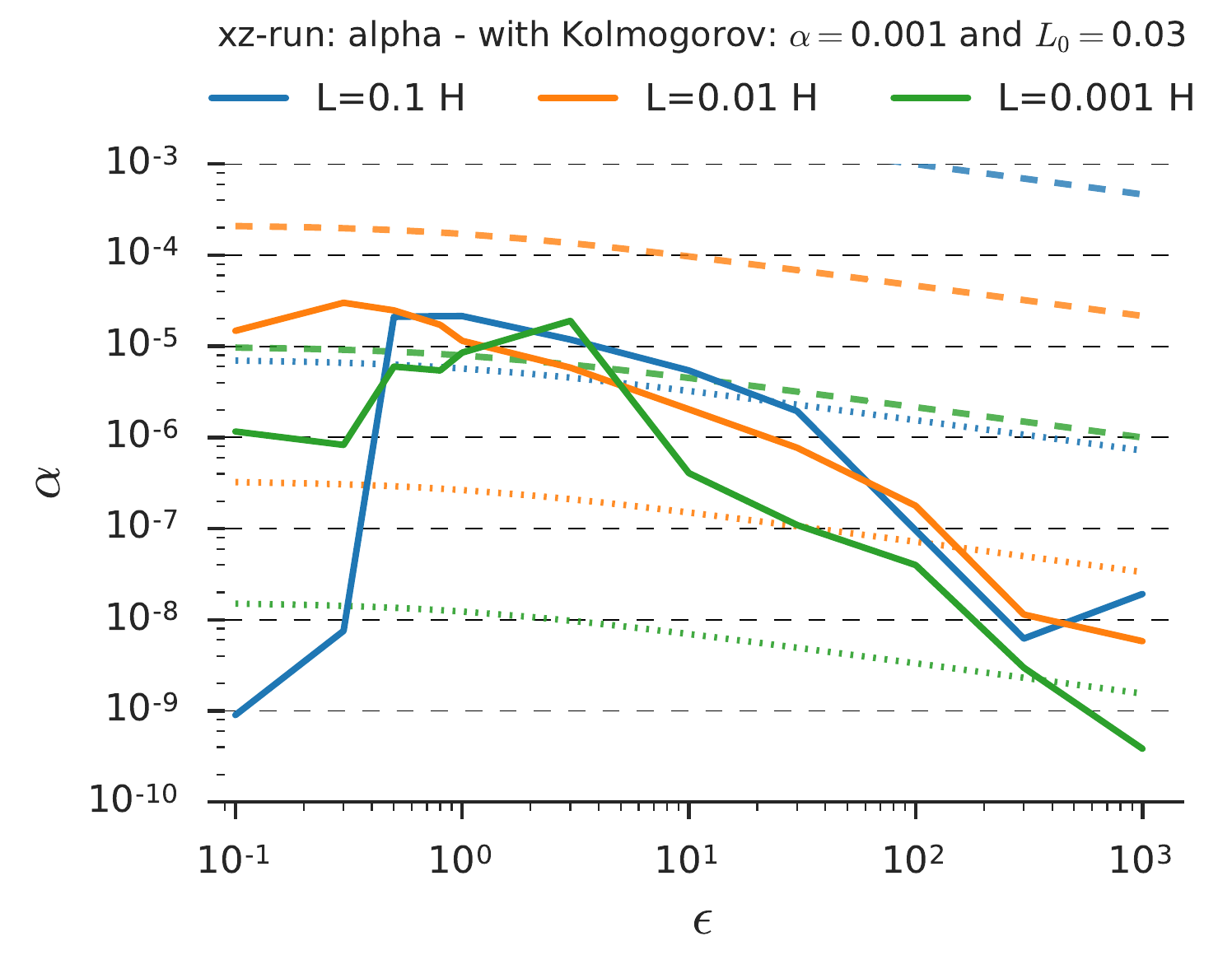}
	\label{fig:01_xz_alpha_kolm_dx}}
	\end{subfigure}\\
	\begin{subfigure}[$\stokes=0.01$ - $r$-$z$ plane]{\includegraphics[width=\columnwidth,
			trim={0.2cm, 0.3cm, 0.2cm, 1.6cm}, clip
			]{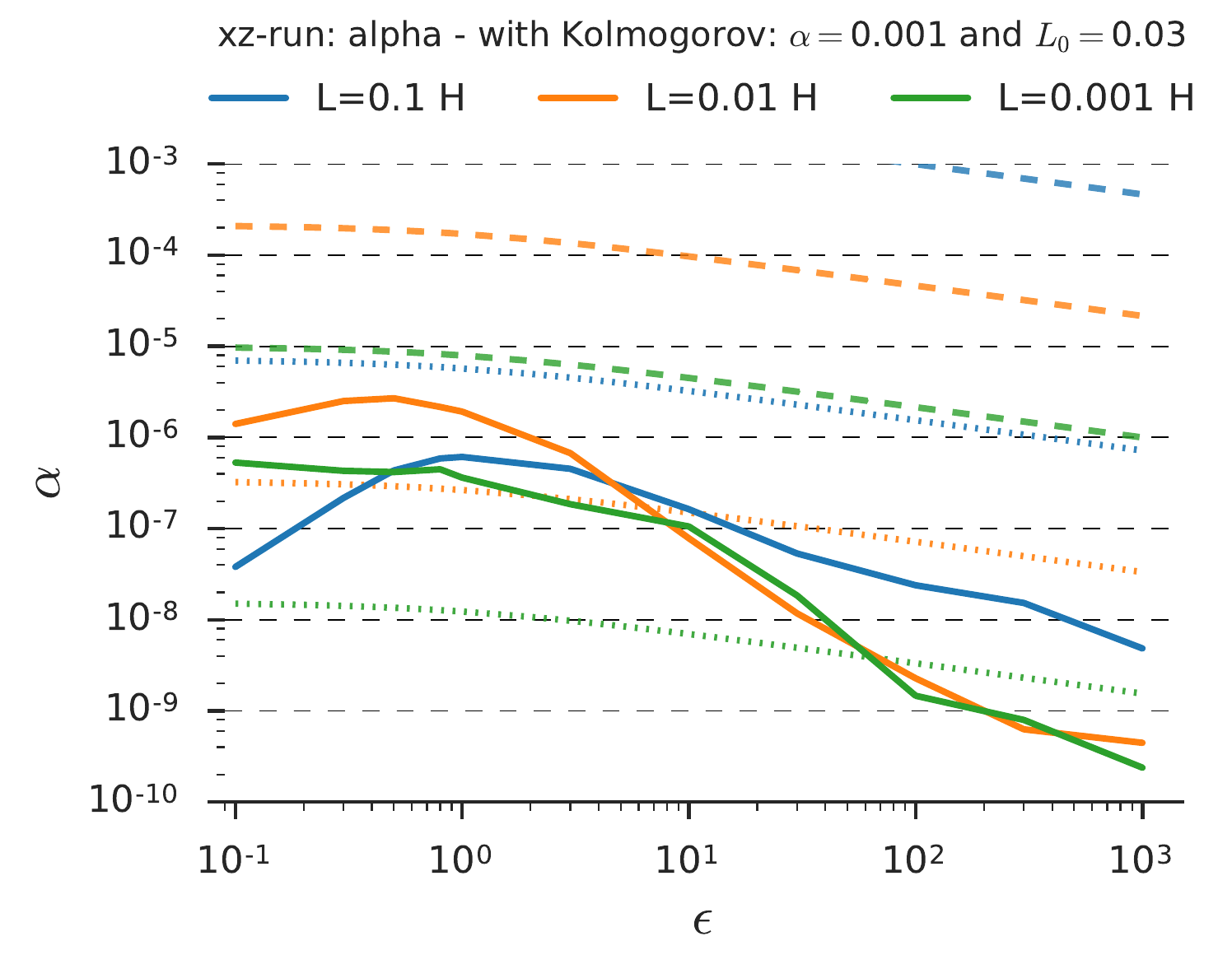}
	\label{fig:001_xz_alpha_kolm_dx}}
	\end{subfigure}
	\caption{Comparison of the measured $\alpha$-turbulence inside the simulations (bold line) with $\alpha_0=1e-3$ gas-only turbulence cascaded down on the scales of the simulation and taking an additional dust load into account. The turbulent viscosity for two scales per simulation are shown, domain size $l' = L_x$ (dashed) and grid size $l' = \de x$ (dotted). Since the injection scale $L_0$ of the isotropic turbulence is smaller than $L=\SI{0.1}{\scaleheight}$, the dashed blue line should not be taken as a serious approximation for an upward directed turbulent cascade.} %
	\label{fig:kolmogorov} %
\end{figure}%
As stated in the main text, the measured $\alpha$-values of the aSI/SI are lower than typical $\alpha$-turbulence measured in simulations of magneto-rotational instability (MRI) or vertical shear instability (VSI). A typical value used for describing gas turbulence in a PPD is $\alpha\approx\n{e-3}$, see \cite{Turner2014} which is not taking the presence of marginally coupled particles into account. In our work we find values of $\alpha<\n{e-4}$ for the SI turbulence. These two values can not directly be compared, as one has to take the Kolmogorov cascade and the additional dust load into account.

The gas turbulence in a PPD can be parametrized as turbulent viscosity $\alpha$. This parameter separates into a gas turbulent velocity $u$ and a characteristic eddy size $L_0$, as argued by \cite{Cuzzi2001}, via
\begin{equation}\label{eq:cuzziAlpha}
u_0 = \sqrt{\alpha_0} \cs \qquad \t{and} \qquad L_0 = \sqrt{\alpha_0} H\,.
\end{equation}
This assumes the turbulence to be isotropic on the scale of $L_0$. This is not necessarily the physical process actually found in MRI or VSI, where also turbulent upwards cascading is possible (vortex formation) and spiral waves radially transport angular momentum, too. But, lets keep this idea of isotropic turbulence in order to derive a comparison.

Now, using this turbulent velocity and letting it undergo its Kolmogorov cascade, see \cite{Kolmogorov1991}, one finds
\begin{equation}\label{eq:kolmogorov}
u^2 \cdot \frac{u}{l} = \frac{u_0^3}{L_0} = C\,,
\end{equation}
where $u^2$ is the kinetic energy, $u/l$ the energy dissipation time scale, and the $C$ is the dissipation rate. Kolmogorov states this rate to be constant over all length scale until the molecular dissipation scale is reached.

But, this neglects the influence of dust, as it is a dust-gas mixture in which the aSI and SI operates. Hence, one has to additionally undergo an energy transfer from pure gas into a dust-gas-mixture and this has to be done in the energy dissipation picture of Kolmogorov. For simplicity, we assume the dust and gas to have the same velocity, i.e. introducing the dust as additional mass contribution onto the gas, equal to assume $\stokes=0$. This means, the cascade starts with the prescription from \eq{eq:kolmogorov} and goes down until it reaches a primed length scale $l'$. On this primed scale, the kinetic energy of a pure gas eddy gets transferred onto a new eddy with $(1+\eps)$ higher mass. In this energy picture, since the density changed from $\rhoGas$ to $\rhoGas+\rhoDust = \rhoGas (1+\eps)$, the Kolmogorov cascade produces a turbulent velocity $u'$ on the scale of $l'=L'_0$, as
\begin{equation}\label{eq:kolmogorovWdust}
\frac{u_0^3}{L_0} = \frac{\sqrt{\alpha_0}^3 \cs^3}{\sqrt{\alpha_0} H} = \alpha_0 \cs^2 \Omega  \mustBeEq \frac{{u'}_0^3}{l'} = {u'}^2 \left(1+\eps\right) \cdot \frac{u'}{l'}\,.
\end{equation}

Now, in order to convert this $u'$ velocity into a turbulent viscosity, one has to be aware of the fact that for scales $l<L_0$ the splitting of $\alpha$ into an equal amount of turbulent gas velocity and eddy length scale is not justified anymore, as the Kolmogorov cascade does not has a constant Richardson number, see \cite{Cuzzi2001}. Hence, the $\alpha$ measured in our simulations needs to be compared with a turbulent viscosity $\nu$ derived from the cascade. In order to get this viscosity, one can separate this viscosity into a length scale and a velocity, and use \eq{eq:kolmogorovWdust}
\begin{equation}
\nu' = l' u' = \left[\frac{l'}{1+\eps} C \right]^{1/3} l' = \left(\frac{\alpha_0 }{1+\eps}\right)^{1/3} {l'}^{4/3}\,.
\end{equation}

This equation now can be used to calculate a turbulent viscosity value that originates from a pure gas turbulence of strength $\alpha_0$, but cascaded down onto the scales of aSI/SI, and is taking the dust load into account. The scale $l'$ that one has to use for this comparison is in the best case $\de x$, and in the worst case $L_x$. \fig{fig:kolmogorov} does this comparison for all simulations with initial turbulent strength of $\alpha_0=\n{e-3}$, i.e. injection on the length scale of $L_0 = \SI{0.03}{\scaleheight}$. Here, only the results for the SI are shown, since the curves for the aSI are very similar. The solid line is the measured $\alpha$-turbulence from the parameter study. The dashed line is $\nu'$ on the scales of $L_x$ for each simulation in its respective color. The dotted line does the same, but on the scales of the respective grid scale $\de x$. One could also do this comparison on the prominent eddy size of our simulations, which is $\lcorr$ as introduced in \app{sec:01_xz_corrL}. But, here we want to show the range of turbulent viscosity that one could expect from a gas turbulent $\alpha$ value inside our simulations.

These plots show that the expected turbulent viscosity parameter $\alpha$, with origin in the turbulent cascade, to be in a rather broad range of values. The reason is that simulation domain size and grid resolution span a range of two orders of magnitude, hence also the viscosity spans two orders of magnitude. Still, for $\stokes=0.1$ the found $\alpha$ by the aSI/SI is well above the best case for cascade down to $l'=\de x$, and sometimes even stronger than what is expected for the worst case, i.e. cascade down to only $l'=L$.

This whole discussion does not take all effects into account. How the particles react to the gas turbulence is only mimicked by an energy transfer on a dust loaded eddy. In reality, particles of different sizes are present and the Kolmogorov cascade is not present in a pure gas form, but in a dust loaded form. Turbulence can only occur as long as the Reynolds criteria is fulfilled; how the dynamic viscosity has to be formulated for the situation of a dusty PPDs remains to be shown. Also, the scales and position where $\alpha_0$ turbulence is actually present might not be the spot were aSI/SI is at work. For example, in a dead zone, the MRI might be active in a layer on top of a dead zone, but the SI could be active within the mid-plane. So the turbulence not only has to cascade down, but the turbulent velocity needs to be advected towards the mid-plane. Also, the fraction of turbulent energy residing in the particles might get lost earlier in the cascade than in our model, since particle can undergo elastic collisions. Lastly, the assumption of equal splitting of $\alpha$ into velocity and length scale might be wrong, as this assumes Richardson numbers around unity, which is not necessarily true.
\section{On collision time-scales}
\label{app:collTime}
The probability for a collision after a time interval $t$ is $P=t v \sigma n$, with number density $n=N_\mathrm{p}/V$, geometrical cross-section $\sigma$ and particle velocity $v$. Then $v\sigma t$ can be interpreted as the space-time cross-section volume and $n$ is the number of particles in a unity volume. The probability for a collision is $P=1$ if $v \sigma t \cdot n= 1$, meaning if the space-time cross-section volume multiplied with the number density is equal to a unity volume. One gets $1=\tau v \sigma n$, with collision time-scale $\tau$ and since $\tau v = \lambda_{\rm free}$ the mean free path can be expressed as 
\begin{equation}
\lambda_{\rm free} = \left(\sqrt{2} \sigma n \right)^{-1} \, ,
\end{equation}
This can be further generalized by using the definition of Stokes number and assuming particles to be in the Epstein regime, see \Eq{eq:stokes}, and \Eq{eq:epstein} can be rewritten as 
\begin{equation}
\tauF = \frac{3 m}{4 \rhoGas \cs A}\, ,
\end{equation}
with $n \cdot m = \rhoDust = \eps \rhoGas$ and $A = \sigma / 4$ for a spherical particle. This results in
\begin{equation}
\lambda_{\rm free} = \frac{\stokes}{6 \eps} \si{\scaleheight} \quad \mathrm{and} \quad \tauColl = \frac{\stokes \si{\scaleheight}}{6 \eps \vrms} \, .
\end{equation}

The length scales on which collision happen in our simulations are of order $\lambda_{\rm free} \left(\eps = 0.1 \dots 1000 \right) \approx \num{e-1} \dots \num{e-5}$ for $\stokes=0.1$, and one order of magnitude less for $\stokes=0.01$. As shown by \cite{Johansen2011}, collisions might promote particle overdensities since particle rms-velocities get damped. But, collisions also alter particle Stokes numbers by compactification that is largely neglected in present works.

In a different approach one might ask if the stopping time $\tauS$ is shorter than the collision time scale. If this is the case, the particles change their momentum due to gas friction faster than due to a collision. Hence, we need to check if $\tauS < \tauColl$:
\begin{equation}
\tauS = \frac{\stokes}{\Omega} < \tauColl = \frac{\stokes \scaleheight}{6\ \epsilon\ \vrms}
\end{equation}
This simplifies to
\begin{equation}
\frac{\vrms}{\cs} < \frac{1}{6 \epsilon},
\end{equation}
which is in our case $10^{-4} < (6\cdot10^3)^{-1}$ and consequently collisions will not drive the particle dynamics but rather friction with the gas does.
\bibliography{PhD} %
\begin{table*}
\section{Detailed lists of simulation runs and results I: aSI}
  \centering\scriptsize
    \begin{tabular}{lccccccccc}
    	\toprule
    	Name              & $\Tmax\left[2\pi\Omega\right]$ &             $\alpha\pm\Delta\alpha$             &             $\dx \pm \Delta\dx$              &    $\sg$     & $\urms\pm\Delta\urms$ &    $\zg$     &    $\lc$     &  $\tauCorr$  &     $\Sc$     \\ \midrule
    	01\_01H\_e0       &             79.58              &  $\left( 2.5 \pm 0.022 \right) \cdot 10^{-8}$   &  $\left( 3.3 \pm 0.9 \right) \cdot 10^{-8}$  & \n{3.13E-04} &     \n{4.55E-02}      & \n{1.41E-02} & \n{1.04E-04} & \n{3.33E-01} & \n{7.76E-01}  \\
    	01\_01H\_e0       &             79.58              &  $\left( 2.1 \pm 0.0095 \right) \cdot 10^{-7}$  & $\left( 3.8 \pm 0.96 \right) \cdot 10^{-7}$  & \n{9.60E-04} &     \n{3.86E-02}      & \n{1.18E-02} & \n{3.94E-04} & \n{4.10E-01} & \n{5.55E-01}  \\
    	01\_01H\_e0       &             79.58              &   $\left( 2.8 \pm 0.02 \right) \cdot 10^{-5}$   & $\left( 1.1 \pm 0.41 \right) \cdot 10^{-4}$  & \n{1.68E-02} &     \n{3.95E-02}      & \n{1.14E-02} & \n{6.43E-03} & \n{3.84E-01} & \n{2.60E-01}  \\
    	01\_01H\_e0       &             79.58              &  $\left( 3.8 \pm 0.019 \right) \cdot 10^{-5}$   & $\left( 1.1 \pm 0.36 \right) \cdot 10^{-4}$  & \n{1.56E-02} &     \n{3.45E-02}      & \n{9.69E-03} & \n{7.01E-03} & \n{4.50E-01} & \n{3.52E-01}  \\
    	01\_01H\_e1       &             79.58              &  $\left( 2.6 \pm 0.018 \right) \cdot 10^{-5}$   & $\left( 1.0 \pm 0.29 \right) \cdot 10^{-4}$  & \n{1.41E-02} &     \n{3.22E-02}      & \n{8.67E-03} & \n{7.17E-03} & \n{5.08E-01} & \n{2.62E-01}  \\
    	01\_01H\_e3       &             79.58              &  $\left( 1.4 \pm 0.0068 \right) \cdot 10^{-5}$  &  $\left( 4.9 \pm 1.2 \right) \cdot 10^{-5}$  & \n{9.57E-03} &     \n{1.92E-02}      & \n{4.74E-03} & \n{5.11E-03} & \n{5.34E-01} & \n{2.83E-01}  \\
    	01\_01H\_e10      &             79.58              &  $\left( 3.7 \pm 0.008 \right) \cdot 10^{-6}$   &  $\left( 1.3 \pm 0.4 \right) \cdot 10^{-5}$  & \n{4.74E-03} &     \n{8.53E-03}      & \n{1.91E-03} & \n{2.84E-03} & \n{5.99E-01} & \n{2.76E-01}  \\
    	01\_01H\_e30      &             79.58              &  $\left( 1.4 \pm 0.0018 \right) \cdot 10^{-6}$  &  $\left( 2.9 \pm 1.4 \right) \cdot 10^{-6}$  & \n{3.38E-03} &     \n{6.41E-03}      & \n{1.10E-03} & \n{8.61E-04} & \n{2.55E-01} & \n{4.96E-01}  \\
    	01\_01H\_e100     &             79.58              &  $\left( 4.3 \pm 0.0062 \right) \cdot 10^{-8}$  &  $\left( 1.9 \pm 0.4 \right) \cdot 10^{-7}$  & \n{5.10E-04} &     \n{8.96E-04}      & \n{2.08E-04} & \n{3.78E-04} & \n{7.41E-01} & \n{2.24E-01}  \\
    	01\_01H\_e300     &             79.58              &  $\left( 1.3 \pm 0.0006 \right) \cdot 10^{-8}$  &  $\left( 6.3 \pm 1.5 \right) \cdot 10^{-8}$  & \n{2.10E-04} &     \n{4.11E-04}      & \n{1.19E-04} & \n{2.98E-04} & \n{1.42E+00} & \n{2.12E-01}  \\
    	01\_01H\_e1000    &             79.58              & $\left( 1.1 \pm 0.00067 \right) \cdot 10^{-9}$  & $\left( 2.3 \pm 0.87 \right) \cdot 10^{-9}$  & \n{4.73E-05} &     \n{1.58E-04}      & \n{8.33E-05} & \n{4.79E-05} & \n{1.01E+00} & \n{5.01E-01}  \\
    	01\_001H\_e0      &             57.82              &   $\left( 9.7 \pm 2.3 \right) \cdot 10^{-10}$   & $\left( 6.3 \pm 2.8 \right) \cdot 10^{-10}$  & \n{5.99E-05} &     \n{4.55E-02}      & \n{1.41E-02} & \n{1.05E-05} & \n{1.75E-01} & \n{1.55E+00}  \\
    	01\_001H\_e0      &              57.4              &   $\left( 9.0 \pm 0.92 \right) \cdot 10^{-9}$   & $\left( 1.2 \pm 0.53 \right) \cdot 10^{-8}$  & \n{2.08E-04} &     \n{3.86E-02}      & \n{1.18E-02} & \n{5.94E-05} & \n{2.85E-01} & \n{7.27E-01}  \\
    	01\_001H\_e0      &             55.97              &  $\left( 6.2 \pm 0.067 \right) \cdot 10^{-8}$   & $\left( 1.1 \pm 0.39 \right) \cdot 10^{-7}$  & \n{5.61E-04} &     \n{3.34E-02}      & \n{1.02E-02} & \n{1.94E-04} & \n{3.45E-01} & \n{5.69E-01}  \\
    	01\_001H\_e0      &             51.82              &  $\left( 5.0 \pm 0.075 \right) \cdot 10^{-6}$   &  $\left( 1.7 \pm 1.1 \right) \cdot 10^{-5}$  & \n{8.35E-03} &     \n{3.05E-02}      & \n{9.37E-03} & \n{2.07E-03} & \n{2.48E-01} & \n{2.91E-01}  \\
    	01\_001H\_e1      &             51.26              &   $\left( 8.6 \pm 0.24 \right) \cdot 10^{-6}$   & $\left( 1.6 \pm 0.93 \right) \cdot 10^{-5}$  & \n{9.66E-03} &     \n{2.79E-02}      & \n{8.82E-03} & \n{1.69E-03} & \n{1.75E-01} & \n{5.26E-01}  \\
    	01\_001H\_e3      &             53.18              &  $\left( 4.4 \pm 0.028 \right) \cdot 10^{-6}$   & $\left( 1.0 \pm 0.46 \right) \cdot 10^{-5}$  & \n{5.63E-03} &     \n{1.58E-02}      & \n{4.40E-03} & \n{1.80E-03} & \n{3.19E-01} & \n{4.34E-01}  \\
    	01\_001H\_e10     &             55.99              &  $\left( 2.0 \pm 0.0031 \right) \cdot 10^{-6}$  &  $\left( 3.4 \pm 1.1 \right) \cdot 10^{-6}$  & \n{3.18E-03} &     \n{7.30E-03}      & \n{1.80E-03} & \n{1.07E-03} & \n{3.37E-01} & \n{5.94E-01}  \\
    	01\_001H\_e30     &             34.34              &  $\left( 5.1 \pm 0.016 \right) \cdot 10^{-7}$   &  $\left( 8.4 \pm 2.2 \right) \cdot 10^{-7}$  & \n{1.45E-03} &     \n{3.22E-03}      & \n{6.99E-04} & \n{5.80E-04} & \n{4.00E-01} & \n{6.07E-01}  \\
    	01\_001H\_e100    &             34.59              &  $\left( 5.6 \pm 0.002 \right) \cdot 10^{-8}$   &  $\left( 1.2 \pm 0.2 \right) \cdot 10^{-7}$  & \n{4.64E-04} &     \n{9.61E-04}      & \n{2.35E-04} & \n{2.51E-04} & \n{5.41E-01} & \n{4.83E-01}  \\
    	01\_001H\_e300    &             34.66              & $\left( 1.1 \pm 0.00081 \right) \cdot 10^{-8}$  & $\left( 1.4 \pm 0.22 \right) \cdot 10^{-8}$  & \n{1.82E-04} &     \n{3.86E-04}      & \n{1.29E-04} & \n{7.87E-05} & \n{4.33E-01} & \n{7.42E-01}  \\
    	01\_001H\_e1000   &             34.68              & $\left( 1.1 \pm 0.00034 \right) \cdot 10^{-9}$  & $\left( 8.7 \pm 2.4 \right) \cdot 10^{-10}$  & \n{6.21E-05} &     \n{1.73E-04}      & \n{9.45E-05} & \n{1.40E-05} & \n{2.26E-01} & \n{1.31E+00}  \\
    	01\_0001H\_e0     &              3.37              &  $\left( -1.9 \pm 0.22 \right) \cdot 10^{-9}$   & $\left( 1.9 \pm 2.5 \right) \cdot 10^{-11}$  & \n{9.88E-06} &     \n{4.55E-02}      & \n{1.41E-02} & \n{1.95E-06} & \n{1.97E-01} & \n{-9.92E+01} \\
    	01\_0001H\_e0     &              6.66              &   $\left( 4.4 \pm 0.75 \right) \cdot 10^{-9}$   & $\left( 1.5 \pm 1.2 \right) \cdot 10^{-10}$  & \n{2.96E-05} &     \n{3.86E-02}      & \n{1.18E-02} & \n{5.10E-06} & \n{1.72E-01} & \n{2.95E+01}  \\
    	01\_0001H\_e0     &             10.01              &  $\left( -1.9 \pm 0.96 \right) \cdot 10^{-9}$   &  $\left( 2.9 \pm 2.0 \right) \cdot 10^{-9}$  & \n{6.65E-05} &     \n{3.34E-02}      & \n{1.01E-02} & \n{4.29E-05} & \n{6.44E-01} & \n{-6.60E-01} \\
    	01\_0001H\_e0     &              6.09              &  $\left( 7.4 \pm 0.031 \right) \cdot 10^{-7}$   &  $\left( 5.6 \pm 3.9 \right) \cdot 10^{-7}$  & \n{1.47E-03} &     \n{2.80E-02}      & \n{8.47E-03} & \n{3.79E-04} & \n{2.58E-01} & \n{1.33E+00}  \\
    	01\_0001H\_e1     &              6.01              &   $\left( 4.8 \pm 0.03 \right) \cdot 10^{-7}$   &  $\left( 7.6 \pm 7.0 \right) \cdot 10^{-7}$  & \n{1.18E-03} &     \n{2.51E-02}      & \n{7.62E-03} & \n{6.43E-04} & \n{5.46E-01} & \n{6.38E-01}  \\
    	01\_0001H\_e3     &              5.86              &   $\left( 8.4 \pm 0.21 \right) \cdot 10^{-7}$   &  $\left( 1.3 \pm 1.2 \right) \cdot 10^{-6}$  & \n{1.92E-03} &     \n{1.34E-02}      & \n{3.89E-03} & \n{6.71E-04} & \n{3.49E-01} & \n{6.48E-01}  \\
    	01\_0001H\_e10    &              3.45              &  $\left( 8.2 \pm 0.035 \right) \cdot 10^{-7}$   &  $\left( 3.7 \pm 1.5 \right) \cdot 10^{-7}$  & \n{1.39E-03} &     \n{5.62E-03}      & \n{1.54E-03} & \n{2.69E-04} & \n{1.93E-01} & \n{2.20E+00}  \\
    	01\_0001H\_e30    &              3.46              &  $\left( 2.9 \pm 0.0076 \right) \cdot 10^{-7}$  & $\left( 1.8 \pm 0.61 \right) \cdot 10^{-7}$  & \n{7.92E-04} &     \n{2.54E-03}      & \n{6.22E-04} & \n{2.33E-04} & \n{2.94E-01} & \n{1.59E+00}  \\
    	01\_0001H\_e100   &              3.48              &  $\left( 6.7 \pm 0.032 \right) \cdot 10^{-8}$   &  $\left( 3.0 \pm 0.9 \right) \cdot 10^{-8}$  & \n{3.43E-04} &     \n{8.67E-04}      & \n{2.30E-04} & \n{8.85E-05} & \n{2.58E-01} & \n{2.20E+00}  \\
    	01\_0001H\_e300   &              3.48              &  $\left( 2.3 \pm 0.0056 \right) \cdot 10^{-9}$  &  $\left( 4.4 \pm 1.2 \right) \cdot 10^{-9}$  & \n{1.09E-04} &     \n{2.69E-04}      & \n{1.18E-04} & \n{4.06E-05} & \n{3.73E-01} & \n{5.19E-01}  \\
    	01\_0001H\_e1000  &              3.49              & $\left( 2.9 \pm 0.0015 \right) \cdot 10^{-10}$  & $\left( 5.2 \pm 2.4 \right) \cdot 10^{-10}$  & \n{4.38E-05} &     \n{1.38E-04}      & \n{8.85E-05} & \n{1.18E-05} & \n{2.69E-01} & \n{5.63E-01}  \\ \bottomrule
    	                  &                                &                                                 &                                              &              &                       &              &              &              &               \\
    	Name       & $\Tmax\left[2\pi\Omega\right]$ &             $\alpha\pm\Delta\alpha$             &             $\dx \pm \Delta\dx$              &    $\sg$     & $\urms\pm\Delta\urms$ &    $\zg$     &    $\lc$     &  $\tauCorr$  &     $\Sc$     \\ \midrule
    	001\_01H\_e0      &             159.15             &  $\left( 1.7 \pm 0.0029 \right) \cdot 10^{-7}$  & $\left( 2.1 \pm 0.48 \right) \cdot 10^{-6}$  & \n{1.39E-03} &     \n{4.55E-02}      & \n{1.14E-02} & \n{1.54E-03} & \n{1.11E+00} & \n{7.87E-02}  \\
    	001\_01H\_e0      &             110.04             &   $\left( 4.0 \pm 0.37 \right) \cdot 10^{-7}$   &  $\left( 8.0 \pm 1.9 \right) \cdot 10^{-6}$  & \n{3.91E-03} &     \n{3.87E-02}      & \n{9.67E-03} & \n{2.05E-03} & \n{5.24E-01} & \n{4.96E-02}  \\
    	001\_01H\_e0      &             108.62             &  $\left( 1.3 \pm 0.009 \right) \cdot 10^{-6}$   & $\left( 1.4 \pm 0.38 \right) \cdot 10^{-5}$  & \n{4.05E-03} &     \n{3.37E-02}      & \n{8.42E-03} & \n{3.49E-03} & \n{8.61E-01} & \n{8.95E-02}  \\
    	001\_01H\_e0      &             159.15             &  $\left( 2.0 \pm 0.0057 \right) \cdot 10^{-6}$  & $\left( 1.9 \pm 0.55 \right) \cdot 10^{-5}$  & \n{5.59E-03} &     \n{2.83E-02}      & \n{7.07E-03} & \n{3.33E-03} & \n{5.95E-01} & \n{1.07E-01}  \\
    	001\_01H\_e1      &             159.15             &  $\left( 7.8 \pm 0.023 \right) \cdot 10^{-7}$   &  $\left( 9.2 \pm 2.0 \right) \cdot 10^{-6}$  & \n{2.85E-03} &     \n{2.51E-02}      & \n{6.28E-03} & \n{3.22E-03} & \n{1.13E+00} & \n{8.48E-02}  \\
    	001\_01H\_e3      &             159.15             &  $\left( 1.3 \pm 0.0018 \right) \cdot 10^{-6}$  & $\left( 1.2 \pm 0.38 \right) \cdot 10^{-5}$  & \n{5.63E-03} &     \n{1.37E-02}      & \n{3.41E-03} & \n{2.18E-03} & \n{3.88E-01} & \n{1.05E-01}  \\
    	001\_01H\_e10     &             159.15             &  $\left( 1.5 \pm 0.011 \right) \cdot 10^{-7}$   & $\left( 2.1 \pm 0.41 \right) \cdot 10^{-6}$  & \n{1.36E-03} &     \n{4.77E-03}      & \n{1.18E-03} & \n{1.54E-03} & \n{1.13E+00} & \n{7.18E-02}  \\
    	001\_01H\_e30     &              80.4              & $\left( 1.3 \pm 0.00081 \right) \cdot 10^{-7}$  & $\left( 1.0 \pm 0.16 \right) \cdot 10^{-6}$  & \n{1.01E-03} &     \n{2.03E-03}      & \n{4.76E-04} & \n{9.95E-04} & \n{9.89E-01} & \n{1.31E-01}  \\
    	001\_01H\_e100    &             54.86              & $\left( 3.2 \pm 0.00063 \right) \cdot 10^{-8}$  & $\left( 2.3 \pm 0.32 \right) \cdot 10^{-7}$  & \n{4.88E-04} &     \n{8.01E-04}      & \n{1.74E-04} & \n{4.62E-04} & \n{9.46E-01} & \n{1.41E-01}  \\
    	001\_01H\_e300    &              36.3              &  $\left( 6.4 \pm 0.0015 \right) \cdot 10^{-9}$  & $\left( 5.4 \pm 0.76 \right) \cdot 10^{-8}$  & \n{2.28E-04} &     \n{3.55E-04}      & \n{7.13E-05} & \n{2.35E-04} & \n{1.03E+00} & \n{1.19E-01}  \\
    	001\_01H\_e1000   &              55.1              & $\left( 6.8 \pm 0.0025 \right) \cdot 10^{-10}$  & $\left( 4.2 \pm 0.65 \right) \cdot 10^{-9}$  & \n{6.22E-05} &     \n{1.02E-04}      & \n{2.22E-05} & \n{6.67E-05} & \n{1.07E+00} & \n{1.63E-01}  \\
    	001\_001H\_e0     &             61.74              &  $\left( 1e+01 \pm 0.54 \right) \cdot 10^{-7}$  &  $\left( 8.7 \pm 4.8 \right) \cdot 10^{-6}$  & \n{7.07E-03} &     \n{4.60E-02}      & \n{1.16E-02} & \n{1.23E-03} & \n{1.75E-01} & \n{1.15E-01}  \\
    	001\_001H\_e0     &             60.31              &  $\left( -1.4 \pm 0.077 \right) \cdot 10^{-6}$  &  $\left( 1.6 \pm 1.1 \right) \cdot 10^{-5}$  & \n{9.35E-03} &     \n{3.97E-02}      & \n{9.96E-03} & \n{1.72E-03} & \n{1.84E-01} & \n{-8.87E-02} \\
    	001\_001H\_e0     &             60.27              &  $\left( 5.1 \pm 0.028 \right) \cdot 10^{-6}$   & $\left( 1.4 \pm 0.93 \right) \cdot 10^{-5}$  & \n{9.29E-03} &     \n{3.47E-02}      & \n{8.67E-03} & \n{1.56E-03} & \n{1.68E-01} & \n{3.51E-01}  \\
    	001\_001H\_e0     &             61.82              &  $\left( -6.3 \pm 0.016 \right) \cdot 10^{-6}$  & $\left( 1.3 \pm 0.77 \right) \cdot 10^{-5}$  & \n{8.95E-03} &     \n{2.93E-02}      & \n{7.30E-03} & \n{1.50E-03} & \n{1.68E-01} & \n{-4.71E-01} \\
    	001\_001H\_e1     &             100.41             &  $\left( 3.5 \pm 0.028 \right) \cdot 10^{-6}$   & $\left( 1.2 \pm 0.77 \right) \cdot 10^{-5}$  & \n{9.22E-03} &     \n{2.67E-02}      & \n{6.71E-03} & \n{1.28E-03} & \n{1.39E-01} & \n{2.94E-01}  \\
    	001\_001H\_e3     &             67.88              &  $\left( 3.3 \pm 0.0076 \right) \cdot 10^{-7}$  &  $\left( 6.1 \pm 3.2 \right) \cdot 10^{-6}$  & \n{4.80E-03} &     \n{1.35E-02}      & \n{3.36E-03} & \n{1.27E-03} & \n{2.64E-01} & \n{5.38E-02}  \\
    	001\_001H\_e10    &             103.12             &  $\left( 7.5 \pm 0.0041 \right) \cdot 10^{-7}$  & $\left( 1.3 \pm 0.52 \right) \cdot 10^{-6}$  & \n{3.09E-03} &     \n{5.57E-03}      & \n{1.37E-03} & \n{4.08E-04} & \n{1.32E-01} & \n{5.99E-01}  \\
    	001\_001H\_e30    &             68.87              &  $\left( 3.1 \pm 0.034 \right) \cdot 10^{-8}$   & $\left( 3.1 \pm 0.91 \right) \cdot 10^{-7}$  & \n{1.13E-03} &     \n{1.97E-03}      & \n{4.92E-04} & \n{2.76E-04} & \n{2.44E-01} & \n{1.00E-01}  \\
    	001\_001H\_e100   &             18.62              &  $\left( 4.3 \pm 0.0056 \right) \cdot 10^{-9}$  & $\left( 4.7 \pm 0.79 \right) \cdot 10^{-8}$  & \n{2.51E-04} &     \n{5.75E-04}      & \n{1.40E-04} & \n{1.88E-04} & \n{7.50E-01} & \n{9.07E-02}  \\
    	001\_001H\_e300   &             36.45              & $\left( 1.5 \pm 0.00013 \right) \cdot 10^{-9}$  & $\left( 1e+01 \pm 1.6 \right) \cdot 10^{-9}$ & \n{9.95E-05} &     \n{2.13E-04}      & \n{4.95E-05} & \n{1.00E-04} & \n{1.01E+00} & \n{1.52E-01}  \\
    	001\_001H\_e1000  &             17.78              & $\left( 4.2 \pm 0.00062 \right) \cdot 10^{-10}$ &  $\left( 1.8 \pm 0.3 \right) \cdot 10^{-9}$  & \n{4.37E-05} &     \n{8.13E-05}      & \n{1.92E-05} & \n{4.07E-05} & \n{9.33E-01} & \n{2.39E-01}  \\
    	001\_0001H\_e0    &              8.28              &  $\left( 1.7 \pm 0.071 \right) \cdot 10^{-6}$   &  $\left( 2.3 \pm 2.5 \right) \cdot 10^{-6}$  & \n{6.59E-03} &     \n{4.59E-02}      & \n{1.13E-02} & \n{3.49E-04} & \n{5.30E-02} & \n{7.47E-01}  \\
    	001\_0001H\_e0    &              5.93              &  $\left( 2.8 \pm 0.033 \right) \cdot 10^{-6}$   &  $\left( 2.1 \pm 1.9 \right) \cdot 10^{-6}$  & \n{4.32E-03} &     \n{3.89E-02}      & \n{9.62E-03} & \n{4.90E-04} & \n{1.13E-01} & \n{1.31E+00}  \\
    	001\_0001H\_e0    &              6.03              &   $\left( 1.3 \pm 0.14 \right) \cdot 10^{-7}$   &  $\left( 1.0 \pm 1.0 \right) \cdot 10^{-6}$  & \n{3.74E-03} &     \n{3.36E-02}      & \n{8.45E-03} & \n{2.76E-04} & \n{7.37E-02} & \n{1.29E-01}  \\
    	001\_0001H\_e0    &              6.17              &   $\left( 7.2 \pm 0.15 \right) \cdot 10^{-7}$   &  $\left( 1.3 \pm 1.3 \right) \cdot 10^{-6}$  & \n{3.48E-03} &     \n{2.80E-02}      & \n{7.02E-03} & \n{3.84E-04} & \n{1.11E-01} & \n{5.36E-01}  \\
    	001\_0001H\_e1    &              3.4               &   $\left( 5.9 \pm 0.33 \right) \cdot 10^{-7}$   &  $\left( 8.9 \pm 6.0 \right) \cdot 10^{-7}$  & \n{2.98E-03} &     \n{2.53E-02}      & \n{6.29E-03} & \n{3.01E-04} & \n{1.01E-01} & \n{6.58E-01}  \\
    	001\_0001H\_e3    &              3.43              &   $\left( 9.5 \pm 0.22 \right) \cdot 10^{-8}$   &  $\left( 5.2 \pm 2.2 \right) \cdot 10^{-7}$  & \n{1.93E-03} &     \n{1.29E-02}      & \n{3.20E-03} & \n{2.72E-04} & \n{1.41E-01} & \n{1.81E-01}  \\
    	001\_0001H\_e10   &             10.64              &   $\left( 5.1 \pm 0.13 \right) \cdot 10^{-8}$   &  $\left( 2.5 \pm 1.0 \right) \cdot 10^{-7}$  & \n{1.26E-03} &     \n{4.99E-03}      & \n{1.20E-03} & \n{2.03E-04} & \n{1.61E-01} & \n{2.01E-01}  \\
    	001\_0001H\_e30   &              6.95              &  $\left( 5.6 \pm 0.011 \right) \cdot 10^{-8}$   &  $\left( 9.6 \pm 3.0 \right) \cdot 10^{-8}$  & \n{7.35E-04} &     \n{1.81E-03}      & \n{4.42E-04} & \n{1.30E-04} & \n{1.77E-01} & \n{5.83E-01}  \\
    	001\_0001H\_e100  &             10.45              &  $\left( 2.9 \pm 0.012 \right) \cdot 10^{-9}$   & $\left( 2.1 \pm 0.96 \right) \cdot 10^{-8}$  & \n{3.01E-04} &     \n{5.95E-04}      & \n{1.47E-04} & \n{6.99E-05} & \n{2.32E-01} & \n{1.37E-01}  \\
    	001\_0001H\_e300  &              6.97              &  $\left( 2.0 \pm 0.0006 \right) \cdot 10^{-9}$  &  $\left( 7.2 \pm 1.6 \right) \cdot 10^{-9}$  & \n{1.21E-04} &     \n{2.20E-04}      & \n{5.43E-05} & \n{5.96E-05} & \n{4.93E-01} & \n{2.79E-01}  \\
    	001\_0001H\_e1000 &             10.46              & $\left( 4.4 \pm 0.0015 \right) \cdot 10^{-10}$  & $\left( 8.5 \pm 2.5 \right) \cdot 10^{-10}$  & \n{4.34E-05} &     \n{8.41E-05}      & \n{2.17E-05} & \n{1.96E-05} & \n{4.52E-01} & \n{5.23E-01}  \\ \bottomrule
    \end{tabular}%
    \caption{2-d simulations in $r$-$\varphi$ extent. Simulation name is constructed from $\stokes$ number, domain size $Lx,y$ and dust-to-gas ratio $\epsInit$.}
    \label{tab:r-phi-runs}
\end{table*}%
\begin{table*}
\section{Detailed lists of simulation runs and results I: aSI}
  \centering\tiny
    \begin{tabular}{lcccccccccc}
    	\toprule
    	Name              & $\Tmax\left[2\pi\Omega\right]$ &             $\alpha\pm\Delta\alpha$             &              $\dx \pm \Delta\dx$               &               $\dz \pm \Delta\dz$               &    $\sg$     & $\urms\pm\Delta\urms$ &    $\zg$     &    $\lc$     &  $\tauCorr$  &     $\Sc$     \\ \midrule
    	01\_01H\_e0       & 79.58                          &  $\left( -1.9 \pm 0.22 \right) \cdot 10^{-9}$   &   $\left( 1.4 \pm 2.6 \right) \cdot 10^{-7}$   & $\left( 3.0 \pm 3.8e+01 \right) \cdot 10^{-8}$  & \n{1.03E-03} &     \n{4.55E-02}      & \n{1.41E-02} & \n{1.35E-04} & \n{1.31E-01} & \n{-1.33E-02} \\
    	01\_01H\_e0       & 79.58                          &  $\left( 1.4 \pm 0.081 \right) \cdot 10^{-8}$   &   $\left( 3.7 \pm 1.5 \right) \cdot 10^{-7}$   & $\left( 7.9 \pm 2.9e+01 \right) \cdot 10^{-8}$  & \n{1.17E-03} &     \n{3.86E-02}      & \n{1.18E-02} & \n{3.17E-04} & \n{2.72E-01} & \n{3.72E-02}  \\
    	01\_01H\_e0       & 79.58                          &  $\left( 1.9 \pm 0.018 \right) \cdot 10^{-5}$   &   $\left( 7.5 \pm 2.4 \right) \cdot 10^{-5}$   &   $\left( 4.9 \pm 1.6 \right) \cdot 10^{-5}$    & \n{1.58E-02} &     \n{3.92E-02}      & \n{1.14E-02} & \n{4.75E-03} & \n{3.01E-01} & \n{2.59E-01}  \\
    	01\_01H\_e0       & 79.58                          &   $\left( 2.5 \pm 0.02 \right) \cdot 10^{-5}$   &   $\left( 6.3 \pm 1.5 \right) \cdot 10^{-5}$   &   $\left( 4.4 \pm 1.3 \right) \cdot 10^{-5}$    & \n{1.57E-02} &     \n{3.42E-02}      & \n{9.78E-03} & \n{4.00E-03} & \n{2.55E-01} & \n{4.03E-01}  \\
    	01\_01H\_e1       & 79.58                          &  $\left( 2.0 \pm 0.005 \right) \cdot 10^{-5}$   &   $\left( 5.6 \pm 1.4 \right) \cdot 10^{-5}$   &   $\left( 4.6 \pm 1.3 \right) \cdot 10^{-5}$    & \n{1.45E-02} &     \n{3.20E-02}      & \n{8.79E-03} & \n{3.90E-03} & \n{2.69E-01} & \n{3.55E-01}  \\
    	01\_01H\_e3       & 79.58                          &  $\left( 1.5 \pm 0.0003 \right) \cdot 10^{-5}$  &  $\left( 2.2 \pm 0.55 \right) \cdot 10^{-5}$   &   $\left( 3.3 \pm 1.0 \right) \cdot 10^{-5}$    & \n{1.19E-02} &     \n{2.60E-02}      & \n{4.69E-03} & \n{1.85E-03} & \n{1.55E-01} & \n{6.99E-01}  \\
    	01\_01H\_e10      & 79.58                          & $\left( 1.0 \pm 0.00035 \right) \cdot 10^{-5}$  &   $\left( 2.1 \pm 1.4 \right) \cdot 10^{-6}$   &   $\left( 7.9 \pm 4.0 \right) \cdot 10^{-5}$    & \n{7.05E-03} &     \n{1.67E-02}      & \n{2.37E-03} & \n{2.91E-04} & \n{4.13E-02} & \n{5.10E+00}  \\
    	01\_01H\_e30      & 79.58                          &  $\left( 4.3 \pm 0.011 \right) \cdot 10^{-6}$   &   $\left( 4.8 \pm 4.7 \right) \cdot 10^{-7}$   &   $\left( 1.7 \pm 0.62 \right) \cdot 10^{-5}$   & \n{3.72E-03} &     \n{1.33E-02}      & \n{1.20E-03} & \n{1.29E-04} & \n{3.46E-02} & \n{8.94E+00}  \\
    	01\_01H\_e100     & 79.58                          & $\left( 2.9 \pm 0.00087 \right) \cdot 10^{-7}$  &   $\left( 8.4 \pm 2.6 \right) \cdot 10^{-8}$   &   $\left( 2.1 \pm 1.1 \right) \cdot 10^{-6}$    & \n{1.05E-03} &     \n{2.35E-03}      & \n{3.43E-04} & \n{7.94E-05} & \n{7.54E-02} & \n{3.42E+00}  \\
    	01\_01H\_e300     & 31.28                          &   $\left( 6.2 \pm 0.01 \right) \cdot 10^{-9}$   &   $\left( 2.4 \pm 1.4 \right) \cdot 10^{-8}$   &   $\left( 1.2 \pm 0.53 \right) \cdot 10^{-7}$   & \n{2.34E-04} &     \n{4.56E-04}      & \n{1.08E-04} & \n{1.02E-04} & \n{4.35E-01} & \n{2.62E-01}  \\
    	01\_01H\_e1000    & 12.63                          &  $\left( 1.9 \pm 0.0036 \right) \cdot 10^{-8}$  &  $\left( 1.8 \pm 0.73 \right) \cdot 10^{-9}$   &   $\left( 5.8 \pm 3.7 \right) \cdot 10^{-8}$    & \n{2.23E-04} &     \n{5.67E-04}      & \n{1.15E-04} & \n{7.96E-06} & \n{3.58E-02} & \n{1.08E+01}  \\
    	01\_001H\_e0      & 79.58                          &  $\left( -1.6 \pm 0.11 \right) \cdot 10^{-6}$   &   $\left( 1.2 \pm 1.0 \right) \cdot 10^{-5}$   & $\left( 7.1 \pm 1.9e+01 \right) \cdot 10^{-7}$  & \n{1.10E-02} &     \n{5.13E-02}      & \n{1.47E-02} & \n{1.05E-03} & \n{9.60E-02} & \n{-1.35E-01} \\
    	01\_001H\_e0      & 34.82                          &   $\left( 4.7 \pm 0.97 \right) \cdot 10^{-9}$   &   $\left( 6.8 \pm 3.9 \right) \cdot 10^{-8}$   &   $\left( 1.3 \pm 0.32 \right) \cdot 10^{-9}$   & \n{1.74E-04} &     \n{3.86E-02}      & \n{1.18E-02} & \n{3.89E-04} & \n{2.23E+00} & \n{6.98E-02}  \\
    	01\_001H\_e0      & 79.58                          &  $\left( 2.1 \pm 0.0083 \right) \cdot 10^{-5}$  &   $\left( 7.3 \pm 5.8 \right) \cdot 10^{-5}$   &   $\left( 3.9 \pm 3.1 \right) \cdot 10^{-6}$    & \n{1.13E-02} &     \n{3.68E-02}      & \n{1.15E-02} & \n{6.48E-03} & \n{5.73E-01} & \n{2.83E-01}  \\
    	01\_001H\_e0      & 67.73                          &  $\left( 7.4 \pm 0.026 \right) \cdot 10^{-5}$   &   $\left( 4.5 \pm 4.4 \right) \cdot 10^{-5}$   &   $\left( 5.1 \pm 3.3 \right) \cdot 10^{-6}$    & \n{1.68E-02} &     \n{3.83E-02}      & \n{1.08E-02} & \n{2.68E-03} & \n{1.60E-01} & \n{1.64E+00}  \\
    	01\_001H\_e1      & 33.98                          &   $\left( 5.6 \pm 0.07 \right) \cdot 10^{-6}$   &   $\left( 3.1 \pm 2.4 \right) \cdot 10^{-5}$   &   $\left( 6.3 \pm 3.5 \right) \cdot 10^{-6}$    & \n{9.61E-03} &     \n{2.78E-02}      & \n{8.74E-03} & \n{3.25E-03} & \n{3.39E-01} & \n{1.80E-01}  \\
    	01\_001H\_e3      & 34.01                          &  $\left( 8.3 \pm 0.0093 \right) \cdot 10^{-6}$  &  $\left( 1.5 \pm 0.95 \right) \cdot 10^{-5}$   &   $\left( 4.9 \pm 2.2 \right) \cdot 10^{-6}$    & \n{7.17E-03} &     \n{1.66E-02}      & \n{4.61E-03} & \n{2.10E-03} & \n{2.93E-01} & \n{5.50E-01}  \\
    	01\_001H\_e10     & 34.22                          &  $\left( 2.3 \pm 0.0019 \right) \cdot 10^{-6}$  &   $\left( 4.0 \pm 1.2 \right) \cdot 10^{-6}$   &   $\left( 1.9 \pm 0.67 \right) \cdot 10^{-6}$   & \n{3.35E-03} &     \n{7.89E-03}      & \n{1.76E-03} & \n{1.18E-03} & \n{3.53E-01} & \n{5.72E-01}  \\
    	01\_001H\_e30     & 67.85                          &  $\left( 1.1 \pm 0.0032 \right) \cdot 10^{-6}$  &   $\left( 5.3 \pm 4.3 \right) \cdot 10^{-7}$   &   $\left( 1.1 \pm 0.43 \right) \cdot 10^{-6}$   & \n{2.26E-03} &     \n{4.98E-03}      & \n{8.76E-04} & \n{2.34E-04} & \n{1.03E-01} & \n{2.12E+00}  \\
    	01\_001H\_e100    & 34.63                          & $\left( 5.0 \pm 0.00089 \right) \cdot 10^{-7}$  &   $\left( 7.8 \pm 5.6 \right) \cdot 10^{-8}$   &   $\left( 1.8 \pm 0.99 \right) \cdot 10^{-6}$   & \n{1.26E-03} &     \n{3.65E-03}      & \n{4.83E-04} & \n{6.21E-05} & \n{4.93E-02} & \n{6.40E+00}  \\
    	01\_001H\_e300    & 34.83                          & $\left( 1.4 \pm 0.00042 \right) \cdot 10^{-8}$  &  $\left( 1.2 \pm 0.25 \right) \cdot 10^{-8}$   &   $\left( 4.7 \pm 0.89 \right) \cdot 10^{-8}$   & \n{2.49E-04} &     \n{5.33E-04}      & \n{1.41E-04} & \n{4.64E-05} & \n{1.86E-01} & \n{1.18E+00}  \\
    	01\_001H\_e1000   & 34.84                          &  $\left( 7.7 \pm 0.0022 \right) \cdot 10^{-9}$  &  $\left( 2.0 \pm 0.66 \right) \cdot 10^{-9}$   &   $\left( 5.7 \pm 2.1 \right) \cdot 10^{-8}$    & \n{1.25E-04} &     \n{3.67E-04}      & \n{1.14E-04} & \n{1.56E-05} & \n{1.25E-01} & \n{3.93E+00}  \\
    	01\_0001H\_e0     & 14.03                          &   $\left( 2.0 \pm 4.4 \right) \cdot 10^{-8}$    &   $\left( 2.1 \pm 1.7 \right) \cdot 10^{-6}$   &  $\left( 4.2 \pm 2e+01 \right) \cdot 10^{-9}$   & \n{1.95E-03} &     \n{4.68E-02}      & \n{1.45E-02} & \n{1.05E-03} & \n{5.39E-01} & \n{9.79E-03}  \\
    	01\_0001H\_e0     & 3.46                           &  $\left( 1.1 \pm 0.089 \right) \cdot 10^{-6}$   & $\left( 7.9 \pm 1.8e+01 \right) \cdot 10^{-7}$ &   $\left( 1.8 \pm 7.1 \right) \cdot 10^{-9}$    & \n{3.54E-03} &     \n{4.18E-02}      & \n{1.31E-02} & \n{2.23E-04} & \n{6.29E-02} & \n{1.40E+00}  \\
    	01\_0001H\_e0     & 14.08                          &   $\left( 9.6 \pm 0.21 \right) \cdot 10^{-6}$   &   $\left( 3.8 \pm 1.1 \right) \cdot 10^{-6}$   & $\left( 3.1 \pm 9.9e+01 \right) \cdot 10^{-10}$ & \n{2.60E-03} &     \n{4.66E-02}      & \n{1.03E-02} & \n{1.47E-03} & \n{5.65E-01} & \n{2.50E+00}  \\
    	01\_0001H\_e0     & 7.42                           &   $\left( 7.5 \pm 0.16 \right) \cdot 10^{-6}$   &   $\left( 3.5 \pm 1.2 \right) \cdot 10^{-6}$   & $\left( 3.4 \pm 8.8e+01 \right) \cdot 10^{-10}$ & \n{2.41E-03} &     \n{3.52E-02}      & \n{9.92E-03} & \n{1.47E-03} & \n{6.09E-01} & \n{2.12E+00}  \\
    	01\_0001H\_e1     & 7.66                           &  $\left( 6.6 \pm 0.052 \right) \cdot 10^{-6}$   &  $\left( 2.9 \pm 0.88 \right) \cdot 10^{-6}$   & $\left( 4.4 \pm 8.3e+01 \right) \cdot 10^{-10}$ & \n{3.02E-03} &     \n{3.31E-02}      & \n{9.22E-03} & \n{9.74E-04} & \n{3.22E-01} & \n{2.26E+00}  \\
    	01\_0001H\_e3     & 8.96                           &  $\left( 1.8 \pm 0.016 \right) \cdot 10^{-5}$   &   $\left( 5.9 \pm 1.2 \right) \cdot 10^{-6}$   & $\left( 2.0 \pm 1.4e+01 \right) \cdot 10^{-10}$ & \n{4.04E-03} &     \n{3.01E-02}      & \n{5.31E-03} & \n{1.47E-03} & \n{3.64E-01} & \n{3.07E+00}  \\
    	01\_0001H\_e10    & 3.46                           &  $\left( 1.6 \pm 0.0077 \right) \cdot 10^{-6}$  &   $\left( 1.4 \pm 1.7 \right) \cdot 10^{-6}$   &   $\left( 1.2 \pm 0.56 \right) \cdot 10^{-7}$   & \n{2.72E-03} &     \n{6.08E-03}      & \n{1.75E-03} & \n{5.22E-04} & \n{1.92E-01} & \n{1.14E+00}  \\
    	01\_0001H\_e30    & 3.47                           &  $\left( 9.0 \pm 0.042 \right) \cdot 10^{-8}$   &   $\left( 3.2 \pm 1.8 \right) \cdot 10^{-7}$   &   $\left( 5.8 \pm 1.8 \right) \cdot 10^{-8}$    & \n{8.92E-04} &     \n{2.46E-03}      & \n{6.42E-04} & \n{3.58E-04} & \n{4.01E-01} & \n{2.80E-01}  \\
    	01\_0001H\_e100   & 3.48                           &   $\left( 7.6 \pm 0.04 \right) \cdot 10^{-8}$   &   $\left( 8.2 \pm 3.5 \right) \cdot 10^{-8}$   &   $\left( 2.3 \pm 0.93 \right) \cdot 10^{-8}$   & \n{5.10E-04} &     \n{1.25E-03}      & \n{2.79E-04} & \n{1.61E-04} & \n{3.16E-01} & \n{9.24E-01}  \\
    	01\_0001H\_e300   & 3.48                           &  $\left( 6.1 \pm 0.0017 \right) \cdot 10^{-9}$  &   $\left( 8.3 \pm 1.8 \right) \cdot 10^{-9}$   &   $\left( 3.0 \pm 0.92 \right) \cdot 10^{-9}$   & \n{1.59E-04} &     \n{3.64E-04}      & \n{1.30E-04} & \n{5.23E-05} & \n{3.30E-01} & \n{7.38E-01}  \\
    	01\_0001H\_e1000  & 6.93                           & $\left( 4.9 \pm 0.0033 \right) \cdot 10^{-10}$  &  $\left( 5.9 \pm 2.8 \right) \cdot 10^{-10}$   &   $\left( 4.1 \pm 1.5 \right) \cdot 10^{-10}$   & \n{6.16E-05} &     \n{1.59E-04}      & \n{9.29E-05} & \n{9.65E-06} & \n{1.57E-01} & \n{8.28E-01}  \\ \bottomrule
    	                  &                                &                                                 &                                                &                                                 &              &                       &              &              &              &               \\
    	Name              & $\Tmax\left[2\pi\Omega\right]$ &             $\alpha\pm\Delta\alpha$             &              $\dx \pm \Delta\dx$               &               $\dz \pm \Delta\dz$               &    $\sg$     & $\urms\pm\Delta\urms$ &    $\zg$     &    $\lc$     &  $\tauCorr$  &     $\Sc$     \\ \midrule
    	001\_01H\_e0      & 159.15                         &  $\left( 4.2 \pm 0.0032 \right) \cdot 10^{-8}$  &   $\left( 6.4 \pm 1.2 \right) \cdot 10^{-7}$   &   $\left( 9.9 \pm 2.4 \right) \cdot 10^{-7}$    & \n{1.12E-03} &     \n{4.55E-02}      & \n{1.14E-02} & \n{5.74E-04} & \n{5.14E-01} & \n{6.61E-02}  \\
    	001\_01H\_e0      & 0.36                           &  $\left( 2.2 \pm 0.0058 \right) \cdot 10^{-7}$  &  $\left( 1.5 \pm 0.36 \right) \cdot 10^{-6}$   &   $\left( 5.0 \pm 3.9 \right) \cdot 10^{-8}$    & \n{1.99E-03} &     \n{3.85E-02}      & \n{9.63E-03} & \n{7.63E-04} & \n{3.83E-01} & \n{1.43E-01}  \\
    	001\_01H\_e0      & 159.15                         &  $\left( 4.2 \pm 0.0076 \right) \cdot 10^{-7}$  &  $\left( 1.9 \pm 0.51 \right) \cdot 10^{-6}$   &   $\left( 4.3 \pm 1.2 \right) \cdot 10^{-6}$    & \n{2.61E-03} &     \n{3.34E-02}      & \n{8.35E-03} & \n{7.36E-04} & \n{2.82E-01} & \n{2.18E-01}  \\
    	001\_01H\_e0      & 159.15                         &  $\left( 6.3 \pm 0.0029 \right) \cdot 10^{-7}$  &  $\left( 2.2 \pm 0.65 \right) \cdot 10^{-6}$   &   $\left( 7.8 \pm 2.2 \right) \cdot 10^{-6}$    & \n{3.16E-03} &     \n{2.80E-02}      & \n{6.97E-03} & \n{6.95E-04} & \n{2.20E-01} & \n{2.89E-01}  \\
    	001\_01H\_e1      & 159.15                         &  $\left( 6.5 \pm 0.0037 \right) \cdot 10^{-7}$  &  $\left( 2.2 \pm 0.66 \right) \cdot 10^{-6}$   &   $\left( 8.1 \pm 2.1 \right) \cdot 10^{-6}$    & \n{3.07E-03} &     \n{2.52E-02}      & \n{6.28E-03} & \n{7.07E-04} & \n{2.30E-01} & \n{2.99E-01}  \\
    	001\_01H\_e3      & 159.15                         &  $\left( 4.8 \pm 0.0014 \right) \cdot 10^{-7}$  &  $\left( 1.5 \pm 0.43 \right) \cdot 10^{-6}$   &   $\left( 8.0 \pm 2.8 \right) \cdot 10^{-6}$    & \n{2.44E-03} &     \n{1.28E-02}      & \n{3.16E-03} & \n{6.06E-04} & \n{2.48E-01} & \n{3.22E-01}  \\
    	001\_01H\_e10     & 15.92                          & $\left( 1.7 \pm 0.00053 \right) \cdot 10^{-7}$  &   $\left( 5.4 \pm 1.0 \right) \cdot 10^{-7}$   &   $\left( 9.5 \pm 1.5 \right) \cdot 10^{-7}$    & \n{1.32E-03} &     \n{4.84E-03}      & \n{1.17E-03} & \n{4.13E-04} & \n{3.14E-01} & \n{3.04E-01}  \\
    	001\_01H\_e30     & 65.45                          & $\left( 5.6 \pm 0.00057 \right) \cdot 10^{-8}$  &  $\left( 1.7 \pm 0.28 \right) \cdot 10^{-7}$   &   $\left( 2.9 \pm 0.47 \right) \cdot 10^{-7}$   & \n{6.69E-04} &     \n{1.90E-03}      & \n{4.40E-04} & \n{2.59E-04} & \n{3.87E-01} & \n{3.25E-01}  \\
    	001\_01H\_e100    & 30.81                          & $\left( 2.5 \pm 0.00012 \right) \cdot 10^{-8}$  &   $\left( 6.4 \pm 1.2 \right) \cdot 10^{-8}$   &   $\left( 8.1 \pm 1.3 \right) \cdot 10^{-8}$    & \n{3.44E-04} &     \n{7.74E-04}      & \n{1.50E-04} & \n{1.86E-04} & \n{5.40E-01} & \n{3.89E-01}  \\
    	001\_01H\_e300    & 7.36                           & $\left( 1.5 \pm 2.4e-05 \right) \cdot 10^{-8}$  &  $\left( 1.3 \pm 0.39 \right) \cdot 10^{-8}$   &   $\left( 3.7 \pm 1.3 \right) \cdot 10^{-8}$    & \n{1.86E-04} &     \n{4.47E-04}      & \n{6.27E-05} & \n{7.20E-05} & \n{3.87E-01} & \n{1.14E+00}  \\
    	001\_01H\_e1000   & 18.92                          & $\left( 4.8 \pm 0.00088 \right) \cdot 10^{-9}$  &  $\left( 2.4 \pm 0.87 \right) \cdot 10^{-9}$   &   $\left( 4.2 \pm 2.1 \right) \cdot 10^{-8}$    & \n{1.04E-04} &     \n{2.35E-04}      & \n{3.08E-05} & \n{2.35E-05} & \n{2.26E-01} & \n{1.99E+00}  \\
    	001\_001H\_e0     & 57.23                          &  $\left( 1.4 \pm 0.0026 \right) \cdot 10^{-6}$  &   $\left( 7.1 \pm 5.7 \right) \cdot 10^{-6}$   &   $\left( 7.1 \pm 7.3 \right) \cdot 10^{-6}$    & \n{6.14E-03} &     \n{4.60E-02}      & \n{1.15E-02} & \n{1.16E-03} & \n{1.89E-01} & \n{1.99E-01}  \\
    	001\_001H\_e0     & 57.55                          & $\left( -1.0 \pm 0.0036 \right) \cdot 10^{-6}$  &   $\left( 1.1 \pm 0.8 \right) \cdot 10^{-5}$   &   $\left( 1.4 \pm 1.3 \right) \cdot 10^{-5}$    & \n{1.09E-02} &     \n{4.03E-02}      & \n{9.95E-03} & \n{9.81E-04} & \n{9.01E-02} & \n{-9.62E-02} \\
    	001\_001H\_e0     & 57.8                           &  $\left( 3.8 \pm 0.014 \right) \cdot 10^{-6}$   &  $\left( 1.1 \pm 0.71 \right) \cdot 10^{-5}$   &   $\left( 1.4 \pm 1.3 \right) \cdot 10^{-5}$    & \n{1.04E-02} &     \n{3.52E-02}      & \n{8.76E-03} & \n{1.01E-03} & \n{9.66E-02} & \n{3.64E-01}  \\
    	001\_001H\_e0     & 57.48                          &  $\left( 2.2 \pm 0.0075 \right) \cdot 10^{-6}$  &   $\left( 8.1 \pm 3.8 \right) \cdot 10^{-6}$   &   $\left( 8.5 \pm 5.6 \right) \cdot 10^{-6}$    & \n{7.54E-03} &     \n{2.89E-02}      & \n{7.24E-03} & \n{1.07E-03} & \n{1.41E-01} & \n{2.78E-01}  \\
    	001\_001H\_e1     & 58.52                          &  $\left( 1.9 \pm 0.0009 \right) \cdot 10^{-6}$  &   $\left( 7.3 \pm 3.3 \right) \cdot 10^{-6}$   &   $\left( 8.9 \pm 5.1 \right) \cdot 10^{-6}$    & \n{6.54E-03} &     \n{2.61E-02}      & \n{6.49E-03} & \n{1.12E-03} & \n{1.72E-01} & \n{2.54E-01}  \\
    	001\_001H\_e3     & 60.94                          &  $\left( 5.4 \pm 0.0019 \right) \cdot 10^{-7}$  &  $\left( 2.2 \pm 0.72 \right) \cdot 10^{-6}$   &   $\left( 3.2 \pm 1.2 \right) \cdot 10^{-6}$    & \n{3.69E-03} &     \n{1.31E-02}      & \n{3.24E-03} & \n{6.07E-04} & \n{1.64E-01} & \n{2.39E-01}  \\
    	001\_001H\_e10    & 62.06                          &  $\left( 6.2 \pm 0.0074 \right) \cdot 10^{-8}$  &  $\left( 2.5 \pm 0.69 \right) \cdot 10^{-7}$   &   $\left( 2.5 \pm 0.88 \right) \cdot 10^{-6}$   & \n{1.46E-03} &     \n{4.77E-03}      & \n{1.18E-03} & \n{1.71E-04} & \n{1.18E-01} & \n{2.49E-01}  \\
    	001\_001H\_e30    & 62.45                          & $\left( 1.1 \pm 0.00027 \right) \cdot 10^{-8}$  &  $\left( 3.8 \pm 0.81 \right) \cdot 10^{-8}$   &   $\left( 5.8 \pm 1.9 \right) \cdot 10^{-7}$    & \n{5.39E-04} &     \n{1.74E-03}      & \n{4.30E-04} & \n{7.12E-05} & \n{1.32E-01} & \n{2.93E-01}  \\
    	001\_001H\_e100   & 34.29                          & $\left( 2.3 \pm 0.00064 \right) \cdot 10^{-9}$  &   $\left( 7.6 \pm 1.1 \right) \cdot 10^{-9}$   &   $\left( 1.8 \pm 0.31 \right) \cdot 10^{-8}$   & \n{1.61E-04} &     \n{5.40E-04}      & \n{1.30E-04} & \n{4.74E-05} & \n{2.95E-01} & \n{3.07E-01}  \\
    	001\_001H\_e300   & 32.53                          &  $\left( 7.2 \pm 0.012 \right) \cdot 10^{-10}$  &   $\left( 1.9 \pm 1.0 \right) \cdot 10^{-9}$   &   $\left( 3.9 \pm 1.9 \right) \cdot 10^{-7}$    & \n{1.12E-04} &     \n{2.19E-04}      & \n{4.87E-05} & \n{1.71E-05} & \n{1.52E-01} & \n{3.77E-01}  \\
    	001\_001H\_e1000  & 2.86                           & $\left( 3.9 \pm 0.00021 \right) \cdot 10^{-10}$ &  $\left( 5.4 \pm 1.4 \right) \cdot 10^{-10}$   &   $\left( 7.3 \pm 2.4 \right) \cdot 10^{-10}$   & \n{3.76E-05} &     \n{8.61E-05}      & \n{1.79E-05} & \n{1.43E-05} & \n{3.79E-01} & \n{7.26E-01}  \\
    	001\_0001H\_e0    & 7.82                           &  $\left( 1.1 \pm 0.023 \right) \cdot 10^{-6}$   &   $\left( 3.1 \pm 4.5 \right) \cdot 10^{-6}$   &  $\left( 8.2 \pm 1e+01 \right) \cdot 10^{-7}$   & \n{5.20E-03} &     \n{4.58E-02}      & \n{1.15E-02} & \n{6.02E-04} & \n{1.16E-01} & \n{3.54E-01}  \\
    	001\_0001H\_e0    & 8.77                           &  $\left( 4.8 \pm 0.022 \right) \cdot 10^{-7}$   &   $\left( 2.0 \pm 2.1 \right) \cdot 10^{-6}$   &   $\left( 6.4 \pm 5.3 \right) \cdot 10^{-7}$    & \n{4.14E-03} &     \n{3.87E-02}      & \n{9.72E-03} & \n{4.75E-04} & \n{1.15E-01} & \n{2.46E-01}  \\
    	001\_0001H\_e0    & 5.7                            &  $\left( 1.5 \pm 0.033 \right) \cdot 10^{-7}$   &   $\left( 1.6 \pm 1.7 \right) \cdot 10^{-6}$   &   $\left( 4.6 \pm 3.6 \right) \cdot 10^{-7}$    & \n{3.68E-03} &     \n{3.36E-02}      & \n{8.43E-03} & \n{4.25E-04} & \n{1.16E-01} & \n{9.51E-02}  \\
    	001\_0001H\_e0    & 5.83                           &  $\left( 1.4 \pm 0.0048 \right) \cdot 10^{-6}$  &   $\left( 1.3 \pm 1.2 \right) \cdot 10^{-6}$   &   $\left( 3.2 \pm 1.8 \right) \cdot 10^{-7}$    & \n{4.81E-03} &     \n{2.83E-02}      & \n{7.09E-03} & \n{2.80E-04} & \n{5.83E-02} & \n{1.03E+00}  \\
    	001\_0001H\_e1    & 5.87                           &   $\left( 8.5 \pm 0.1 \right) \cdot 10^{-8}$    &   $\left( 1.5 \pm 2.3 \right) \cdot 10^{-6}$   &   $\left( 3.7 \pm 2.2 \right) \cdot 10^{-7}$    & \n{3.36E-03} &     \n{2.52E-02}      & \n{6.35E-03} & \n{4.43E-04} & \n{1.32E-01} & \n{5.69E-02}  \\
    	001\_0001H\_e3    & 6.18                           &  $\left( 8.6 \pm 0.056 \right) \cdot 10^{-8}$   &   $\left( 6.9 \pm 5.0 \right) \cdot 10^{-7}$   &   $\left( 1.8 \pm 0.83 \right) \cdot 10^{-7}$   & \n{2.06E-03} &     \n{1.27E-02}      & \n{3.16E-03} & \n{3.34E-04} & \n{1.62E-01} & \n{1.25E-01}  \\
    	001\_0001H\_e10   & 6.3                            &  $\left( 3.5 \pm 0.051 \right) \cdot 10^{-8}$   &   $\left( 3.7 \pm 2.2 \right) \cdot 10^{-7}$   &   $\left( 9.6 \pm 4.5 \right) \cdot 10^{-8}$    & \n{1.19E-03} &     \n{4.79E-03}      & \n{1.18E-03} & \n{3.09E-04} & \n{2.60E-01} & \n{9.53E-02}  \\
    	001\_0001H\_e30   & 6.46                           &  $\left( 2.3 \pm 0.0063 \right) \cdot 10^{-8}$  &   $\left( 6.3 \pm 2.1 \right) \cdot 10^{-8}$   &   $\left( 6.6 \pm 3.3 \right) \cdot 10^{-8}$    & \n{5.90E-04} &     \n{1.80E-03}      & \n{4.35E-04} & \n{1.07E-04} & \n{1.81E-01} & \n{3.71E-01}  \\
    	001\_0001H\_e100  & 6.49                           &  $\left( 1.9 \pm 0.0019 \right) \cdot 10^{-9}$  &   $\left( 6.1 \pm 1.6 \right) \cdot 10^{-9}$   &   $\left( 9.5 \pm 2.2 \right) \cdot 10^{-9}$    & \n{1.74E-04} &     \n{5.38E-04}      & \n{1.33E-04} & \n{3.50E-05} & \n{2.01E-01} & \n{3.11E-01}  \\
    	001\_0001H\_e300  & 6.61                           & $\left( 1.0 \pm 0.00012 \right) \cdot 10^{-9}$  &  $\left( 3.2 \pm 0.42 \right) \cdot 10^{-9}$   &   $\left( 2.8 \pm 0.56 \right) \cdot 10^{-9}$   & \n{1.05E-04} &     \n{2.16E-04}      & \n{5.25E-05} & \n{3.05E-05} & \n{2.91E-01} & \n{3.20E-01}  \\
    	001\_0001H\_e1000 & 6.47                           & $\left( 2.9 \pm 0.0011 \right) \cdot 10^{-10}$  &  $\left( 6.5 \pm 1.3 \right) \cdot 10^{-10}$   &   $\left( 7.2 \pm 1.8 \right) \cdot 10^{-10}$   & \n{4.76E-05} &     \n{9.09E-05}      & \n{2.23E-05} & \n{1.37E-05} & \n{2.88E-01} & \n{4.39E-01}  \\ \bottomrule
    \end{tabular}%
    \caption{2-d simulations in $r$-$z$ extent. Simulation name is constructed from $\stokes$ number, domain size $Lx,z$ and dust-to-gas ratio $\epsInit$.}
    \label{tab:r-z-runs}
\end{table*}%
\end{document}